\documentclass[longauth]{aa} 

\usepackage{ amssymb }
\usepackage{graphicx}
\usepackage[colorlinks=true, allcolors=blue]{hyperref}
\usepackage{multirow}
\usepackage[switch]{lineno}
\usepackage{txfonts}
\begin{document}

   \title{Very-high-energy gamma-ray intranight variability from BL~Lacertae during the extreme flaring state of 2022}

   \titlerunning{VHE gamma-ray intranight variability from BL Lac}

\author{
\fontsize{8.45}{8.45}\selectfont
K.~Abe\inst{1} \and
S.~Abe\inst{2} \and
A.~Abhishek\inst{3} \and
F.~Acero\inst{4,5} \and
A.~Aguasca-Cabot\inst{6} \and
I.~Agudo\inst{7} \and
C.~Alispach\inst{8} \and
D.~Ambrosino\inst{9} \and
F.~Ambrosino\inst{10} \and
L.~A.~Antonelli\inst{10}
% ...
}
\institute{
Department of Physics, Tokai University, 4-1-1, Kita-Kaname, Hiratsuka, Kanagawa 259-1292, Japan
\and Institute for Cosmic Ray Research, University of Tokyo, 5-1-5, Kashiwa-no-ha, Kashiwa, Chiba 277-8582, Japan
\and INFN and Università degli Studi di Siena, Dipartimento di Scienze Fisiche, della Terra e dell'Ambiente (DSFTA), Sezione di Fisica, Via Roma 56, 53100 Siena, Italy
\and Université Paris-Saclay, Université Paris Cité, CEA, CNRS, AIM, F-91191 Gif-sur-Yvette Cedex, France
\and FSLAC IRL 2009, CNRS/IAC, La Laguna, Tenerife, Spain
\and Departament de Física Quàntica i Astrofísica, Institut de Ciències del Cosmos, Universitat de Barcelona, IEEC-UB, Martí i Franquès, 1, 08028, Barcelona, Spain
\and Instituto de Astrofísica de Andalucía-CSIC, Glorieta de la Astronomía s/n, 18008, Granada, Spain
\and Department of Astronomy, University of Geneva, Chemin d'Ecogia 16, CH-1290 Versoix, Switzerland
\and INFN Sezione di Napoli, Via Cintia, ed. G, 80126 Napoli, Italy
\and INAF - Osservatorio Astronomico di Roma, Via di Frascati 33, 00040, Monteporzio Catone, Italy
% ...
}

   \date{Received September 15, 1996; accepted March 16, 1997}

  \abstract
   {BL Lacertae (BL Lac), the archetypal blazar of its subclass and one of the most studied blazars of recent decades, has undergone a series of major multi-wavelength outbursts since 2020, resulting in its highest {recorded} $\gamma$-ray flare to date between September and November 2022, together with those from August 2021 and October 2024.}
   {We characterised the $\gamma$-ray and multi-wavelength emission and spectral energy distribution (SED) of BL Lac, as well as their evolution during the major and extended $\gamma$-ray and multi-wavelength flare that occurred between September and November 2022.}
   {We evaluated the variability of the flare, focusing on {the nights of} October 20 and November 13, 2022 when clear intranight very-high-energy (VHE, $E>100$~GeV) $\gamma$-ray variability was observed. We modelled the $\gamma$-ray and broadband SEDs during periods of stable emission identified with a Bayesian block analysis and interpreted the flare's evolution in terms of the variability in the relativistic particles and the jet's physical parameters.}
   {During this flare, the VHE emission shows an average flux of 0.23 Crab units (C.U.) above 200 GeV and a variability amplitude of more than a factor of ten. We observe intranight flux-doubling variations as fast as $\sim$8 minutes during {the nights of October 20 and November 13, 2022 with maximum fluxes of 4.4 C.U. above 100 GeV and 2.8 C.U. above 200 GeV}. The spectral analysis reveals a transition of the X-ray emission from the high- to the low-energy SED peak and a shift in the $\gamma$-ray peak towards higher energies. We interpret the broadband emission within a leptonic two-zone model in which intranight variability is explained as magnetic reconnection in a compact region closely orientated with the line of sight, while variations in the relativistic electron distributions and the injection of freshly accelerated particles explain the weekly scale variations. }
   {}


\maketitle

\section{Introduction}
Relativistic jets launched from blazars, which are active galactic nuclei (AGNs) with their jets orientated towards our line of sight, are extremely efficient and natural particle accelerators that emit across the entire the electromagnetic spectrum. In fact, they are among the few sources that can emit up to TeV energies. Blazars are also highly variable on all timescales. Depending on the properties of their optical spectrum \citep[absence or presence of broad emission lines; see][]{stickel1991}, they can be classified into two main sub-classes: BL Lac {type} objects and flat spectrum radio quasars (FSRQs) \citep{urry1995}.

The broadband spectral energy distribution (SED) of blazars, extending from radio to $\gamma$-ray energies, has a characteristic double-peak shape. The low-energy {component} is produced by synchrotron radiation \citep[see][]{1981ApJ...243..700K}. The mechanisms invoked to explain the high-energy {component} fall {into two groups}. An interpretation based on leptonic processes explains this emission through inverse Compton (IC) scattering with low-energy photons, either from synchrotron radiation via synchrotron self Compton \citep[SSC; see][]{1981ApJ...243..700K, 1992ApJ...397L...5M} scattering or from radiation from outside the jet via external Compton \citep[EC; see][]{1994ApJS...90..945D} scattering. However, hadronic-based interpretations are also used in some cases and rely on processes such as proton synchrotron radiation or photo-pion production to explain the high-energy emission of blazars \citep[see, e.g.][]{mannheim1992,aharonian2000}.

BL Lacertae (hereafter BL Lac), the {archetype} of the BL Lac blazar subclass located at redshift $z=0.069$ \citep{miller1978}, historically shows remarkable variability with several episodes of minute- {to} hour-scale variations \citep[see, for instance][]{abeysekara2018,magic2019,jorstad2022}, making it one of the most monitored and studied blazars to date. It is classified as an intermediate-synchrotron-peaked (ISP) blazar according to the peak frequency of its synchrotron emission ($10^{14}~\text{Hz} < \nu_{\text{syn}} < 10^{15}~\text{Hz}$). However, it is known to transition between low-synchrotron-peaked (LSP; $\nu_{\text{syn}} < 10^{14}~\text{Hz}$) and ISP states, depending on its activity \citep[see, e.g.
][]{ackermann2011,nilsson2018}. It has been detected several times in the very-high-energy (VHE; $E>100$~GeV) $\gamma$-ray band, always during {multi-wavelength} flaring emission states \citep{albert2007,arlen2013}. 
These periods of variability on very short timescales imply very compact emitting regions involving the most energetic particle acceleration processes, which in some cases challenge the theoretical models used to interpret the broadband emission of blazars \citep{henri2006}.

Here we report the results of an extreme flare observed from BL Lac by the Large-Sized Telescope prototype (LST-1), which took place between September and November 2022  and was triggered after the detection of enhanced activity by the \textit{Fermi} satellite \citep{lamura2022}. 
{This flare is one of the three brightest events ever detected in VHEs from BL Lac, together with similar events that occurred in 2021 and 2024. It is also the longest flaring period ever recorded for this source in the VHE domain, lasting about three months, with significant variations on day-, hour-, and minute-timescales.  Several multi-wavelength facilities also covered this remarkable flare, providing a rich dataset for interpreting the multi-wavelength emission and variability of BL Lac during this period (see Sect. \ref{sec2}). In this work, we evaluate and interpret the VHE $\gamma$-ray and broadband evolution of the flare. We characterise the minute-scale $\gamma$-ray variability observed during this period (Sect. \ref{sec4}), which sets important constraints on the nature of the emitting region. We evaluate the spectral behaviour of the source (Sect. \ref{sec5}), and model its broadband emission within AGN radiative models (Sect. \ref{sec6}). We discuss and summarise the main results in Sects.~\ref{sec7} and~\ref{sec8}.

\section{Observations and data reduction}\label{sec2}
                            
\subsection{Very-high-energy $\gamma$-ray data: LST-1}
The LST-1 is the first of the four Large-Sized Telescopes to be constructed at the northern site of the Cherenkov Telescope Array Observatory \citep[CTAO;][]{acharyya2019}, located at the Roque de los Muchachos Observatory on the Canary Island of La Palma (Spain). The LST-1 is currently performing commissioning and scientific observations, following the completion of its construction in October 2018. Its large collection area (23 m diameter mirror) and highly sensitive camera, composed of 1855 high quantum efficiency photomultiplier tubes, allows detection of faint Cherenkov light flashes {from $\gamma$-ray showers down to energies} $\simeq$20~GeV \citep[see][]{abe2023}. 

The LST-1 observed BL Lac for 23 nights between 21 September and 25 November 2022 in wobble observing mode \citep{fomin1994}, providing $\sim$36 hours of data after standard data quality cuts in a zenith range between 10$^{\circ}$ and 60$^{\circ}$. We analysed raw LST-1 data with version v0.10.18 of the \texttt{cta-lstchain} software \citep{lopez-coto2022,moralejo2025}, including data cleaning, image parametrisation, $\gamma$-hadron separation, and reconstruction of the direction and energy of the recorded events. 
Among this dataset, $\sim$29~hours were taken under dark-sky  conditions {(no increased background light contamination from the Moon)}, using standard tail-cut cleaning with}
thresholds of 8 and 4 photoelectrons (p.e.) for the core and neighbour pixels of the Cherenkov shower image {and an intensity cut of 50 p.e.} \citep{abe2023}. The remaining $\sim$7 hours of observation covered different moon brightness levels. 
These data were analysed with cleaning thresholds {ranging from} 10 to 16 p.e.  for core pixels and 5 to 8 p.e. for neighbour pixels, and with increased intensity cuts between 63 and 194 {p.e.} {due to noise introduced by the~Moon}, extracted for each night-sky background (NSB) level. The event classification used a Random Forest trained using Monte Carlo (MC) simulations of proton- and $\gamma$-like events at a declination of 34.74$^{\circ}$, close to that of BL~Lac {(42.28$^{\circ}$)}. We produced MC simulations to match the NSB level of~our data, accounting for additional noise relative to dark observations. 
For event selection, we applied energy-dependent gammaness {(gamma-likeness score)}
and $\theta$ ({angular distance} of the detected event {with respect to the nominal source position}) cuts, optimised to retain 70\% of the {simulated MC} $\gamma$-ray events.

We performed the high-level analysis with version 1.2 of \texttt{Gammapy} \citep{donath2023,aguasca-Cabot2023,gammapy:zenodo-1.2}. We characterised the background using an OFF region with a wobble angle of 0.4$^{\circ}$ and at a reflected position of 180$^{\circ}$ from BL Lac. We also applied a {limit} on the effective area $A_{eff}$ >5\% of the maximum area. We calculated the statistical significance following Eq. (17) from \cite{li1983}. 
We estimated the light-curve energy threshold to be $\sim$175~GeV, calculated from the {maximum of the expected event distribution from the} MC simulations and weighted with BL Lac's spectrum for the highest NSB level and zenith distance of $40- 50$$^{\circ}$ to match the data. Therefore, we used a minimum energy of 200 GeV {for light curve reconstruction}. We further verified the energy threshold on the data against background asymmetries in the field of view (FoV). We observed a non-uniformity in the event distribution, particularly along the altitude axis. This effect increases as the telescope pointing is further from the zenith and becomes clearly visible at zenith distances >40$^{\circ}$. Examining the energy distribution of events in the source region and the OFF regions, we observe clear differences in the energy threshold for some observations. Although these differences occur mainly at energies below our analysis threshold, we also performed the spectral analysis using background counts estimated with the \texttt{BAccMod} package\footnote{\url{https://github.com/mdebony/BAccMod}}. This package extracts the background distribution using the full FoV after removing the source region and can thus account for asymmetries. As expected, we observe differences in the extracted spectra {below} 200 GeV for some time periods, but the spectra agree {above} 200 GeV, validating our results at higher energies. We show the VHE $\gamma$-ray light curve in the top panel of Fig.~\ref{fig:mwl_lcs}. We highlight October 20 and November 13, 2022 as two remarkably bright nights, when significances of $\sim$49$\sigma$ and $\sim$30$\sigma$ were reached in only 2.5 and 1.5~hours of observation, respectively. We evaluated the minute-scale variability of these two nights in Sect.~\ref{sec4}.
For October 20, consisting only of {dark-sky} data, we adopted a threshold of 100~GeV, obtained from the {dark-sky} MC simulations at zenith angles $\sim$20-40$^{\circ}$. 
We characterised the spectrum and SED with a log-parabola model, including a term for extragalactic background light (EBL) absorption using the model from \cite{saldana-lopez2021}. We present the spectral analysis in detail in Sect.~\ref{sec5}.

\subsection{High-energy $\gamma$-ray data: \textit{Fermi}-LAT}
We analysed high-energy (HE, $E>100$~MeV) data taken by the pair-conversion Large Area Telescope (LAT) on board the \textit{Fermi} satellite. The LAT operates in sky-survey mode over an energy range of 20~MeV and $\sim$1~TeV, with its best sensitivity {around} a few GeV \citep{atwood2009}. We defined a 20$^{\circ}$-radius region of interest (ROI) centred on BL Lac. We downloaded all data from 21 September to 30 November, 
selecting the Pass8 \texttt{P8R3\_SOURCE\_V3} events between 100~MeV and 300~GeV using the \textsc{Fermitools} standard software. We applied a zenith angle cut of 90$^{\circ}$ to minimise contamination from the Earth's limb. We also applied the recommended versions of the Galactic diffuse emission model and the isotropic component for event selection\footnote{\url{https://fermi.gsfc.nasa.gov/ssc/data/access/lat/BackgroundModels.html}}.

To account for emission from sources near BL Lac, we derived a model that includes all sources within the ROI, plus all sources located in an additional annular region of radius 10$^{\circ}$. We constructed this model using one year of data 
and a binned likelihood analysis, using the second release of the \textit{Fermi}-LAT Fourth Source Catalogue (4FGL-DR2) as a reference \citep{abdollahi2020,ballet2020}. We left the spectral parameters (index and normalisation) of all variable sources within the ROI and with a test statistics {\citep[TS;][]{mattox1996}} >25 ($\sim$5$\sigma$ significance) as free parameters, based on the 4FGL catalogue of LAT variability. We fixed the sources' parameters not fulfilling these conditions to the catalogue values. We also left the normalisation of the diffuse components free. The final model converged after removing all sources with TS<4 (<2$\sigma$).

We obtained the HE $\gamma$-ray light curve by analysing the data within the chosen time window using an unbinned likelihood analysis of events in 12-hour bins. For each bin, we assumed a power-law spectral shape for BL~Lac, leaving the spectral index and normalisation free. We fixed the spectral parameters of the other sources to the model value, except for two highly variable and bright sources with significant detections (TS>25) located within <15$^{\circ}$ of BL Lac (4FGL J2244.2+4057 and 4FGL J2311.0+3425).  Fig.~\ref{fig:mwl_lcs} shows the HE $\gamma$-ray light curve.

\subsection{X-ray data: \textit{Swift}-XRT}
\label{subsec:XRT_analysis}
The scientific payload on board the \textit{Neil Gehrels Swift} observatory includes the X-Ray Telescope \citep[XRT;][]{Burrows2005}, and the Optical/UV Telescope \citep[UVOT;][]{Roming2005}.
In standard operation mode, these instruments perform simultaneous observations on the same targets.
In this project, we included 50 \textit{Swift}-XRT and \textit{Swift}-UVOT observations of BL Lac, carried out between March 2022 and January 2023.

\textit{Swift}-XRT is sensitive to X-rays in the $0.2-10\,\text{keV}$ energy range. We downloaded the reduced XRT data from the UK Swift Science Data Centre \citep{Evans2009}.
We performed the spectral analysis of the reduced data using XSPEC v12.13.1 package through \texttt{PyXspec} v2.1.2. 
We analysed the XRT data in the 0.3 to 10 keV energy range.
We primarily modelled the intrinsic source emission in this range with a power-law function, and to test for intrinsic spectral curvature associated with the transition between the synchrotron and IC bumps, we also employed a log-parabola model.
We fitted the observed spectra with the intrinsic emission models folded with the {\fontfamily{cmtt}\selectfont tbabs}\footnote{\url{https://heasarc.gsfc.nasa.gov/xanadu/xspec/manual/node273.html}} Galactic absorption term, as implemented in XSPEC.
Following \cite{Raiteri2009}, we fixed the neutral hydrogen column density toward BL Lac at $n_{\text{H}}=3.4\times 10^{21}\,\text{cm$^{-2}$}$ and used this value to correct the X-ray fluxes and SEDs for Galactic extinction effects.
We employed the $F-$test to assess whether the log-parabola model is preferred over the power-law model, using a $3\sigma$ significance level threshold.
We summarise the results of the X-ray analysis in Appendix~\ref{appendix_c}.

\subsection{Ultraviolet data: \textit{Swift}-UVOT}
\label{sec2.4}

We analysed the UVOT data using the {\fontfamily{cmtt}\selectfont uvotimsum} and {\fontfamily{cmtt}\selectfont uvotsource} tasks within the HEAsoft 6.32 package, along with the latest (v20240201) Swift/UVOT calibration database (CALDB) files.
The UVOT obtained photometric observations on BL Lac through the optical \emph{vv, bb, uu} and UV \emph{w1, m2, w2} filters \citep{Poole2008}.
We estimated the observed fluxes in the UVOT filters using {\fontfamily{cmtt}\selectfont uvotsource}, adopting a circular source region with a radius of 5$\arcsec$ centred on BL Lac and a circular background region with a 10$\arcsec$ radius in a nearby source-free region.
We corrected the observed fluxes for Galactic extinction effects using $\text{E(B$-$V)} = 0.2802 \pm 0.0119$ \citep{S&F2011}, following the standard procedure described in 
\cite{Fitzpatrick1999}.
We estimated the host galaxy contributions in the UVOT photometric filters \emph{vv, bb, uu, w1, m2}, and \emph{w2}, given the adopted aperture, to be 2.89, 1.30, 0.36, 0.026, 0.020, and 0.017 mJy, respectively \citep{Raiteri2013}.
We accounted for both the host galaxy contribution and Galactic extinction effects to reconstruct the intrinsic BL Lac spectral points in the optical-UV band.
Table \ref{tab:appendix_table_uvot_results} lists the \textit{Swift}-UVOT data from the 2022 high state.

\subsection{Optical data}
{We obtained optical observations in the $BVRI$ Johnson-Cousins bands at Sierra Nevada Observatory (SNO) as part of the TOP-MAPCAT photo-polarimetric blazar monitoring programme \citep{agudo2012}. We reduced the data using the IOP4 software \citep{escudero2024,escudero2024_software}.}
We also obtained observations in the $R$ band as part of the Tuorla blazar monitoring programme\footnote{\url{https://tuorlablazar.utu.fi/}} using the 80 cm Joan Oró Telescope at Montsec Astronomical Observatory (Spain),
and reduced the data as detailed in \cite{nilsson2018}. Additional data were obtained in the $BVRI$ bands with the IAC80 telescope at the Teide Observatory. We coordinated observations in the $RV$ optical bands with the  Las Cumbres Observatory robotic telescope network \citep{brown2013}.

We also included data from the public all-sky survey programmes and databases: the All-Sky Automated Survey for Supernovae \citep[ASAS-SN;][]{shappee2014,kochanek2017} in the Sloan $g$ band, and the Zwicky Transient Facility \citep[ZTF;][]{bellm2019} in the Sloan $gr$ bands. We converted these Sloan magnitudes into the Johnson-Cousins photometric system using the transformations from Lupton (2005)\footnote{\url{https://classic.sdss.org/dr4/algorithms/sdssUBVRITransform.php\#Lupton2005}}. 

\begin{figure*}
    \sidecaption
    \includegraphics[width=1.44\columnwidth]{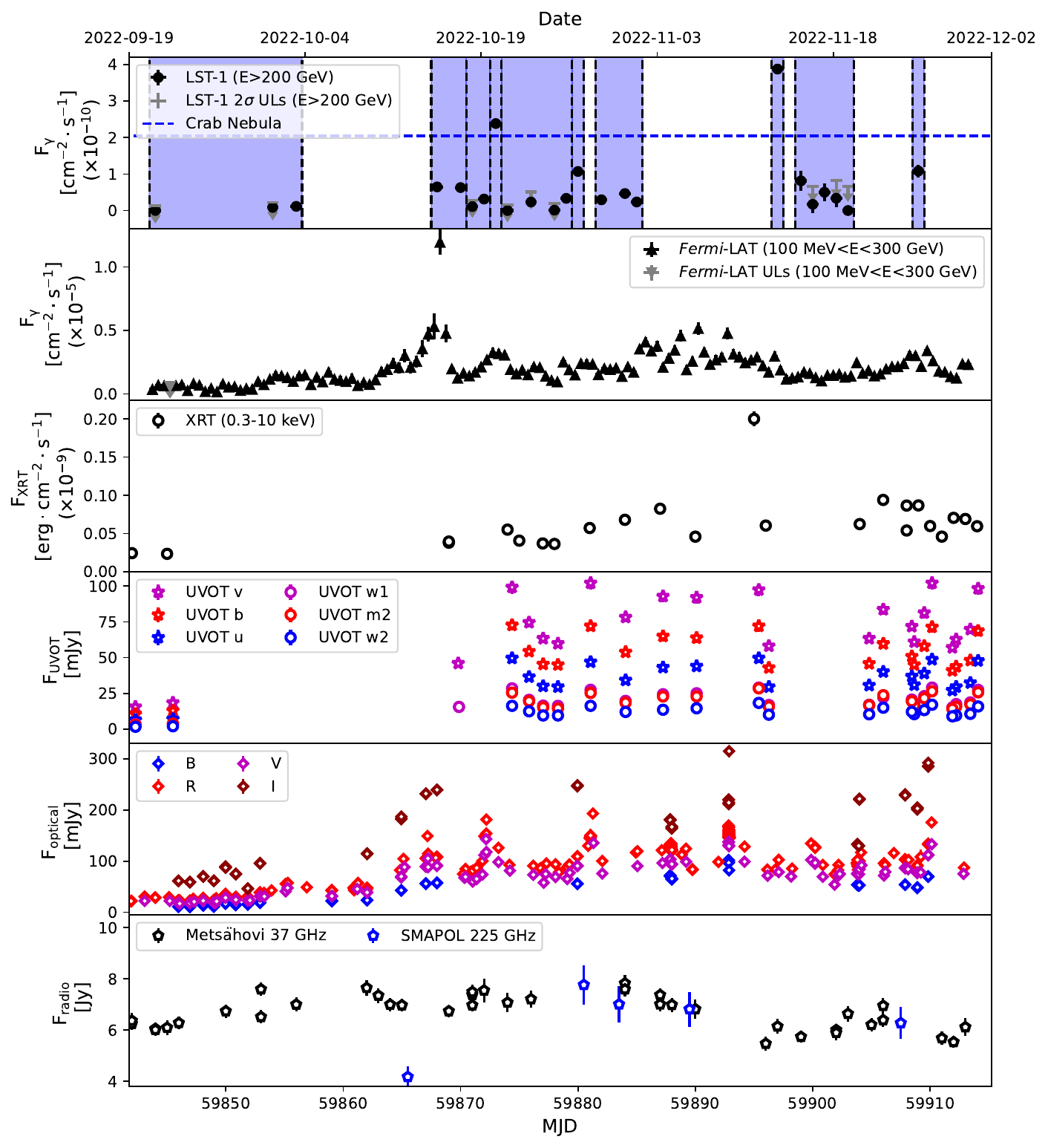}
    \caption{Multi-wavelength light curves of BL Lac between 21 September and 30 November 2022. \textit{From top to bottom}: LST-1 {night-wise} VHE $\gamma$-ray {light curve} above 200 GeV; \textit{Fermi}-LAT {12-hour binned} HE $\gamma$-ray light curves between 100 MeV and 300 GeV; XRT X-ray {light curve} between 0.3 and 10 keV; UVOT optical-UV {light curves} in the available filters; optical {light curves} in the $BVRI$ bands; and 37-GHz and 225-GHz radio light curves. The dashed blue line in the top panel indicates the Crab nebula flux above 200 GeV {\citep{aleksic2016}}. The blue contours, delimited by dashed black lines, indicate the Bayesian blocks identified in the LST-1 light curve.}
    \label{fig:mwl_lcs}
\end{figure*}

We subtracted host galaxy contributions to the observed optical emission within our aperture \citep[see][]{Raiteri2009}, and we corrected all data for Galactic extinction reddening. For this, we adopted the extinction values $A_{B}=1.192$, $A_{V}=0.901$, $A_{R}=0.713$, and $A_{I}=0.495$ 
reported by \cite{S&F2011}. Fig.~\ref{fig:mwl_lcs} presents the optical flux density light curves.

\subsection{Radio-millimetric data}
The 13.7-m diameter Mets\"{a}hovi radio telescope obtained 37-GHz radio band observations as part of its long-term AGN monitoring programme \footnote{\url{https://www.aalto.fi/en/metsahovi-radio-observatory/active-galaxies}} \citep{teraesanta1998}. 
Additionally, we compiled five observations in the 225-GHz band, obtained with the Submillimeter Array \citep[SMA;][]{ho2004} at Mauna Kea, Hawaii. The SMA is an eight-element 6-metre-diameter dish radio interferometer with two orthogonally polarised receivers, which enables the measurement of both the total and polarised flux of astrophysical objects. These receivers are inherently linearly polarised; however, they can be converted to circular using quarter-wave plates installed on the polarimeter \citep{marrone2008}. These observations came from the SMA Monitoring of AGNs with Polarisation (SMAPOL) programme \citep{myserlis2025}, which monitors $\gamma$-ray loud blazars, including BL Lac, on a bi-weekly cadence. 
We calibrated the data using the \texttt{MIR}\footnote{\url{https://lweb.cfa.harvard.edu/~cqi/mircook.html}} package following standard procedures.

\section{VHE intranight variability}\label{sec4}
\begin{figure*}
\centering
\includegraphics[width=0.43\textwidth]{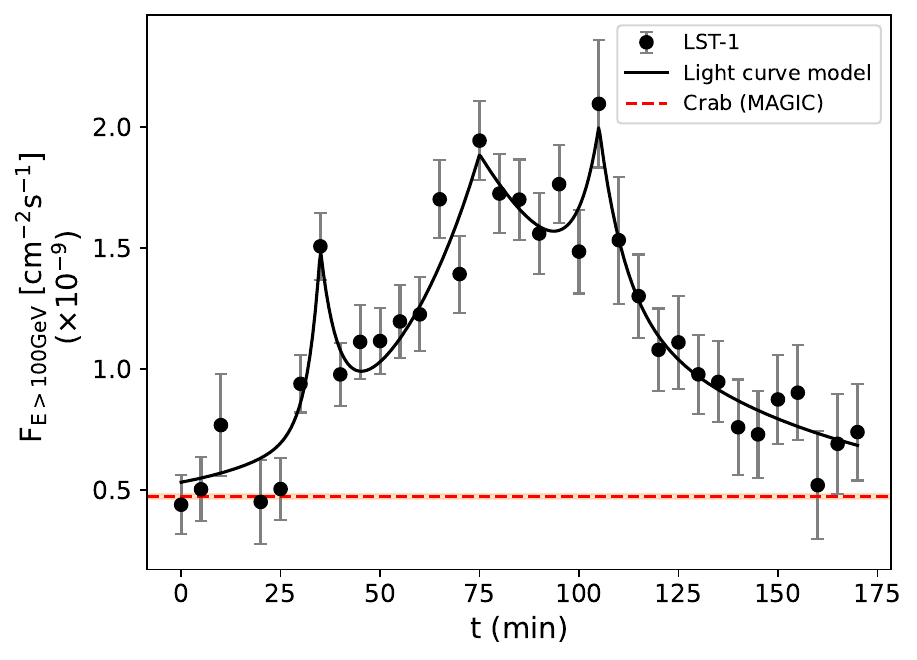}
\includegraphics[width=0.415\textwidth]{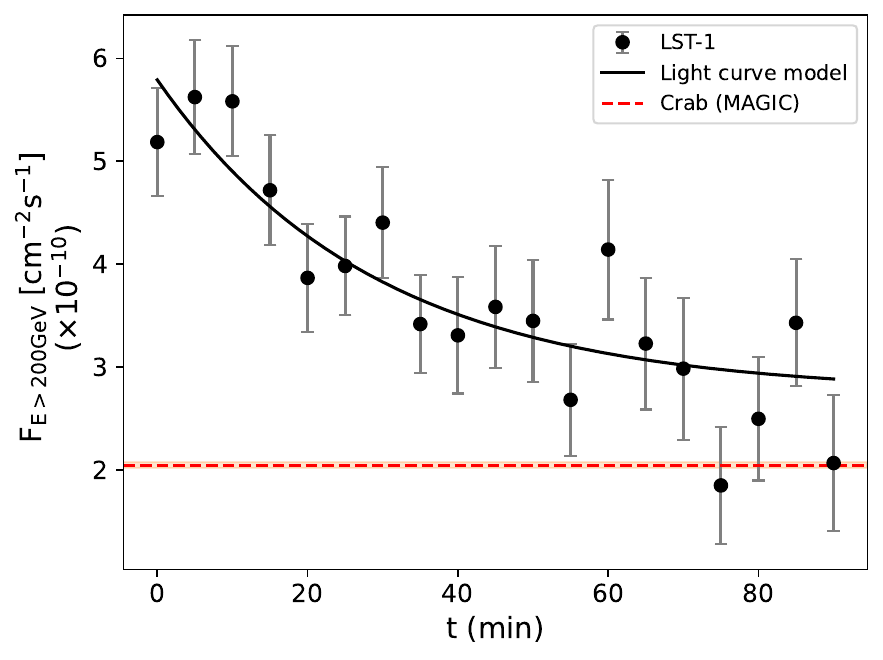}
\caption{Intranight light curves of BL Lac. \textit{Left:} 5-minute binned light curve during the night of October 20 above 100~GeV. \textit{Right:} {5}-minute binned light curve during the night of November 13 above 200~GeV. Black points indicate LST-1 measurements. Solid black lines represent the light curve models. The horizontal dashed red lines indicate the flux of the Crab nebula as measured by the MAGIC telescopes \citep{aleksic2016}.}
\label{fig:intranight_lcs}
\end{figure*}

In addition to day-to-week variability, we evaluated the VHE $\gamma$-ray intranight variability, {a} feature {less} often observed in blazars \citep[other examples are e.g.][]{aharonian2007,magic2020}. Its detection places significant constrains {on the size of the emitting region and} on the interpretation of the emission, potentially favouring leptonic models, as hadrons usually require longer acceleration times {and have longer cooling times by a factor of $\sim (m_{p}/m_{e})^{3}$}. We evaluated the nights with the highest VHE $\gamma$-ray flux: October 20 and November 13. Fig.~\ref{fig:intranight_lcs} shows the 5-minute binned light curves.

\subsection{October 20 flare}

We evaluated the fastest variability timescales observed during the night of October 20. For this night, we considered the integrated flux above 100 GeV, as all data were taken under {dark-sky} conditions and zenith distances of $\sim$20-40$^{\circ}$. The left panel of Fig.~\ref{fig:intranight_lcs} shows that the LST-1 data well sample the rise and decay of the flare, varying from $\sim$1.0~C.U. to $\sim$4.4~C.U. over  $\sim$$1-1.5$~hours.
This well-defined rise and decay profile of the flare allowed us to model the light curve with a rising and decaying exponential function, commonly used for modelling {light curves \citep{norris1996}, including those from} blazar flares \citep[see e.g.][]{aharonian2007}, as
\begin{equation}
F(t) = F_b + \left\{ \begin{array}{lc} F_{0}e^{-(t_{peak}-t)/t_{r}}: & t<t_{peak}; \\ \\ 
F_{0}e^{-(t-t_{peak})/t_{d}}: & t>t_{peak};  \end{array} \right.
\label{eq:exponential_rise_decay}.
\end{equation}
Here, $F_{0}$ represents the flux of the flare at time $t_{peak}$, 
$F_{b}$ is the baseline flux fixed at $(0.45 \pm 0.09) \times 10^{-9}$~cm$^{-2}$~s$^{-1}$ from the stable bins at the start and end of the flare, and $t_{r}$ and $t_{d}$ are the timescales of the flare rise and decay variations by a factor of $e$. We modelled the 5-minute binned light curve of October 20 as a sum of three 
possible substructures during the development of the flare, with peak times at $t_{max,1} = 35.0 \pm 4.8$~min, $t_{max,2} = 75.0 \pm 3.7$~min, and $t_{max,3} = 105.0 \pm 8.2$~min, considering the first flux point as $t_{0} = 0$~min. 
We also considered the simpler cases of {one or two components}and compared the goodness of fit in each case. We measured $\chi^{2}/\text{d.o.f.}$ values for the triple-, double-, and single-exponential profiles of $18.0/22$, $42.6/25$, and $54.5/28$, which yields significances of $4.2\sigma$ and $4.7\sigma$ for the triple-exponential model over the double- and single-exponential models, respectively, {using Wilks' theorem \citep{wilks1938}. We also verified that adding a fourth component does not improve the fit, showing a preference  of only 0.2$\sigma$} over the three-component model. Therefore, we adopted the model based on the sum of three exponential profiles defined by Eq.~(\ref{eq:exponential_rise_decay}).

We show the fitted function in the left panel of Fig.~\ref{fig:intranight_lcs}, and we report the calculated rise and decay timescales of the exponential profile in Table~\ref{tab:intranight_variability}. These timescales can be used to estimate the doubling flux variability timescale during the rise and decay, which represents the fastest significant variability timescales, as 
\begin{equation}
\tau_{rise/fall}=t_{r/d}\times \ln{(2)}.
\label{eq:doubling_flux_rise_decay}
\end{equation}
With this approach, we resolved variability timescales of $\sim$18~min with a confidence level of >$5\sigma$, and detected faster hints between 2.2~min and 4.6~min, albeit with a significance $\lesssim$2$\sigma$.

\begin{table*}
\centering
\caption{VHE $\gamma$-ray intranight variability analysis, showing the shortest variability timescales derived from rise and decay light curve modelling and doubling flux criterion.}
\label{tab:intranight_variability}
\resizebox{\textwidth}{!}{%
\begin{tabular}{ccccccccc}
\hline
\multirow{2}{*}{Date} & $F_b$ & $F_0$ & $t_{max}$ & $t_{r}$ & $t_{d}$ & $\tau_{rise}$ & $\tau_{fall}$ & $t_{doubling}$ \\ 
 & [10$^{-9}$ $\times$ cm$^{-2}$ s$^{-1}$] & [10$^{-9}$ $\times$ cm$^{-2}$ s$^{-1}$] & [min] & [min] & [min] & [min] & [min] & [min] \\ \hline
\multirow{3}{*}{October 20} & \multirow{3}{*}{$0.45 \pm 0.09$} & $0.75 \pm 0.15$ & $35.0 \pm 4.8$ & $3.2 \pm 1.8$ & $4.7 \pm 2.5$ & $2.2 \pm 1.2$ & $3.3 \pm 1.7$ & \multirow{3}{*}{$8.3 \pm 2.5$} \\ 
&  & $1.43 \pm 0.10$ & $75.0 \pm 3.7$ & $26.1 \pm 4.7$ & $52.4 \pm 7.7$ & $18.1 \pm 3.3$ & $36.3 \pm 5.3$  &  \\ 
&  & $0.75 \pm 0.25$ & $105.0 \pm 8.2$ & $6.1 \pm 3.7$ & $6.7 \pm 4.2$ & $4.2 \pm 2.6$ & $4.6 \pm 2.9$  &  \\ \hline
November 13 & $0.27 \pm 0.05$ & $0.30 \pm 0.04$ & 0 & -- & $28.9 \pm 5.9$ & -- & $20.0 \pm 4.1$ & -- \\ \hline
\end{tabular}
}
\end{table*}

We also estimated the fastest variability timescales using the doubling-time criterion for flux variations between consecutive temporal bins, following \cite{zhang1999}:
\begin{equation}
t_{doubling,i}=\frac{F_{i}+F_{i+1}}{2}\frac{t_{i+1}-t_{i}}{|F_{i+1}-F_{i}|},
\label{eq:doubling_flux_bin}
\end{equation}
where $F_{i+1}$ and $F_{i}$ are the fluxes at times $t_{i+1}$ and $t_{i}$. We considered only  timescales with variability significance $>$2$\sigma$. We list the results derived with this approach in the last column of Table~\ref{tab:intranight_variability}. This yields a significant doubling flux timescale of $\tau_{min} = 8.3 \pm 2.5$~min.

Causality arguments imply that very fast variability originates in compact jet regions. The observed fastest variability timescales thus constrain the size of the emitting region responsible for these variations as 
\begin{equation}
R\leq \frac{\tau_{min} c \delta }{1+z},
\label{eq:emitting_region_size}
\end{equation}
where $R$ is the radius of the emitting region, $\tau_{min}$ represents the fastest variability timescale observed, $c$ is the speed of light, $\delta$ is the Doppler factor of the region, and $z$ corresponds to the redshift of the source {\citep[see][]{dondi1995,tavecchio2010}}. We adopted a conservative approach, using the shortest timescale with a significance >$3\sigma$ {before trials},
namely $\tau_{min} = 8.3 \pm 2.5$~min. With this timescale and considering a range of Doppler factor values $\delta = 10-50$, the size of the emitting region responsible for the fast variability is constrained to $R\leq (1.4 \pm 0.4) \times 10^{14}$~cm to $R\leq (7.0 \pm 2.1) \times 10^{14}$~cm. 
The $2.2-6.7$ min timescales, if confirmed, imply a region a factor $\sim$1.5 to $\sim$4 times more compact. Nevertheless, due to their low significance and the typical systematic errors for LST-1 flux measurements \citep[see][]{abe2023}, these timescales remain unconfirmed. 

\subsection{November 13 flare}
We performed the same analysis for the night of November 13, when we also observed {intranight} variability in the VHE $\gamma$-ray emission of BL Lac. In this case, owing to the higher energy threshold introduced by moonlight, we considered the flux above 200 GeV. During this night, the emission decayed from $\sim$2.8 C.U. to $\sim$1.0 C.U. over 1.5 hours. We show the corresponding 5-minute binned light curve in the right panel of Fig.~\ref{fig:intranight_lcs}. The LST-1 covered only the decay of the VHE $\gamma$-ray flare, which we modelled with a decaying exponential function as
\begin{equation}
F(t)=F_{b}+F_{0}e^{-(t_{peak}-t)/t_{d}},
\label{eq:exponential_decay}
\end{equation}
following the form of Eq. (\ref{eq:exponential_rise_decay}). We used $F_{b}= (0.27 \pm 0.05) \times 10^{-9}$~cm$^{-2}$~s$^{-1}$, where the VHE flux appeared to stabilise after the flare. We also report the results for this night in Table~\ref{tab:intranight_variability}. 
Using the estimated timescale $\tau_{min} = 20.0 \pm 4.1$~min, we again constrained the size of the emitting region with Eq.~(\ref{eq:emitting_region_size}) for typical Doppler factors $\delta = 10-50$. This constrains the size between $R \leq (3.4 \pm 0.7) \times 10^{14}$~cm and $R \leq (16.8 \pm 3.5) \times 10^{14}$~cm. In this case, the doubling flux criterion yields no significant minute-scale variability.

\section{Characterisation of the HE-VHE $\gamma$-ray spectrum}\label{sec5}

\begin{table*}
\centering
\caption{Joint LAT and LST-1 $\gamma$-ray intrinsic spectrum at an energy $E_0=1$~GeV, corrected for EBL absorption for each BB.}
\label{tab:joint_spectra}
\begin{tabular}{ccccccc}
\hline
& Period & $f_{0}$ & \multirow{2}{*}{$\alpha$} & \multirow{2}{*}{$\beta$} & $E_{peak}$ & $\nu F_{\nu}^{max}$\\
 & [MJD] & [GeV$^{-1}$ cm$^{-2}$ s$^{-1}$] &  &  & [GeV] & [erg~cm$^{-2}$~s$^{-1}$] \\ \hline
BB1 & 59843.5 -- 59856.5 & $(6.68 \pm 0.58) \times 10^{-8}$ & $1.92 \pm 0.07$ & $0.13 \pm 0.02$ & 1.4 & $ (1.09 \pm 0.11) \times 10^{-10}$ \\ 
BB2 & 59867.5 -- 59870.5 & $(2.86 \pm 0.50) \times 10^{-7}$ & $1.86 \pm 0.08$ & $0.10 \pm 0.01$ & 1.9 & $ (4.81 \pm 0.85) \times 10^{-10}$ \\ 
BB3 & 59870.5 -- 59872.5 & $(2.41 \pm 0.46) \times 10^{-7}$ & $1.80 \pm 0.13$ & $0.14 \pm 0.03$ & 1.9 & $ (4.14 \pm 0.90) \times 10^{-10}$ \\ 
BB4 & 59872.5 -- 59873.5 & $(4.66 \pm 0.86) \times 10^{-7}$ & $1.53 \pm 0.06$ & $0.14 \pm 0.01$ & 5.5 & $ (1.11 \pm 0.12) \times 10^{-9}$ \\ 
BB5 & 59873.5 -- 59879.5 & $(1.97 \pm 0.16) \times 10^{-7}$ & $1.99 \pm 0.06$ & $0.13 \pm 0.02$ & 1.1 & $(3.15 \pm 0.27) \times 10^{-10}$ \\ 
BB6 & 59879.5 -- 59880.5 & $(2.62 \pm 0.24) \times 10^{-7}$ & $1.81 \pm 0.08$ & $0.11 \pm 0.01$ & 2.3 & $(4.58 \pm 0.45) \times 10^{-10}$ \\ 
BB7 & 59881.5 -- 59885.5 & $(3.03 \pm 0.18) \times 10^{-7}$ & $1.91 \pm 0.04$ & $0.11 \pm 0.01$ & 1.6 & $(4.95 \pm 0.35) \times 10^{-10}$ \\ 
BB8 & 59896.5 -- 59897.5 & $(4.26 \pm 0.88) \times 10^{-7}$ & $1.56 \pm 0.07$ & $0.11 \pm 0.01$ & 6.5 & $(1.05 \pm 0.15) \times 10^{-9}$ \\ 
BB9 & 59898.5 -- {59903.5} & $(1.71 \pm 0.12) \times 10^{-7}$ & $1.86 \pm 0.05$ & $0.11 \pm 0.01$ & 1.9 & $(2.86 \pm 0.24) \times 10^{-10}$ \\ 
BB10 & {59908.5} -- {59909.5} & $(3.48 \pm 0.30) \times 10^{-7}$ & $1.84 \pm 0.05$ & $0.10 \pm 0.01$ & 2.3 & $(5.94 \pm 0.62) \times 10^{-10}$ \\ \hline
\end{tabular}
\end{table*}

To evaluate possible $\gamma$-ray spectral variability patterns, we studied the evolution of the $\gamma$-ray spectrum during the flare. For this, we applied a Bayesian block (BB) analysis on the VHE $\gamma$-ray light curve using the algorithm developed by \cite{wagner2022}\footnote{\url{https://github.com/swagner-astro/lightcurves}}, identifying ten BBs of comparable $\gamma$-ray activity with a false-positive rate of 1\% (see Fig.~\ref{fig:mwl_lcs}). Then, using the LAT and LST-1 data, we characterised the high-energy SED peak for the different identified periods from 100~MeV up to a few TeV. We performed a joint fit of the data from both instruments with \texttt{Gammapy}, combining multi-instrument data at the event level, while accounting for instrument response functions (IRFs), backgrounds, and other effects such as {EBL} absorption \citep{nievas2025}. 

For the \textit{Fermi}-LAT data for each BB, we filtered {out} all sources in the model with TS<10. We left free the parameters of all sources within a circular region of 10$^{\circ}$ radius centred at the position of BL Lac and with TS>30, while for sources within 5$^{\circ}$ of BL Lac with TS<30, we left only the spectral normalisation as a free parameter. For the remaining sources, we fixed all parameters to the catalogue values. We then performed a pre-fit of the \textit{Fermi}-LAT data to remove weak sources still included in the model for which the {expected} counts were <5.
Finally, we combined the LAT and LST-1 datasets for each BB and calculated the complete $\gamma$-ray spectrum and SED peak, incorporating the response of each instrument into the analysis.

\begin{figure}
    \centering
    \includegraphics[width=\linewidth]{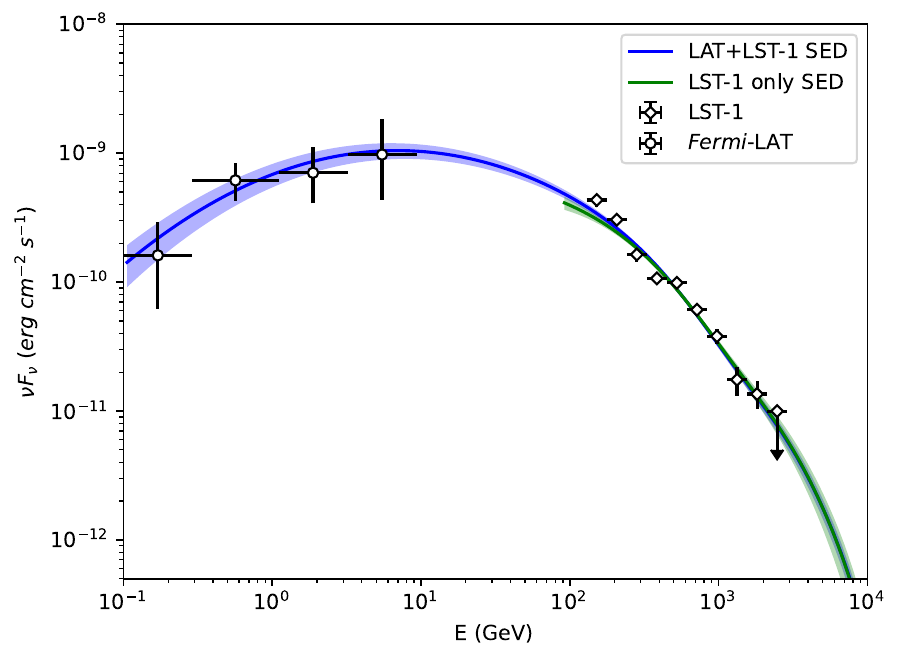}
    \caption{Joint LAT-LST-1 SED for BB8 using a log-parabola plus EBL model. White circles and squares correspond to LAT and LST-1 spectral points, respectively. Down-pointing arrows indicate  flux upper limits. The blue and red lines and bowtie shapes show the best fits of the joint LAT plus LST-1 and LST-1-only data, respectively.}
    \label{fig:joint_fit_bb4}
\end{figure}

We tested several spectral shapes of the HE-VHE emission, including log-parabola, power-law with exponential cut-off and log-parabola with exponential cut-off spectral models. We find that the first function provides the most accurate fit to our data. Therefore, we used a model consisting of a log-parabola function plus a term accounting for EBL absorption at redshift $z=0.069$, following the model from \cite{saldana-lopez2021},
\begin{equation}
\phi(E)=\phi_{0}(E/E_{0})^{-\alpha -\beta \log(E/E_{0})}e^{-\alpha_{EBL} \tau(E,z)}.
\label{eq:logpar_ebl}
\end{equation}
To directly compare the different spectra, we adopted $E_0=1$~GeV for all periods. Table~\ref{tab:joint_spectra} lists the best-fit spectral parameters. Fig.~\ref{fig:joint_fit_bb4} shows an example of the joint SED for the eighth BB (November 13). 
Figs.~\ref{fig:joint_sed_all1} and \ref{fig:joint_sed_all2} show the remaining SEDs.
{Due to the minimal variability in spectral curvature, we tested a log-parabola function with $\beta$ fixed at $0.12$, based on the average value over the entire period. The results were consistent within uncertainties with those reported in Table~\ref{tab:joint_spectra}. }

The complete characterisation of the high-energy SED peak performed in Sect~\ref{sec5} allowed us to investigate spectral variability patterns during the evolution of the flare. The spectral index at 1~GeV remains stable around $\alpha \sim 1.90$, with nearly all BBs consistent within errors at $\sim$2$\sigma$. {Nevertheless, spectral index variations are particularly noticeable} on the nights of October 20 ($\alpha = 1.53$) and November 13 ($\alpha = 1.56$), indicating that particles are accelerated to considerably higher energies during these flares. This is also reflected in the high-energy peak shift, as reported in Table~\ref{tab:joint_spectra}. The IC peak shift is evident for the four brightest BBs, namely the nights of October 20 and 27, and November 13 and 25, which are the only ones with $E_{peak} > 2$~GeV. The average value of $E_{peak}$ is below 2 GeV, while during these nights we observe shifts of up to {5.5}~GeV on October 20, {6.5}~GeV on November 13, and {2.3}~GeV on October 27 and November 25. The Spearman correlation coefficients between $E_{peak}$ and $\alpha$, and between $E_{peak}$ and $f_0$, indicate a strong correlation ($\rho \sim 0.85$ and $\rho \sim 0.75$, respectively), suggesting a possible harder-when-brighter behaviour. However, this correlation is dominated by the two brightest nights and  does not account for uncertainties.

\section{Broadband SEDs}\label{sec6}

\begin{figure}
\centering
\includegraphics[width=\columnwidth]
{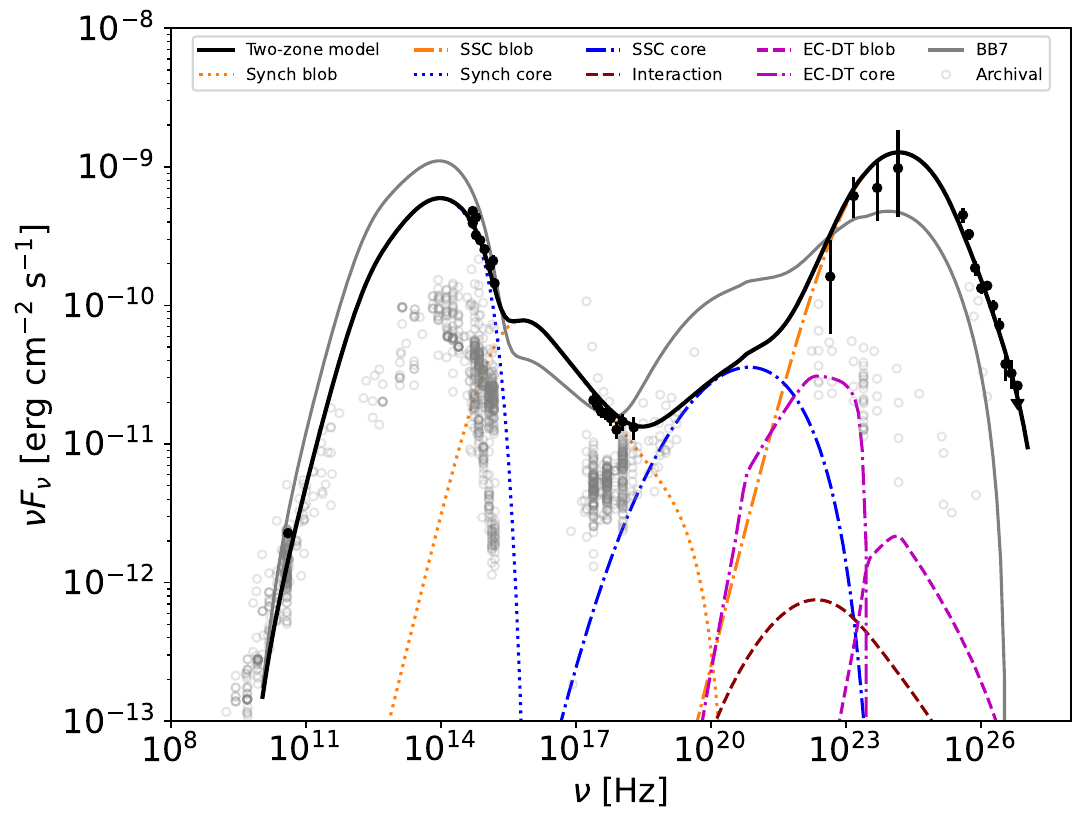}
\caption{Broadband SED of BL Lac for BB8. Black points and arrows indicate flux points and upper limits obtained from the different instruments. The solid black line shows the fitted co-spatial two-zone SSC model. The dotted and dash-dotted blue lines show the synchrotron and SSC emission of the core, respectively. The dotted and dash-dotted orange lines show the synchrotron and SSC emission of the blob. The dash-dotted and dashed magenta lines show the EC emission from the core and the blob, respectively. The dashed dark red line shows the emission resulting from the interaction between the core and the blob. {The solid grey line shows the SED of the previous block.} Historical data retrieved from the SSDC database (\url{https://tools.ssdc.asi.it/SED/}) are shown with grey open points.}
\label{fig:sed_model_bb8}
\end{figure}

We modelled the evolution of the broadband emission of BL Lac during the high state of 2022. We considered ten broadband SEDs, corresponding to each of the BBs identified above.
Fig.~\ref{fig:sed_model_bb8} shows, as an example, the model for {the eighth BB}. We present all fitted SED models in Figs.~\ref{fig:sed_models} and \ref{fig:sed_models2}. We note that due to the absence of strictly simultaneous \textit{Swift} data ($\pm$0.5 days with respect to the LST-1 observations) to BBs 3, 4, 6, 8, and 9, we adopted a window of $\pm$1.5 days for these periods to ensure complete multi-wavelength coverage of the SEDs. 

\begin{table*}
\caption{SED modelling parameters for the two-zone SSC model describing the broadband emission during each BB.}
\centering  
\label{tab:sedparams}
\resizebox{\textwidth}{!}{
\begin{tabular}{cccccccccccccc}
\hline
  (1) & (2) & (3) & (4) & (5) & (6) & (7) & (8) & (9) & (10) & (11) & (12) & (13) & (14)\\
  \multirow{2}{*}{Epoch} & \multirow{2}{*}{Component} & $\gamma_{\text{min}}$ & $\gamma_{\text{b}}$ & $\gamma_{\text{max}}$ & \multirow{2}{*}{$n_1$} & \multirow{2}{*}{$n_2$} & B & K & R & \multirow{2}{*}{$\Gamma$} & \multirow{2}{*}{$\delta$} & \multirow{2}{*}{$U'_{B}/U'_{e}$} & $t_{SSC}$ \\
  
   &  & ($\times 10^{3}$) &($\times 10^{4}$) &($\times 10^{5}$) &  &  & [G] & [$\times 10^{-2}$\,cm$^{-3}$] & [$\times 10^{15}$\,cm] & & & & [min] \\ 
   \hline
 \multirow{2}{*}{BB1} 
      & Blob & 3.00 & 0.90 & 0.17 & 2.0 & 2.8 & 0.50 & 3.00 & 0.3 & 33 & 33 & $0.31 \times 10^{-2}$ & 35.0 \\
      & Core & 0.27 & 0.20 & 0.05 & 2.2 & 3.4 & 0.20 & 0.01 & 100 & 20 & 20 & 1.64 & -- \\
\hline
\multirow{2}{*}{BB2} 
      & Blob & 2.50 & 1.00 & 1.50 & 2.8 & 3.8 & 0.50 & 3.00 & 0.3 & 33 & 33 & $0.13 \times 10^{-2}$ & 15.3 \\
      & Core & 0.30 & 0.23 & 0.05 & 1.9 & 2.9 & 0.20 & 0.01 & 100 & 20 & 20 & 1.54 & -- \\
\hline
  \multirow{2}{*}{BB3} 
      & Blob & 3.00 & 0.95 & 0.80 & 2.9 & 3.7 & 0.50 & 3.00 & 0.3 & 33 & 33 & $0.17 \times 10^{-2}$ & 19.2 \\
      & Core & {0.25} & 0.23 & 0.05 & {2.0} & 2.9 & 0.20 & 0.01 & 100 & 20 & 20 & {0.75} & -- \\
\hline
\multirow{2}{*}{BB4}
      & Blob & 2.50 & 0.95 & 1.80 & 2.6 & 3.5 & 0.50 & 5.00 & 0.3 & 33 & 33 & $0.11 \times 10^{-2}$ & 11.1 \\
      & Core & 0.30 & 0.19 & 0.07 & 2.0 & 2.8 & 0.20 & 0.01 & 100 & 20 & 20 & 2.00 & -- \\
\hline
\multirow{2}{*}{BB5}
      & Blob & 3.20 & 1.00 & 0.35 & 2.8 & 3.8 & 0.50 & 3.00 & 0.3 & 33 & 33 & $0.17 \times 10^{-2}$ & 17.9 \\
      & Core & {0.25} & 0.20 & 0.07 & {2.4} & 3.4 & 0.20 & 0.01 & 100 & 20 & 20 & {0.79} & -- \\
\hline
\multirow{2}{*}{BB6}
      & Blob & 2.00 & 0.95 & {1.10} & 2.7 & 3.4 & 0.50 & 3.00 & 0.3 & 33 & 33 & $0.13 \times 10^{-2}$ & {16.2} \\
      & Core & {0.50} & 0.25 & 0.07 & {2.4} & 3.7 & 0.20 & 0.01 & 100 & 20 & 20 & {0.88} & -- \\
\hline
\multirow{2}{*}{BB7}
      & Blob & 2.70 & 1.05 & 1.00 & 2.7 & 3.6 & 0.50 & 3.00 & 0.3 & 33 & 33 & $0.13 \times 10^{-2}$ & 12.3 \\
      & Core & 0.30 & 0.23 & 0.07 & 2.2 & 3.3 & 0.20 & 0.01 & 100 & 20 & 20 & 1.19 & -- \\
\hline
\multirow{2}{*}{BB8}
      & Blob & 4.00 & 1.00 & 3.80 & 2.1 & 3.8 & 0.50 & 7.50 & 0.3 & 33 & 33 & $0.11 \times 10^{-2}$ & 8.8 \\
      & Core & 0.30 & 0.18 & 0.07 & 1.9 & 3.0 & 0.20 & 0.01 & 100 & 20 & 20 & 2.56 & -- \\
\hline
\multirow{2}{*}{BB9}
      & Blob & 3.60 & 1.00 & {0.80} & 2.7 & {3.6} & 0.50 & 3.00 & 0.3 & 33 & 33 & $0.20 \times 10^{-2}$ & {17.5} \\
      & Core & {0.15} & 0.19 & 0.06 & {2.3} & 2.7 & 0.20 & 0.01 & 100 & 20 & 20 & {0.85} & -- \\
\hline
\multirow{2}{*}{BB10}
      & Blob & 2.00 & 1.05 & 1.00 & 2.6 & 3.6 & 0.50 & 3.00 & 0.3 & 33 & 33 & $0.11 \times 10^{-2}$ & 12.1 \\
      & Core & 0.28 & 0.20 & 0.06 & 2.3 & 3.0 & 0.20 & 0.01 & 100 & 20 & 20 & 1.45 & -- \\
\hline
\end{tabular}
}
\tablefoot{Columns: (1) Observation campaign or state. (2) Emission region. (3), (4), and (5) Minimum, break, and maximum electron Lorentz factors. (6) and (7) Electron distribution slopes below and above $\gamma_{b}$. (8) Magnetic field. (9) Electron density. (10) Emission-region size. (11) Bulk Lorentz factor. (12) Doppler factor. (13) Ratio between the magnetic-field energy density and the relativistic-electron energy density. (14) Blob SSC cooling time at $\gamma_b$.}
\end{table*}

We considered a leptonic-only origin for the broadband emission.
We used version v0.4.0 of the \textsc{python} package \texttt{agnpy} \citep{nigro2022,nigro2023}. 
We find that SSC models based on a single emitting region are unable to accurately describe the broadband emission of BL Lac during this period.
One-zone models based on EC scattering present an alternative to explain VHE $\gamma$-ray emission and have previously been invoked for BL Lac {due to the presence of a small and weak broad line region (BLR),} which provides external low-energy photons as seeds for $\gamma$-ray production \citep{magic2019}. 
However, we find that this solution is insufficient to reproduce the broadband emission and variability during the 2022 high state, particularly given the tight constraints on the size of the emitting region imposed by the intranight variability. 
We therefore considered an SSC model based on two emitting regions \citep{tavecchio2011}, which has frequently been used to model the multi-wavelength emission of this source \citep[see e.g.][]{magic2019,sahakyan2022}.
We denote the two regions as `core' and `blob', with $R_{core}>R_{blob}$.
We consider the regions to be co-spatial and interacting, motivated by the high correlation of the broadband emission from the optical band up to $\gamma$ rays {(Appendix~\ref{sec3.2})}.
Fig.~9b of \cite{magic2019} illustrates a schematic of this layout.
Each region is characterised by a magnetic field $B$, a Lorentz factor $\Gamma$, a Dopper factor $\delta$,
and an electron population described by a broken power-law distribution with respect to the electron Lorentz factor $\gamma$,
\begin{equation}
N(\gamma) = \left\{ \begin{array}{lc} K(\gamma/\gamma_{b})^{-n_{1}}: & \gamma_{min}<\gamma<\gamma_{b} \\ \\  K(\gamma/\gamma_{b})^{-n_{2}}: & \gamma_{b}<\gamma<\gamma_{max}  \end{array} \right.
\label{eq:broken_PL_electrons},
\end{equation}
where $K$ is the normalisation of the distribution, $\gamma_{min}$ and $\gamma_{max}$ are the minimum and maximum values of the Lorentz factor, and $n_{1}$ and $n_{2}$ are the spectral indices below and above the break of the distribution $\gamma_{b}$, respectively. We constrained the size of the blob, responsible for the intranight VHE $\gamma$-ray variability, using the estimates presented in Sect.~\ref{sec4}. {Additionally, we incorporated an EC contribution from external photon fields in the SED model, namely the BLR, the accretion disc, and the dusty torus, since these components have previously been used to reproduce the broad IC SED peak \citep[see e.g.][]{khatoon2024}. 
Following \cite{ghisellini2009, magic2019}, we adopted $L_{BLR}=2.5 \times 10^{42}$~erg~s$^{-1}$ and $R_{BLR}=2 \times 10^{16}$~cm for the BLR, which implies $L_{disc}=2.5 \times 10^{43}$~erg~s$^{-1}$ and $R_{torus}=4 \times 10^{17}$~cm using the typical relations between BLR, disc, and torus \citep[see][]{kaspi2007,cleary2007}. We assumed a torus covering factor $\Xi_{DT}=0.1$ and a BLR reprocessing factor (based on the H$_{\alpha}$ line) $\xi_{H \alpha}=0.1$. For the dusty torus, we assumed a typical temperature of 1000~K. We adopted a black hole mass $M_{BH}=10^{8.7}M_{\odot}$ \citep{ghisellini2010}.
We considered the core located at a distance $d_{core}=6.5 \times 10^{17}$~cm from the central engine and the blob at $d_{blob}=6.2 \times 10^{17}$~cm.
We accounted for possible $\gamma \gamma$ absorption at GeV-TeV energies and the EBL absorption at VHE. Our analysis shows that $\gamma \gamma$ absorption due to external radiation fields is negligible.}

{Multi-zone models of blazar SEDs are characterised by a high degree of degeneracy due to the relatively large number of parameters.
This degeneracy generally prevents the use of standard fitting routines based on cost-function minimisation for effectively constraining the parameter space.
In such cases, it is necessary to adopt a physically motivated scenario as a guiding framework.
In this work, we investigated a scenario in which the evolution of the flare is driven by magnetic reconnection, which is strongly supported by the minute-scale variability detected at VHE with LST-1. 
To fit the SED of each BB, we used the parameters of the previous block as initial parameters, iterating until we achieved a satisfactory representation of the SED. To reduce the number of free parameters, we kept the magnetic fields $B$, the emitting region sizes $R$, the Lorentz factor $\Gamma$, and the Doppler factor $\delta$ fixed, adopting values consistent with the limit values set by the variability analysis. {Additionally, we minimised changes in the physical parameters in the core region between adjacent BBs to reflect the slowly evolving nature of the core, unless the time interval between the centre of two adjacent BBs exceeded two days, corresponding to the light-crossing timescale of the core region in our reference frame. In contrast, the blob is considered as a newly formed plasmoid originating from an unresolved magnetic reconnection layer in the core, whose parameters can vary considerably depending on the reconnection layer.} Therefore, we based our SED evolution on changes in the electron distribution of each region, namely $K$, $\gamma_{min}$, $\gamma_{b}$, $\gamma_{max}$, $n_1$, and $n_2$. The parameter changes were implemented in a manner consistent with the flare evolution timescales.
Figs.~\ref{fig:sed_models} and \ref{fig:sed_models2} show the resulting models for each BB, and Table~\ref{tab:sedparams} summarises the corresponding parameters.
We discuss the consistency of these models with a magnetic reconnection scenario in Sect.~\ref{sec7}.
}

The leptonic origin of the emission considered here is supported by recent results obtained by the Imaging X-ray Polarimetry Explorer (IXPE) satellite, based on the analysis of the broadband polarised emission of BL Lac \citep[see][]{middei2023,peirson2023}. Within a {leptonic} framework in which the X-ray spectrum is dominated by the high-energy IC scattering component, the X-ray polarisation {degree $\Pi_X$} is expected to be substantially lower than that of the {optical band, which originates from electron synchrotron radiation} \citep{liodakis2019,peirson2019,zhang2024}. \cite{middei2023} measured an upper limit of $\Pi_{X}<12.6-14.2$\% for two IXPE observations of BL Lac during May and July 2022, consistent with this scenario given the optical polarisation degree.

Moreover, during strong flares in which the soft X-ray emission has a significant synchrotron radiation contribution, as indicated by a much flatter X-ray spectrum (as in the present case, e.g. BB3 compared to BB8 when the source was at its brightest), we expect a significantly higher polarisation degree of soft X-ray photons than at hard X-ray energies, and even higher than the optical polarisation degree \citep{liodakis2019,peirson2019,zhang2024}. A third IXPE observation of BL Lac, performed two days after the last LST-1 observation (27-29 November) resulted in a measured X-ray polarisation degree of $\Pi_{X}=27.6\% \pm 11.6\%$ over the entire X-ray band. After splitting the analysis into two bins corresponding to the soft and hard X-ray regimes, almost all the polarised emission appears in the soft X-ray bin, with a polarisation degree of $\Pi_{X}=27.6^{+5.6}_{-7.9}\%$. \cite{peirson2023} estimated that the synchrotron emission comprises 44\% of the total flux in the soft X-ray band. After accounting only for synchrotron emission and assuming that the remaining 56\% of the flux originates from unpolarised SSC emission, these authors calculated the X-ray polarisation fraction to be $\sim$49\%, much higher than that in optical frequencies ($\sim$13\%), and well within the expectations for an SSC scenario \citep{peirson2019}. Therefore, the leptonic SED modelling performed here is consistent with the expected nature of the broadband emission based on both spectral and multi-wavelength polarisation analyses. 

A final IXPE campaign on BL Lac during November 2023 resulted in an extremely high optical polarisation of $47.5$\%, not accompanied by high X-ray polarisation, yielding a 3$\sigma$ upper limit of 7.4\%.  BL Lac behaved as an LSP during this period, placing the X-ray emission in the high-energy SED component. These results strongly favour SSC multi-zone or energy-stratified models \citep[see][]{zhang2024,agudo2025}, leaving little room for a hadronic contribution, except within the framework of a hybrid model, as shown by the detailed broadband analysis of \cite{liodakis2025}. However, as shown by these authors, hybrid models expect a hardening of the VHE spectrum to reproduce the multi-wavelength polarisation signatures, which is not observed in the present data.

\section{Discussion}\label{sec7}

\subsection{$\gamma$-ray and X-ray spectral behaviour}
X-ray spectral trends also appear in BL Lac prior to the 2022 high state \cite[e.g.][]{Weaver2020}. Through extensive SED modelling, \cite{sahakyan2022} attribute the synchrotron component emerging in X-rays during periods of high state to a second emission zone located beyond the BLR, distinct from the region that dominates the emission in the optical-UV band. Emission from this second region can shift the synchrotron peak towards higher frequencies, reaching  $\gtrsim$10$^{15}-10^{16}$~Hz, well within the high-synchrotron-peaked (HSP) range. BL Lac is well known for transitioning between the LSP regime, primarily during low flux states, to an IBL-HBL behaviour during active phases \citep[see e.g.][]{ackermann2011,nilsson2018}. Since approximately 2019, the source has undergone several years of very high activity, culminating in its three brightest flares ever recorded in 2021, 2022, and 2024. Therefore, the X-ray spectral variability trends observed in the dataset analysed here can be attributed to the emergence of synchrotron emission from freshly accelerated particles in a secondary emission region along the jet. The shift of the X-ray emission from the high- to low-energy peak of the SED may also explain the lower correlation between the hard X-ray and optical flux compared to the soft X-ray flux, as reported in {Appendix~\ref{sec3.2}}. Furthermore, in our model, the emission from the secondary region (blob) also dominates in the HE and VHE $\gamma$-ray bands. Consequently, these freshly accelerated particles are responsible for the intense flaring activity detected with \textit{Fermi}-LAT and LST-1 via SSC, as well as for the observed brightening and shifting of the $\gamma$-ray spectrum towards higher energies (see Sect.~\ref{sec5}).

\subsection{Week-scale evolution of the broadband emission}

We successfully modelled the broadband emission and evolution of the flare as an SSC two-zone model, including the EC contribution. {The flare begins with emission dominated by the slowly evolving core, extending up to X-ray energies. Plasmoids (blobs) can develop within highly magnetised current sheets in the core via magnetic reconnection. These blobs are primarily responsible for the gamma-ray emission. At this early stage, EC scattering also contributes to the gamma-ray emission. As magnetic reconnection progresses, particles in the blob are accelerated to higher energies, increasing the blob's contribution in the X-ray band. Our spectral fitting reveals several major ejections of plasmoids and/or blobs 
 in the emission region, specifically BB4, BB6, BB8, and BB10. During each of these episodes, a newly formed blob dominates the gamma-ray emission and may contribute significantly at other wavelengths, depending on the blob parameters. Each episode is followed by a relatively low-flux period during which the newly formed blob cools and loses its energy. Because each blob forms independently, their parameters are expected to vary considerably, whereas the core evolves slowly, so its physical parameters remain largely unchanged between different episodes.}

{This scenario is consistent with the correlation analysis presented in {Appendix~\ref{sec3.2}}, where the high degree of correlation between different bands is attributed to the co-spatial nature of the broadband emission, with the exception of the lower correlation observed in the radio and VHE bands. For the radio band, it is well established that the inner regions of the jet are highly opaque to radio emission, which typically originates from an outer part \citep{max-moerbeck2014}. Several studies report long delays of hundreds of days between the radio and multi-wavelength emission of blazars \citep[see e.g.][]{liodakis2018}. We could not  test this hypothesis as we focused on a shorter period. Nevertheless, such behaviour could explain the lower correlation observed in this band. We note, however, that the radio band has a lower observing cadence than the optical band, which can also bias the measured correlation. For the VHE band, the lower correlation can be explained by the shorter cooling timescales of the highest-energy electrons responsible for the TeV emission. However, as with the radio band, the lower observing cadence of the VHE data may introduce systematic effects in the correlation analysis that dilute the correlation in the TeV emission.}

With the configuration adopted in Sect.~\ref{sec6}, the low-energy photons from dusty torus dominate the EC contribution. This EC emission has previously been used to model the broadband emission of BL Lac \citep{khatoon2024}. In the present flare, due to the very bright state of the source during most of the period, we find that the EC emission is only relevant during the early stages of the outburst, when the source is faintest (see the model for BB1 in Fig.~\ref{fig:sed_models}). We also considered possible emission arising from the interaction between the core and blob regions, that is, low-energy photons from one region being IC-scattered by electrons in the other region. However, this contribution was found to be negligible because of the compactness of the blob as a result of the size constraints derived from the VHE minute-scale variability. We also evaluated the absorption $\gamma \gamma$ and found it to be negligible up to TeV energies, where the source is not detected significantly and we estimate only upper limits.

Overall, the parameters are consistent with the typical values derived from a large sample of blazars \citep{ghisellini2010}. The core region primarily drives the emission from radio to UV wavelengths and contributes significantly to the hard X-ray emission during periods where the X-ray emission is transitioning from the low- to high-energy peak of the SED. During the early stages of the flare, when the soft X-rays also lie within the IC peak, the core dominates both the soft and hard X-ray emission. {Changes in the parameters between adjacent BBs are relatively small, given that the time interval between consecutive blocks generally exceeds the light-crossing time of the core, reflecting its slow evolution.}
Conversely, during periods when BL Lac fully transitions from LSP to ISP and the X-ray emission is entirely produced by synchrotron radiation, the low-energy emission from the blob becomes dominant in the X-ray band. {The $\gamma_{min}$ values of the blob are at the high end of normal values. \cite{zech2021} interpret such high $\gamma_{min}$ values in other BL Lac objects as resulting from thermal coupling between protons and electrons within a weakly magnetised shock, where protons transfer energy to electrons within the shock layer.}

The primary driver of variability on day-to-week timescales is the evolution of the particle populations within each region during the flare. The blob's electron distribution behaviour of the blob is particularly noticeable during the brightest nights (BB4 and BB8), when it clearly reaches  higher energies, consistent with the IC peak shift towards higher energies reported in Sect.~\ref{sec5}.
Furthermore, for these two nights, an injection of freshly accelerated particles was also necessary to account for the remarkably high $\gamma$-ray emission. The blob is also characterised {by} higher magnetisation and Doppler boosting than the core, with $B_{blob}=0.5$~G, $\delta_{blob}=33$, and $B_{core}=0.2$~G, $\delta_{core}=20$. This, combined with the equipartition values of the core, which are close to equipartition but still $U'_B/U'_e>$1, and its occasionally very hard spectral index, is consistent with a magnetic reconnection scenario \citep[see, e.g.][]{jormanainen2023}, as discussed in detail in Sect.~\ref{sec7.3}. The equipartition values of the blob are far from $\sim$1, with the electron energy density dominating on the order of magnitude of $U'_{e}\sim 10^2 U'_{B}$. This limitation is commonly encountered in such models \citep[see][]{tavecchio2016}. This may result from adopting a an overly simplified representation of relativistic jets while attempting to capture all observational features, including, week-, day-, and minute-scale variability. Nevertheless, the models presented here can accurately describe the evolution of the flare while providing a reasonable solution to observational features such as multi-wavelength correlations and week-, day-, and minute-scale VHE variability.

\subsection{Origin of the minute-scale VHE variability}\label{sec7.3}

The extremely fast, minute-scale variability requires a compact region with extreme particle acceleration (the blob in our fitting model) in addition to the blazar zone (the core in our model) , which governs the day-to-week-scale variability. Such a physical setup is often attributed to {the mini-jet scenario under magnetic reconnection \citep{giannios2009,petropoulou2016}. 
The blob also requires a particularly high Lorentz factor to prevent VHE $\gamma$-ray attenuation within the compact blob.
The optical depth for VHE $\gamma$-rays is given by $\tau_{\gamma\gamma}\sim (1+z)^2 \sigma_T L_{target} E_{obs}/(72\pi m_e^2 c^6 \delta^6 \tau_{min})$ \citep{ackermann2016}, where $E_{obs}>200~\rm{GeV}$ is the observed VHE photon energy, $L_{target}$ is the optical luminosity providing the target photons for VHE $\gamma$-rays, and $\tau_{min}$ is the shortest observed variability timescale. We find that the Doppler factor $\delta\gtrsim 30$ ensures that the compact blob is optically thin to the observed VHE photons. This range of Doppler factor values is typically higher than those commonly observed in blazars. Hence, the mini-jet scenario is particularly relevant here to solve this problem and prevent VHE $\gamma$-ray attenuation within the compact blob.} In the mini-jet scenario, the terminal bulk Lorentz factor of the plasmoid can approximately equal the magnetisation factor of the reconnection site in the co-moving frame \citep{sironi2016}. Table~\ref{tab:sedparams} shows that the magnetisation factor in the core, estimated by $U'_B/U'_e$, is generally $\gtrsim 1$. The bulk Lorentz factor of the blob relative to the core can be approximated by $\delta_{blob}/\delta_{core}$ (assuming both are close to the line of sight), which is comparable to the magnetisation factor. Although our fitting model and the above estimates are rather simplified, they remain consistent with the mini-jet within a magnetic reconnection scenario \citep{nalewajko2011}.

A plasmoid merger in magnetic reconnection presents an alternative scenario \citep{zhang2022}. Previous particle-in-cell (PIC) simulations show that the SSC can exhibit much higher and faster variability than the synchrotron and EC, owing to the highly concentrated electrons freshly accelerated at the plasmoid merging site. The magnetisation factor at such sites can drop below one, even though the reconnection site has an overall magnetisation greater than one, consistent with the fitting parameters derived for the blob \citep{zhang2022}. In principle, such collisions and mergers of plasma structures can also occur in magnetised turbulence \citep{zhang2023}. However, the very high intrinsic X-ray polarisation \citep[$\sim$49\%, see][]{peirson2023}, measured just two days after the last flare, disfavours a turbulent environment.

In addition to the rapid variability, the very hard particle spectrum in the core ($n_1\lesssim 2.0$) also indicates a magnetic reconnection origin. Although we fitted the radio-to-optical spectrum by the synchrotron emission from the core, the large-scale jet can often contribute significantly to the radio emission. Consequently, the radio-to-optical emission intrinsic to the core in our model may require an even harder particle spectrum. It is generally accepted that only magnetic reconnection in a highly magnetised environment can produce a particle spectrum harder than~two \citep{guo2016,guo2021}. Interestingly, since the correlated quasi-periodic flux and polarisation variations detected by \citet{jorstad2022} suggest kink instabilities in BL Lac, and the very high optical polarisation degree reported by \citet{agudo2025} indicates the relaxation of a magnetic spring in the jet, combined with the fast variability observed in this work, it is plausible that the intense activity of BL Lac since approximately 2020 is associated with a giant eruption of magnetised plasma from the central engine, likely arising from a magnetically-arrested-disc jet scenario.

{Within this framework, our two-zone model essentially consists of a segment of the kinked jet (the core) and plasmoids (the blob) produced by magnetic reconnection, which is periodically triggered by the kinked jet. As they are co-spatial and physically connected, all bands above the synchrotron self-absorption limit should exhibit some degree of correlation. However, the production of plasmoids and their mergers during reconnection is somewhat random, as demonstrated in numerical simulations \citep{sironi2016,zhang2022}. Consequently, the VHE emission, which is fully dominated by the blob, may appear less correlated with the optical emission, which is fully dominated by the core.} {Moreover, the models presented here propose a simplified picture of the magnetic reconnection scenario. As the flare develops, plasmoids grow and can be ejected from the reconnection layer under various physical conditions, as indicated by recent PIC simulations \citep{sironi2016,zhang2024}. Although a fully numerical treatment of plasma dynamics during reconnection is beyond the scope of this paper, we modelled these plasmoids as a single blob with highly variable physical parameters, as demonstrated in several previous studies \citep[e.g.][]{christie2019}. We note that the two-zone model exhibits significant parameter degeneracy; consequently, our fitting parameters may not be unique.}

{Furthermore, we calculated the synchrotron and SSC cooling timescales of the blob in the observer frame for the electrons responsible for the GeV-TeV $\gamma$-ray emission. We estimated the synchrotron cooling timescales as \citep[see e.g.][]{abe2025}}
\begin{equation}
t_{sync}=\frac{(1+z)}{\delta_{blob}}\frac{6 \pi m_ec}{\sigma_T B^2 \gamma_e},
\end{equation}
where $m_e$ is the electron mass, $\sigma_T$ is the Thomson cross-section, $B$ is the magnetic field of the blob, and $\gamma_e$ is the Lorentz factor of the electrons. {We calculated the SSC cooling timescales  of the blob in the observer frame as $t_{SSC} = \frac{(1+z)}{\delta_{blob}} \, \gamma_e / \left( \frac{d\gamma_e}{dt} \right)_{SSC}$, with $\left( \frac{d\gamma_e}{dt} \right)_{SSC}$ expressed in the blob reference frame as \citep{2005MNRAS.363..954M}}
\begin{equation}
\left( \frac{d\gamma_e}{dt} \right)_{SSC} = \frac{4 c \sigma_T}{3} \gamma_e ^ 2 \int f_{KN}(4 \epsilon \gamma_e) \, \epsilon \, n_{ph}(\epsilon) \, d\epsilon \, ,
\end{equation}
{Here, $\epsilon$ denotes the photon energy in the rest frame of the blob in units of the electron rest-mass energy, $\epsilon = h \nu'/m_e c^2$, with  $\nu'$ representing  the photon frequency in the blob frame. The function $f_{KN}$ is defined by Eq. (C5) in \cite{2005MNRAS.363..954M}, $n_{ph}(\epsilon)$ is the synchrotron photon number density in the blob rest frame, calculated as in \cite{abe2025}.
We find that the SSC cooling timescales of the blob are shorter than the synchrotron timescales, consistent with the Compton dominance of $5$–$10$ inferred from the blob SED.
Table~\ref{tab:sedparams} lists the SSC cooling timescales of the blob at $\gamma_e=\gamma_{b}$. 
}For the two nights with intranight variability, our models predict SSC cooling timescales of the order of a few minutes for electrons radiating at GeV-TeV energies, in agreement with the minute-scale variability observed. Therefore, considering all of the above, a plausible scenario is that the observed VHE $\gamma$ rays flares are produced by freshly accelerated particles via magnetic reconnection with {cooling dominated by SSC processes.}

{We note that the scenario proposed here assumed a magnetically-dominated region. However, an alternative scenario, based on a two-zone model with a weakly magnetised plasma, can still offer a good fitting to the SED, provided that the blob has a higher Lorentz factor than the core, and exceeds the expected minimum Lorentz factor obtained from the TeV fast variability estimated above. The most natural explanation for a higher Lorentz factor in the small blob relative to the overall blazar zone is magnetic reconnection. Nevertheless, owing to the high degeneracy of blazar SED models (as noted in Sect.~\ref{sec6}), other combinations of parameters can yield valid representations of the SED and its evolution, possibly relying on non-magnetically dominated scenarios. } 

\section{Conclusions}\label{sec8}

BL Lac exhibited one of its three brightest $\gamma$-ray and multi-wavelength flares on record between September and November 2022. We characterised its HE and VHE emission, which extended up to a few TeV during the flares' brightest periods. A key feature of this event is the remarkably fast VHE $\gamma$-ray flux variability, with flux doubling timescales as short as 8.3 min, which sets tight constraints on the size of the VHE emitting region. This represents one of the fastest flux variations ever detected at VHEs from blazars.

The HE-VHE spectrum shows a shift in the peak towards higher energies during the most intense stages of the flare. Accompanying this behaviour, we report a softer-when-brighter evolution of the X-ray spectrum resulting from a clear transition of the X-ray emission from IC-dominated to synchrotron-dominated. Consequently, the source transitions from LSP to ISP, as previously observed in BL Lac \citep{nilsson2018}.

We modelled broadband emission within a leptonic scenario, motivated by recent IXPE results \citep{middei2023,peirson2023,agudo2025,liodakis2025}. One-zone models were unsuccessful in explaining the broadband emission. Therefore, we employed a co-spatial two-zone SSC model, also incorporating the EC contribution from the dusty torus. Within this scenario, the jet evolution is driven by changes in the electron populations of both regions, with injections of freshly accelerated electrons triggering the brightest outbursts. The observed minute-scale variability constrains the size of the compact blob region to $\lesssim4.5 \times 10^{14}$~cm for the adopted Doppler factor, $\delta_{blob}=33$, consistent with the value used in our model. This blob region is therefore responsible for the intranight $\gamma$-ray variability, which can be associated with plasmoids forming during a magnetic reconnection event.

   \bibliographystyle{aa}
   \bibliography{biblio.bib}

\begin{thebibliography}{94}
\expandafter\ifx\csname natexlab\endcsname\relax\def\natexlab#1{#1}\fi

\bibitem[{{Abdollahi} {et~al.}(2020){Abdollahi}, {Acero}, {Ackermann}, {Ajello}, {Atwood}, {Axelsson}, {Baldini}, {Ballet}, {Barbiellini}, {Bastieri}, {Becerra Gonzalez}, {Bellazzini}, {Berretta}, {Bissaldi}, {Blandford}, {Bloom}, {Bonino}, {Bottacini}, {Brandt}, {Bregeon}, {Bruel}, {Buehler}, {Burnett}, {Buson}, {Cameron}, {Caputo}, {Caraveo}, {Casandjian}, {Castro}, {Cavazzuti}, {Charles}, {Chaty}, {Chen}, {Cheung}, {Chiaro}, {Ciprini}, {Cohen-Tanugi}, {Cominsky}, {Coronado-Bl{\'a}zquez}, {Costantin}, {Cuoco}, {Cutini}, {D'Ammando}, {DeKlotz}, {de la Torre Luque}, {de Palma}, {Desai}, {Digel}, {Di Lalla}, {Di Mauro}, {Di Venere}, {Dom{\'\i}nguez}, {Dumora}, {Fana Dirirsa}, {Fegan}, {Ferrara}, {Franckowiak}, {Fukazawa}, {Funk}, {Fusco}, {Gargano}, {Gasparrini}, {Giglietto}, {Giommi}, {Giordano}, {Giroletti}, {Glanzman}, {Green}, {Grenier}, {Griffin}, {Grondin}, {Grove}, {Guiriec}, {Harding}, {Hayashi}, {Hays}, {Hewitt}, {Horan}, {J{\'o}hannesson}, {Johnson}, {Kamae}, {Kerr}, {Kocevski}, {Kovac'evic'},
  {Kuss}, {Landriu}, {Larsson}, {Latronico}, {Lemoine-Goumard}, {Li}, {Liodakis}, {Longo}, {Loparco}, {Lott}, {Lovellette}, {Lubrano}, {Madejski}, {Maldera}, {Malyshev}, {Manfreda}, {Marchesini}, {Marcotulli}, {Mart{\'\i}-Devesa}, {Martin}, {Massaro}, {Mazziotta}, {McEnery}, {Mereu}, {Meyer}, {Michelson}, {Mirabal}, {Mizuno}, {Monzani}, {Morselli}, {Moskalenko}, {Negro}, {Nuss}, {Ojha}, {Omodei}, {Orienti}, {Orlando}, {Ormes}, {Palatiello}, {Paliya}, {Paneque}, {Pei}, {Pe{\~n}a-Herazo}, {Perkins}, {Persic}, {Pesce-Rollins}, {Petrosian}, {Petrov}, {Piron}, {Poon}, {Porter}, {Principe}, {Rain{\`o}}, {Rando}, {Razzano}, {Razzaque}, {Reimer}, {Reimer}, {Remy}, {Reposeur}, {Romani}, {Saz Parkinson}, {Schinzel}, {Serini}, {Sgr{\`o}}, {Siskind}, {Smith}, {Spandre}, {Spinelli}, {Strong}, {Suson}, {Tajima}, {Takahashi}, {Tak}, {Thayer}, {Thompson}, {Tibaldo}, {Torres}, {Torresi}, {Valverde}, {Van Klaveren}, {van Zyl}, {Wood}, {Yassine}, \& {Zaharijas}}]{abdollahi2020}
{Abdollahi}, S., {Acero}, F., {Ackermann}, M., {et~al.} 2020, \apjs, 247, 33

\bibitem[{{Abe} {et~al.}(2023){Abe}, {Abe}, {Abe}, {Aguasca-Cabot}, {Agudo}, {Alvarez Crespo}, {Antonelli}, {Aramo}, {Arbet-Engels}, {Arcaro}, {Artero}, {Asano}, {Aubert}, {Baktash}, {Bamba}, {Baquero Larriva}, {Baroncelli}, {Barres de Almeida}, {Barrio}, {Batkovic}, {Baxter}, {Becerra Gonz{\'a}lez}, {Bernardini}, {Bernardos}, {Bernete Medrano}, {Berti}, {Bhattacharjee}, {Biederbeck}, {Bigongiari}, {Bissaldi}, {Blanch}, {Bonnoli}, {Bordas}, {Borghese}, {Bulgarelli}, {Burelli}, {Buscemi}, {Cardillo}, {Caroff}, {Carosi}, {Cassol}, {Cauz}, {Ceribella}, {Chai}, {Cheng}, {Chiavassa}, {Chikawa}, {Chytka}, {Cifuentes}, {Contreras}, {Cortina}, {Costantini}, {D'Amico}, {Dalchenko}, {De Angelis}, {de Bony de Lavergne}, {De Lotto}, {de Menezes}, {Deleglise}, {Delgado}, {Delgado Mengual}, {della Volpe}, {Dellaiera}, {Depaoli}, {Di Piano}, {Di Pierro}, {Di Tria}, {Di Venere}, {D{\'\i}az}, {Dominik}, {Dominis Prester}, {Donini}, {Dorner}, {Doro}, {Els{\"a}sser}, {Emery}, {Escudero}, {Fallah Ramazani}, {Ferrara},
  {Ferrarotto}, {Fiasson}, {Freixas Coromina}, {Fr{\"o}se}, {Fukami}, {Fukazawa}, {Garcia}, {Garcia L{\'o}pez}, {Gasbarra}, {Gasparrini}, {Geyer}, {Giesbrecht Paiva}, {Giglietto}, {Giordano}, {Giro}, {Gliwny}, {Godinovic}, {Grau}, {Green}, {Green}, {Gunji}, {Hackfeld}, {Hadasch}, {Hahn}, {Hashiyama}, {Hassan}, {Hayashi}, {Heckmann}, {Heller}, {Herrera Llorente}, {Hirotani}, {Hoffmann}, {Horns}, {Houles}, {Hrabovsky}, {Hrupec}, {Hui}, {H{\"u}tten}, {Iarlori}, {Imazawa}, {Inada}, {Inome}, {Ioka}, {Iori}, {Ishio}, {Iwamura}, {Jacquemont}, {Jimenez Martinez}, {Jurysek}, {Kagaya}, {Karas}, {Katagiri}, {Kataoka}, {Kerszberg}, {Kobayashi}, {Kong}, {Kubo}, {Kushida}, {Lainez}, {Lamanna}, {Lamastra}, {Le Flour}, {Linhoff}, {Longo}, {L{\'o}pez-Coto}, {L{\'o}pez-Moya}, {L{\'o}pez-Oramas}, {Loporchio}, {Lorini}, {Luque-Escamilla}, {Majumdar}, {Makariev}, {Mandat}, {Manganaro}, {Manic{\`o}}, {Mannheim}, {Mariotti}, {Marquez}, {Marsella}, {Mart{\'\i}}, {Martinez}, {Mart{\'\i}nez}, {Mart{\'\i}nez}, {Marusevec},
  {Mas-Aguilar}, {Maurin}, {Mazin}, {Mestre Guillen}, {Micanovic}, {Miceli}, {Miener}, {Miranda}, {Mirzoyan}, {Mizuno}, {Molero Gonzalez}, {Molina}, {Montaruli}, {Monteiro}, {Moralejo}, {Morcuende}, {Morselli}, {Mrakovcic}, {Murase}, {Nagai}, {Nagataki}, {Nakamori}, {Nickel}, {Nievas}, {Nishijima}, {Noda}, {Nosek}, {Nozaki}, {Ohishi}, {Ohtani}, {Oka}, {Okazaki}, {Okumura}, {Orito}, {Otero-Santos}, {Palatiello}, {Paneque}, {Pantaleo}, {Paoletti}, {Paredes}, {Pech}, {Pecimotika}, {Peresano}, {P{\'e}rez}, {Pietropaolo}, {Pirola}, {Plard}, {Podobnik}, {Poireau}, {Polo}, {Pons}, {Prandini}, {Prast}, {Principe}, {Priyadarshi}, {Prouza}, {Rando}, {Rhode}, {Rib{\'o}}, {Rizi}, {Rodriguez Fernandez}, {Ruiz}, {Saito}, {Sakurai}, {Sanchez}, {{\v{S}}ari{\'c}}, {Sato}, {Saturni}, {Schleicher}, {Schmuckermaier}, {Schubert}, {Schussler}, {Schweizer}, {Seglar Arroyo}, {Silvia}, {Sitarek}, {Sliusar}, {Spolon}, {Stri{\v{s}}kovi{\'c}}, {Strzys}, {Suda}, {Sunada}, {Tajima}, {Takahashi}, {Takahashi}, {Takata}, {Takeishi}, {Tam},
  {Tanaka}, {Tateishi}, {Tejedor}, {Temnikov}, {Terada}, {Terauchi}, {Terzic}, {Teshima}, {Tluczykont}, {Tokanai}, {Torres}, {Travnicek}, {Truzzi}, {Tutone}, {Uhlrich}, {Vacula}, {Vallania}, {van Scherpenberg}, {V{\'a}zquez Acosta}, {Verguilov}, {Viale}, {Vigliano}, {Vigorito}, {Vitale}, {Voutsinas}, {Vovk}, {Vuillaume}, {Walter}, {Will}, {Yamamoto}, {Yamazaki}, {Yoshida}, {Yoshikoshi}, {Zywucka}, {Bernl{\"o}hr}, {Gueta}, {Kosack}, {Maier}, \& {Watson}}]{abe2023}
{Abe}, H., {Abe}, K., {Abe}, S., {et~al.} 2023, \apj, 956, 80

\bibitem[{{Abe} {et~al.}(2025){Abe}, {Abe}, {Abhir}, {Abhishek}, {Acciari}, {Aguasca-Cabot}, {Agudo}, {Aniello}, {Ansoldi}, {Antonelli}, {Arbet Engels}, {Arcaro}, {Asano}, {Babi{\'c}}, {Barres de Almeida}, {Barrio}, {Barrios-Jim{\'e}nez}, {Batkovi{\'c}}, {Baxter}, {Becerra Gonz{\'a}lez}, {Bednarek}, {Bernardini}, {Bernete}, {Berti}, {Besenrieder}, {Bigongiari}, {Biland}, {Blanch}, {Bonnoli}, {Bo{\v{s}}njak}, {Bronzini}, {Burelli}, {Campoy-Ordaz}, {Carosi}, {Carosi}, {Carretero-Castrillo}, {Castro-Tirado}, {Cerasole}, {Ceribella}, {Chilingarian}, {Cifuentes}, {Colombo}, {Contreras}, {Cortina}, {Covino}, {D'Ammando}, {D'Amico}, {Da Vela}, {Dazzi}, {De Angelis}, {De Lotto}, {de Menezes}, {Delfino}, {Delgado}, {Delgado Mendez}, {Di Pierro}, {Di Tria}, {Di Venere}, {Dinesh}, {Dominis Prester}, {Donini}, {Dorner}, {Doro}, {Eisenberger}, {Elsaesser}, {Escudero}, {Fari{\~n}a}, {Foffano}, {Font}, {Fr{\"o}se}, {Fukazawa}, {Garc{\'\i}a L{\'o}pez}, {Garczarczyk}, {Gasparyan}, {Gaug}, {Giesbrecht Paiva}, {Giglietto},
  {Giordano}, {Gliwny}, {Gradetzke}, {Grau}, {Green}, {Green}, {G{\"u}nther}, {Hadasch}, {Hahn}, {Hassan}, {Heckmann}, {Herrera Llorente}, {Hrupec}, {Imazawa}, {Israyelyan}, {Itokawa}, {Jim{\'e}nez Mart{\'\i}nez}, {Jim{\'e}nez Quiles}, {Jormanainen}, {Kankkunen}, {Kayanoki}, {Kerszberg}, {Khachatryan}, {Kluge}, {Kobayashi}, {Konrad}, {Kouch}, {Kubo}, {Kushida}, {L{\'a}inez}, {Lamastra}, {Lindfors}, {Lombardi}, {Longo}, {L{\'o}pez-Coto}, {L{\'o}pez-Moya}, {L{\'o}pez-Oramas}, {Loporchio}, {Lorini}, {Luli{\'c}}, {Lyard}, {Majumdar}, {Makariev}, {Maneva}, {Manganaro}, {Mangano}, {Mannheim}, {Mariotti}, {Mart{\'\i}nez}, {Maru{\v{s}}evec}, {Mas-Aguilar}, {Mazin}, {Menchiari}, {Mender}, {Miceli}, {Miranda}, {Mirzoyan}, {Molero Gonz{\'a}lez}, {Molina}, {Mondal}, {Moralejo}, {Nakamori}, {Nanci}, {Neustroev}, {Nickel}, {Nievas Rosillo}, {Nigro}, {Nikoli{\'c}}, {Nilsson}, {Nishijima}, {Njoh Ekoume}, {Noda}, {Nozaki}, {Okumura}, {Paiano}, {Paneque}, {Paoletti}, {Paredes}, {Pavleti{\'c}}, {Peresano}, {Persic}, {Pihet},
  {Pirola}, {Podobnik}, {Prada Moroni}, {Prandini}, {Principe}, {Rhode}, {Rib{\'o}}, {Rico}, {Righi}, {Sahakyan}, {Saito}, {Saturni}, {Schmitz}, {Schmuckermaier}, {Schubert}, {Sciaccaluga}, {Silvestri}, {Sitarek}, {Sliusar}, {Sobczynska}, {Stamerra}, {Stri{\v{s}}kovi{\'c}}, {Strom}, {Strzys}, {Suda}, {Tajima}, {Takahashi}, {Takeishi}, {Temnikov}, {Terauchi}, {Terzi{\'c}}, {Teshima}, {Tutone}, {Ubach}, {van Scherpenberg}, {Vazquez Acosta}, {Ventura}, {Verna}, {Viale}, {Vigliano}, \& {Vigorito}}]{abe2025}
{Abe}, K., {Abe}, S., {Abhir}, J., {et~al.} 2025, \aap, 697, A172

\bibitem[{{Abeysekara} {et~al.}(2018){Abeysekara}, {Benbow}, {Bird}, {Brantseg}, {Brose}, {Buchovecky}, {Buckley}, {Bugaev}, {Connolly}, {Cui}, {Daniel}, {Falcone}, {Feng}, {Finley}, {Fortson}, {Furniss}, {Gillanders}, {Gunawardhana}, {H{\"u}tten}, {Hanna}, {Hervet}, {Holder}, {Hughes}, {Humensky}, {Johnson}, {Kaaret}, {Kar}, {Kertzman}, {Krennrich}, {Lang}, {Lin}, {McArthur}, {Moriarty}, {Mukherjee}, {O'Brien}, {Ong}, {Otte}, {Park}, {Petrashyk}, {Pohl}, {Pueschel}, {Quinn}, {Ragan}, {Reynolds}, {Richards}, {Roache}, {Rulten}, {Sadeh}, {Santander}, {Sembroski}, {Shahinyan}, {Wakely}, {Weinstein}, {Wells}, {Wilcox}, {Williams}, {Zitzer}, {VERITAS Collaboration}, {Jorstad}, {Marscher}, {Lister}, {Kovalev}, {Pushkarev}, {Savolainen}, {Agudo}, {Molina}, {G{\'o}mez}, {Larionov}, {Borman}, {Mokrushina}, {Tornikoski}, {L{\"a}hteenm{\"a}ki}, {Chamani}, {Enestam}, {Kiehlmann}, {Hovatta}, {Smith}, \& {Pontrelli}}]{abeysekara2018}
{Abeysekara}, A.~U., {Benbow}, W., {Bird}, R., {et~al.} 2018, \apj, 856, 95

\bibitem[{Acero {et~al.}(2024)Acero, Bernete, Biederbeck, Djuvsland, Donath, Feijen, Fröse, Galelli, Khélifi, Konrad, Kornecki, Linhoff, McKee, Mender, Morcuende, Olivera-Nieto, Pintore, Punch, Regeard, Remy, Sinha, Stapel, Streil, Terrier, \& Unbehaun}]{gammapy:zenodo-1.2}
Acero, F., Bernete, J., Biederbeck, N., {et~al.} 2024, Gammapy v1.2: Python toolbox for gamma-ray astronomy

\bibitem[{{Acharyya} {et~al.}(2019){Acharyya}, {Agudo}, {Ang{\"u}ner}, {Alfaro}, {Alfaro}, {Alispach}, {Aloisio}, {Alves Batista}, {Amans}, {Amati}, {Amato}, {Ambrosi}, {Antonelli}, {Aramo}, {Armstrong}, {Arqueros}, {Arrabito}, {Asano}, {Ashkar}, {Balazs}, {Balbo}, {Balmaverde}, {Barai}, {Barbano}, {Barkov}, {Barres de Almeida}, {Barrio}, {Bastieri}, {Becerra Gonz{\'a}lez}, {Becker Tjus}, {Bellizzi}, {Benbow}, {Bernardini}, {Bernardos}, {Bernl{\"o}hr}, {Berti}, {Berton}, {Bertucci}, {Beshley}, {Biasuzzi}, {Bigongiari}, {Bird}, {Bissaldi}, {Biteau}, {Blanch}, {Blazek}, {Boisson}, {Bonanno}, {Bonardi}, {Bonavolont{\'a}}, {Bonnoli}, {Bordas}, {B{\"o}ttcher}, {Bregeon}, {Brill}, {Brown}, {Br{\"u}gge}, {Brun}, {Bruno}, {Bulgarelli}, {Bulik}, {Burton}, {Burtovoi}, {Busetto}, {Cameron}, {Canestrari}, {Capalbi}, {Caproni}, {Capuzzo-Dolcetta}, {Caraveo}, {Caroff}, {Carosi}, {Casanova}, {Cascone}, {Cassol}, {Catalani}, {Catalano}, {Cauz}, {Cerruti}, {Chaty}, {Chen}, {Chernyakova}, {Chiaro}, {Cie{\'s}lar}, {Colak},
  {Conforti}, {Congiu}, {Contreras}, {Cortina}, {Costa}, {Costantini}, {Cotter}, {Cristofari}, {Cumani}, {Cusumano}, {D'A{\'\i}}, {D'Ammando}, {Dangeon}, {Da Vela}, {Dazzi}, {De Angelis}, {De Caprio}, {de C{\'a}ssia dos Anjos}, {De Frondat}, {de Gouveia Dal Pino}, {De Lotto}, {De Martino}, {de Naurois}, {de O{\~n}a Wilhelmi}, {de Palma}, {de Souza}, {Del Santo}, {Delgado}, {della Volpe}, {Di Girolamo}, {Di Pierro}, {Di Venere}, {D{\'\i}az}, {Diebold}, {Djannati-Ata{\"\i}}, {Dmytriiev}, {Dominis Prester}, {Donini}, {Dorner}, {Doro}, {Dournaux}, {Ebr}, {Ekoume}, {Els{\"a}sser}, {Emery}, {Falceta-Goncalves}, {Fedorova}, {Fegan}, {Feng}, {Ferrand}, {Fiandrini}, {Fiasson}, {Filipovic}, {Fioretti}, {Fiori}, {Flis}, {Fonseca}, {Fontaine}, {Freixas Coromina}, {Fukami}, {Fukui}, {Funk}, {F{\"u}{\ss}ling}, {Gaggero}, {Galanti}, {Garcia L{\'o}pez}, {Garczarczyk}, {Gascon}, {Gasparetto}, {Gaug}, {Ghalumyan}, {Gianotti}, {Giavitto}, {Giglietto}, {Giordano}, {Giroletti}, {Gironnet}, {Glicenstein}, {Gnatyk}, {Goldoni},
  {Gonz{\'a}lez}, {Gonz{\'a}lez}, {Gourgouliatos}, {Grabarczyk}, {Granot}, {Green}, {Greenshaw}, {Grondin}, {Gueta}, {Hadasch}, {Hassan}, {Hayashida}, {Heller}, {Hervet}, {Hinton}, {Hiroshima}, {Hnatyk}, {Hofmann}, {Horvath}, {Hrabovsky}, {Hrupec}, {Humensky}, {H{\"u}tten}, {Inada}, {Iocco}, {Ionica}, {Iori}, {Iwamura}, {Jamrozy}, {Janecek}, {Jankowsky}, {Jean}, {Jouvin}, {Jurysek}, {Kaaret}, {Kadowaki}, {Karkar}, {Kerszberg}, {Kh{\'e}lifi}, {Kieda}, {Kimeswenger}, {Klu{\'z}niak}, {Knapp}, {Kn{\"o}dlseder}, {Kobayashi}, {Koch}, {Kocot}, {Komin}, {Kong}, {Kowal}, {Krause}, {Kubo}, {Kushida}, {Kushwaha}, {La Parola}, {La Rosa}, {Lallena Arquillo}, {Lang}, {Lapington}, {Le Blanc}, {Lefaucheur}, {Leigui de Oliveira}, {Lemoine-Goumard}, {Lenain}, {Leto}, {Lico}, {Lindfors}, {Lohse}, {Lombardi}, {Longo}, {Lopez}, {L{\'o}pez}, {Lopez-Oramas}, {L{\'o}pez-Coto}, {Loporchio}, {Luque-Escamilla}, {Lyard}, {Maccarone}, {Mach}, {Maggio}, {Majumdar}, {Malaguti}, {Mallamaci}, {Mandat}, {Maneva}, {Manganaro}, {Mangano},
  {Marculewicz}, {Mariotti}, {Mart{\'\i}}, {Mart{\'\i}nez}, {Mart{\'\i}nez}, {Mart{\'\i}nez-Huerta}, {Masuda}, {Maxted}, {Mazin}, {Meunier}, {Meyer}, {Micanovic}, {Millul}, {Minaya}, {Mitchell}, {Mizuno}, {Moderski}, {Mohrmann}, {Montaruli}, {Moralejo}, {Morcuende}, {Morlino}, {Morselli}, {Moulin}, {Mukherjee}, {Munar}, {Mundell}, {Murach}, {Nagai}, {Nagayoshi}, {Naito}, {Nakamori}, {Nemmen}, {Niemiec}, {Nieto}, {Nievas Rosillo}, {Niko{\l}ajuk}, {Ninci}, {Nishijima}, {Noda}, {Nosek}, {N{\"o}the}, {Nozaki}, {Ohishi}, {Ohtani}, {Okumura}, {Ong}, {Orienti}, {Orito}, {Ostrowski}, {Otte}, {Ou}, {Oya}, {Pagliaro}, {Palatiello}, {Palatka}, {Paoletti}, {Paredes}, {Pareschi}, {Parmiggiani}, {Parsons}, {Patricelli}, {Pe'er}, {Pech}, {Pe{\~n}il Del Campo}, {P{\'e}rez-Romero}, {Perri}, {Persic}, {Petrucci}, {Petruk}, {Pfrang}, {Piel}, {Pietropaolo}, {Pohl}, {Polo}, {Poutanen}, {Prandini}, {Produit}, {Prokoph}, {Prouza}, {Przybilski}, {P{\"u}hlhofer}, {Punch}, {Queiroz}, {Quirrenbach}, {Rain{\`o}}, {Rando}, {Razzaque},
  {Reimer}, {Renault-Tinacci}, {Renier}, {Ribeiro}, {Rib{\'o}}, {Rico}, {Rieger}, {Rizi}, {Rodriguez Fernandez}, {Rodriguez-Ramirez}, {Rodr{\'\i}-guez V{\'a}zquez}, {Romano}, {Romeo}, {Roncadelli}, {Rosado}, {Rowell}, {Rudak}, {Rugliancich}, {Rulten}, {Sadeh}, {Saha}, {Saito}, {Sakurai}, {Salesa Greus}, {Sangiorgi}, {Sano}, {Santander}, {Santangelo}, {Santos-Lima}, {Sanuy}, {Satalecka}, {Saturni}, {Sawangwit}, {Schlenstedt}, {Schovanek}, {Schussler}, {Schwanke}, {Sciacca}, {Scuderi}, {Sedlaczek}, {Seglar-Arroyo}, {Sergijenko}, {Seweryn}, {Shalchi}, {Shellard}, {Siejkowski}, {Sillanp{\"a}{\"a}}, {Sinha}, {Sironi}, {Sliusar}, {Slowikowska}, {Sol}, {Specovius}, {Spencer}, {Spengler}, {Stamerra}, {Stani{\v{c}}}, {Stawarz}, {Stefanik}, {Stolarczyk}, {Straumann}, {Suomijarvi}, {{\'S}wierk}, {Szepieniec}, {Tagliaferri}, {Tajima}, {Tam}, {Tavecchio}, {Taylor}, {Tejedor}, {Temnikov}, {Terzic}, {Testa}, {Tibaldo}, {Todero Peixoto}, {Tokanai}, {Tomankova}, {Tonev}, {Torres}, {Tosti}, {Tosti}, {Tothill}, {Toussenel},
  {Tovmassian}, {Travnicek}, {Trichard}, {Umana}, {Vagelli}, {Valentino}, {Vallage}, {Vallania}, {Valore}, {Vandenbroucke}, {Varner}, {Vasileiadis}, {Vassiliev}, {V{\'a}zquez Acosta}, {Vecchi}, {Vercellone}, {Vergani}, {Vettolani}, {Viana}, {Vigorito}, {Vink}, {Vitale}, {Voelk}, {Vollhardt}, {Vorobiov}, {Wagner}, {Walter}, {Werner}, {White}, {Wierzcholska}, {Will}, {Williams}, {Wischnewski}, {Yang}, {Yoshida}, {Yoshikoshi}, {Zacharias}, {Zampieri}, {Zavrtanik}, {Zavrtanik}, {Zdziarski}, {Zech}, {Zechlin}, {Zenin}, {Zhdanov}, {Zimmer}, \& {Zorn}}]{acharyya2019}
{Acharyya}, A., {Agudo}, I., {Ang{\"u}ner}, E.~O., {et~al.} 2019, Astroparticle Physics, 111, 35

\bibitem[{{Ackermann} {et~al.}(2011){Ackermann}, {Ajello}, {Allafort}, {Antolini}, {Atwood}, {Axelsson}, {Baldini}, {Ballet}, {Barbiellini}, {Bastieri}, {Bechtol}, {Bellazzini}, {Berenji}, {Blandford}, {Bloom}, {Bonamente}, {Borgland}, {Bottacini}, {Bouvier}, {Bregeon}, {Brigida}, {Bruel}, {Buehler}, {Burnett}, {Buson}, {Caliandro}, {Cameron}, {Caraveo}, {Casandjian}, {Cavazzuti}, {Cecchi}, {Charles}, {Cheung}, {Chiang}, {Ciprini}, {Claus}, {Cohen-Tanugi}, {Conrad}, {Costamante}, {Cutini}, {de Angelis}, {de Palma}, {Dermer}, {Digel}, {Silva}, {Drell}, {Dubois}, {Escande}, {Favuzzi}, {Fegan}, {Ferrara}, {Finke}, {Focke}, {Fortin}, {Frailis}, {Fukazawa}, {Funk}, {Fusco}, {Gargano}, {Gasparrini}, {Gehrels}, {Germani}, {Giebels}, {Giglietto}, {Giommi}, {Giordano}, {Giroletti}, {Glanzman}, {Godfrey}, {Grenier}, {Grove}, {Guiriec}, {Gustafsson}, {Hadasch}, {Hayashida}, {Hays}, {Healey}, {Horan}, {Hou}, {Hughes}, {Iafrate}, {J{\'o}hannesson}, {Johnson}, {Johnson}, {Kamae}, {Katagiri}, {Kataoka}, {Kn{\"o}dlseder},
  {Kuss}, {Lande}, {Larsson}, {Latronico}, {Longo}, {Loparco}, {Lott}, {Lovellette}, {Lubrano}, {Madejski}, {Mazziotta}, {McConville}, {McEnery}, {Michelson}, {Mitthumsiri}, {Mizuno}, {Moiseev}, {Monte}, {Monzani}, {Moretti}, {Morselli}, {Moskalenko}, {Murgia}, {Nakamori}, {Naumann-Godo}, {Nolan}, {Norris}, {Nuss}, {Ohno}, {Ohsugi}, {Okumura}, {Omodei}, {Orienti}, {Orlando}, {Ormes}, {Ozaki}, {Paneque}, {Parent}, {Pesce-Rollins}, {Pierbattista}, {Piranomonte}, {Piron}, {Pivato}, {Porter}, {Rain{\`o}}, {Rando}, {Razzano}, {Razzaque}, {Reimer}, {Reimer}, {Ritz}, {Rochester}, {Romani}, {Roth}, {Sanchez}, {Sbarra}, {Scargle}, {Schalk}, {Sgr{\`o}}, {Shaw}, {Siskind}, {Spandre}, {Spinelli}, {Strong}, {Suson}, {Tajima}, {Takahashi}, {Takahashi}, {Tanaka}, {Thayer}, {Thayer}, {Thompson}, {Tibaldo}, {Tinivella}, {Torres}, {Tosti}, {Troja}, {Uchiyama}, {Vandenbroucke}, {Vasileiou}, {Vianello}, {Vitale}, {Waite}, {Wallace}, {Wang}, {Winer}, {Wood}, {Wood}, \& {Zimmer}}]{ackermann2011}
{Ackermann}, M., {Ajello}, M., {Allafort}, A., {et~al.} 2011, \apj, 743, 171

\bibitem[{{Ackermann} {et~al.}(2016){Ackermann}, {Anantua}, {Asano}, {Baldini}, {Barbiellini}, {Bastieri}, {Becerra Gonzalez}, {Bellazzini}, {Bissaldi}, {Blandford}, {Bloom}, {Bonino}, {Bottacini}, {Bruel}, {Buehler}, {Caliandro}, {Cameron}, {Caragiulo}, {Caraveo}, {Cavazzuti}, {Cecchi}, {Cheung}, {Chiang}, {Chiaro}, {Ciprini}, {Cohen-Tanugi}, {Costanza}, {Cutini}, {D'Ammando}, {de Palma}, {Desiante}, {Digel}, {Di Lalla}, {Di Mauro}, {Di Venere}, {Drell}, {Favuzzi}, {Fegan}, {Ferrara}, {Fukazawa}, {Funk}, {Fusco}, {Gargano}, {Gasparrini}, {Giglietto}, {Giordano}, {Giroletti}, {Grenier}, {Guillemot}, {Guiriec}, {Hayashida}, {Hays}, {Horan}, {J{\'o}hannesson}, {Kensei}, {Kocevski}, {Kuss}, {La Mura}, {Larsson}, {Latronico}, {Li}, {Longo}, {Loparco}, {Lott}, {Lovellette}, {Lubrano}, {Madejski}, {Magill}, {Maldera}, {Manfreda}, {Mayer}, {Mazziotta}, {Michelson}, {Mirabal}, {Mizuno}, {Monzani}, {Morselli}, {Moskalenko}, {Nalewajko}, {Negro}, {Nuss}, {Ohsugi}, {Orlando}, {Paneque}, {Perkins}, {Pesce-Rollins},
  {Piron}, {Pivato}, {Porter}, {Principe}, {Rando}, {Razzano}, {Razzaque}, {Reimer}, {Scargle}, {Sgr{\`o}}, {Sikora}, {Simone}, {Siskind}, {Spada}, {Spinelli}, {Stawarz}, {Thayer}, {Thompson}, {Torres}, {Troja}, {Uchiyama}, {Yuan}, \& {Zimmer}}]{ackermann2016}
{Ackermann}, M., {Anantua}, R., {Asano}, K., {et~al.} 2016, \apjl, 824, L20

\bibitem[{Aguasca-Cabot {et~al.}(2023)Aguasca-Cabot, Donath, Feijen, Gr\'eaux, Giunti, Kh\'elifi, Linhoff, Mender, Mohrmann, Nigro, Olivera-Nieto, Pintore, Regeard, Remy, Sinha, Streil, \& Terrier}]{aguasca-Cabot2023}
Aguasca-Cabot, A., Donath, A., Feijen, K., {et~al.} 2023, Gammapy: Python toolbox for gamma-ray astronomy

\bibitem[{{Agudo} {et~al.}(2025){Agudo}, {Liodakis}, {Otero-Santos}, {Middei}, {Marscher}, {Jorstad}, {Zhang}, {Li}, {Di Gesu}, {Romani}, {Kim}, {Fenu}, {Marshall}, {Pacciani}, {Escudero Pedrosa}, {Aceituno}, {Ag{\'\i}s-Gonz{\'a}lez}, {Bonnoli}, {Casanova}, {Morcuende}, {Piirola}, {Sota}, {Kouch}, {Lindfors}, {McCall}, {Jermak}, {Steele}, {Borman}, {Grishina}, {Hagen-Thorn}, {Kopatskaya}, {Larionova}, {Morozova}, {Savchenko}, {Shishkina}, {Troitskiy}, {Troitskaya}, {Vasilyev}, {Zhovtan}, {Myserlis}, {Gurwell}, {Keating}, {Rao}, {Kang}, {Lee}, {Kim}, {Cheong}, {Jeong}, {Angelakis}, {Kraus}, {Blinov}, {Maharana}, {Bachev}, {Jormanainen}, {Nilsson}, {Fallah Ramazani}, {Casadio}, {Fuentes}, {Traianou}, {Thum}, {G{\'o}mez}, {Antonelli}, {Bachetti}, {Baldini}, {Baumgartner}, {Bellazzini}, {Bianchi}, {Bongiorno}, {Bonino}, {Brez}, {Bucciantini}, {Capitanio}, {Castellano}, {Cavazzuti}, {Chen}, {Ciprini}, {Costa}, {De Rosa}, {Del Monte}, {Di Lalla}, {Di Marco}, {Donnarumma}, {Doroshenko}, {Dov{\v{c}}iak}, {Ehlert},
  {Enoto}, {Evangelista}, {Fabiani}, {Ferrazzoli}, {Garc{\'\i}a}, {Gunji}, {Hayashida}, {Heyl}, {Iwakiri}, {Kaaret}, {Karas}, {Kislat}, {Kitaguchi}, {Kolodziejczak}, {Krawczynski}, {La Monaca}, {Latronico}, {Maldera}, {Manfreda}, {Marin}, {Marinucci}, {Massaro}, {Matt}, {Mitsuishi}, {Mizuno}, {Muleri}, {Negro}, {Ng}, {O'Dell}, {Omodei}, {Oppedisano}, {Papitto}, {Pavlov}, {Peirson}, {Perri}, {Pesce-Rollins}, {Petrucci}, {Pilia}, {Possenti}, {Poutanen}, {Puccetti}, {Ramsey}, {Rankin}, {Ratheesh}, {Roberts}, {Sgr{\`o}}, {Slane}, {Soffitta}, {Spandre}, {Swartz}, {Tamagawa}, {Tavecchio}, {Taverna}, {Tawara}, {Tennant}, {Thomas}, {Tombesi}, {Trois}, {Tsygankov}, {Turolla}, {Vink}, {Weisskopf}, {Wu}, {Xie}, \& {Zane}}]{agudo2025}
{Agudo}, I., {Liodakis}, I., {Otero-Santos}, J., {et~al.} 2025, \apjl, 985, L15

\bibitem[{{Agudo} {et~al.}(2012){Agudo}, {Molina}, {G{\'o}mez}, {Marscher}, {Jorstad}, \& {Heidt}}]{agudo2012}
{Agudo}, I., {Molina}, S.~N., {G{\'o}mez}, J.~L., {et~al.} 2012, in International Journal of Modern Physics Conference Series, Vol.~8, International Journal of Modern Physics Conference Series, 299--302

\bibitem[{{Aharonian} {et~al.}(2007){Aharonian}, {Akhperjanian}, {Bazer-Bachi}, {Behera}, {Beilicke}, {Benbow}, {Berge}, {Bernl{\"o}hr}, {Boisson}, {Bolz}, {Borrel}, {Boutelier}, {Braun}, {Brion}, {Brown}, {B{\"u}hler}, {B{\"u}sching}, {Bulik}, {Carrigan}, {Chadwick}, {Clapson}, {Chounet}, {Coignet}, {Cornils}, {Costamante}, {Degrange}, {Dickinson}, {Djannati-Ata{\"\i}}, {Domainko}, {Drury}, {Dubus}, {Dyks}, {Egberts}, {Emmanoulopoulos}, {Espigat}, {Farnier}, {Feinstein}, {Fiasson}, {F{\"o}rster}, {Fontaine}, {Funk}, {Funk}, {F{\"u}{\ss}ling}, {Gallant}, {Giebels}, {Glicenstein}, {Gl{\"u}ck}, {Goret}, {Hadjichristidis}, {Hauser}, {Hauser}, {Heinzelmann}, {Henri}, {Hermann}, {Hinton}, {Hoffmann}, {Hofmann}, {Holleran}, {Hoppe}, {Horns}, {Jacholkowska}, {de Jager}, {Kendziorra}, {Kerschhaggl}, {Kh{\'e}lifi}, {Komin}, {Kosack}, {Lamanna}, {Latham}, {Le Gallou}, {Lemi{\`e}re}, {Lemoine-Goumard}, {Lenain}, {Lohse}, {Martin}, {Martineau-Huynh}, {Marcowith}, {Masterson}, {Maurin}, {McComb}, {Moderski}, {Moulin}, {de
  Naurois}, {Nedbal}, {Nolan}, {Olive}, {Orford}, {Osborne}, {Ostrowski}, {Panter}, {Pedaletti}, {Pelletier}, {Petrucci}, {Pita}, {P{\"u}hlhofer}, {Punch}, {Ranchon}, {Raubenheimer}, {Raue}, {Rayner}, {Renaud}, {Ripken}, {Rob}, {Rolland}, {Rosier-Lees}, {Rowell}, {Rudak}, {Ruppel}, {Sahakian}, {Santangelo}, {Saug{\'e}}, {Schlenker}, {Schlickeiser}, {Schr{\"o}der}, {Schwanke}, {Schwarzburg}, {Schwemmer}, {Shalchi}, {Sol}, {Spangler}, {Stawarz}, {Steenkamp}, {Stegmann}, {Superina}, {Tam}, {Tavernet}, {Terrier}, {van Eldik}, {Vasileiadis}, {Venter}, {Vialle}, {Vincent}, {Vivier}, {V{\"o}lk}, {Volpe}, {Wagner}, {Ward}, \& {Zdziarski}}]{aharonian2007}
{Aharonian}, F., {Akhperjanian}, A.~G., {Bazer-Bachi}, A.~R., {et~al.} 2007, \apjl, 664, L71

\bibitem[{{Aharonian}(2000)}]{aharonian2000}
{Aharonian}, F.~A. 2000, \na, 5, 377

\bibitem[{{Albert} {et~al.}(2007){Albert}, {Aliu}, {Anderhub}, {Antoranz}, {Armada}, {Baixeras}, {Barrio}, {Bartko}, {Bastieri}, {Becker}, {Bednarek}, {Berger}, {Bigongiari}, {Biland}, {Bock}, {Bordas}, {Bosch-Ramon}, {Bretz}, {Britvitch}, {Camara}, {Carmona}, {Chilingarian}, {Coarasa}, {Commichau}, {Contreras}, {Cortina}, {Costado}, {Curtef}, {Danielyan}, {Dazzi}, {De Angelis}, {Delgado}, {de los Reyes}, {De Lotto}, {Domingo-Santamar{\'\i}a}, {Dorner}, {Doro}, {Errando}, {Fagiolini}, {Ferenc}, {Fern{\'a}ndez}, {Firpo}, {Flix}, {Fonseca}, {Font}, {Fuchs}, {Galante}, {Garc{\'\i}a-L{\'o}pez}, {Garczarczyk}, {Gaug}, {Giller}, {Goebel}, {Hakobyan}, {Hayashida}, {Hengstebeck}, {Herrero}, {H{\"o}hne}, {Hose}, {Hsu}, {Jacon}, {Jogler}, {Kosyra}, {Kranich}, {Kritzer}, {Laille}, {Lindfors}, {Lombardi}, {Longo}, {L{\'o}pez}, {L{\'o}pez}, {Lorenz}, {Majumdar}, {Maneva}, {Mannheim}, {Mansutti}, {Mariotti}, {Mart{\'\i}nez}, {Mazin}, {Merck}, {Meucci}, {Meyer}, {Miranda}, {Mirzoyan}, {Mizobuchi}, {Moralejo}, {Nilsson},
  {Ninkovic}, {O{\~n}a-Wilhelmi}, {Otte}, {Oya}, {Paneque}, {Panniello}, {Paoletti}, {Paredes}, {Pasanen}, {Pascoli}, {Pauss}, {Pegna}, {Persic}, {Peruzzo}, {Piccioli}, {Poller}, {Prandini}, {Puchades}, {Raymers}, {Rhode}, {Rib{\'o}}, {Rico}, {Rissi}, {Robert}, {R{\"u}gamer}, {Saggion}, {S{\'a}nchez}, {Sartori}, {Scalzotto}, {Scapin}, {Schmitt}, {Schweizer}, {Shayduk}, {Shinozaki}, {Shore}, {Sidro}, {Sillanp{\"a}{\"a}}, {Sobczynska}, {Stamerra}, {Stark}, {Takalo}, {Temnikov}, {Tescaro}, {Teshima}, {Tonello}, {Torres}, {Turini}, {Vankov}, {Vitale}, {Wagner}, {Wibig}, {Wittek}, {Zandanel}, {Zanin}, \& {Zapatero}}]{albert2007}
{Albert}, J., {Aliu}, E., {Anderhub}, H., {et~al.} 2007, \apjl, 666, L17

\bibitem[{{Aleksi{\'c}} {et~al.}(2016){Aleksi{\'c}}, {Ansoldi}, {Antonelli}, {Antoranz}, {Babic}, {Bangale}, {Barcel{\'o}}, {Barrio}, {Becerra Gonz{\'a}lez}, {Bednarek}, {Bernardini}, {Biasuzzi}, {Biland}, {Bitossi}, {Blanch}, {Bonnefoy}, {Bonnoli}, {Borracci}, {Bretz}, {Carmona}, {Carosi}, {Cecchi}, {Colin}, {Colombo}, {Contreras}, {Corti}, {Cortina}, {Covino}, {Da Vela}, {Dazzi}, {De Angelis}, {De Caneva}, {De Lotto}, {de O{\~n}a Wilhelmi}, {Delgado Mendez}, {Dettlaff}, {Dominis Prester}, {Dorner}, {Doro}, {Einecke}, {Eisenacher}, {Elsaesser}, {Fidalgo}, {Fink}, {Fonseca}, {Font}, {Frantzen}, {Fruck}, {Galindo}, {Garc{\'\i}a L{\'o}pez}, {Garczarczyk}, {Garrido Terrats}, {Gaug}, {Giavitto}, {Godinovi{\'c}}, {Gonz{\'a}lez Mu{\~n}oz}, {Gozzini}, {Haberer}, {Hadasch}, {Hanabata}, {Hayashida}, {Herrera}, {Hildebrand}, {Hose}, {Hrupec}, {Idec}, {Illa}, {Kadenius}, {Kellermann}, {Knoetig}, {Kodani}, {Konno}, {Krause}, {Kubo}, {Kushida}, {La Barbera}, {Lelas}, {Lemus}, {Lewandowska}, {Lindfors}, {Lombardi},
  {Longo}, {L{\'o}pez}, {L{\'o}pez-Coto}, {L{\'o}pez-Oramas}, {Lorca}, {Lorenz}, {Lozano}, {Makariev}, {Mallot}, {Maneva}, {Mankuzhiyil}, {Mannheim}, {Maraschi}, {Marcote}, {Mariotti}, {Mart{\'\i}nez}, {Mazin}, {Menzel}, {Miranda}, {Mirzoyan}, {Moralejo}, {Munar-Adrover}, {Nakajima}, {Negrello}, {Neustroev}, {Niedzwiecki}, {Nilsson}, {Nishijima}, {Noda}, {Orito}, {Overkemping}, {Paiano}, {Palatiello}, {Paneque}, {Paoletti}, {Paredes}, {Paredes-Fortuny}, {Persic}, {Poutanen}, {Prada Moroni}, {Prandini}, {Puljak}, {Reinthal}, {Rhode}, {Rib{\'o}}, {Rico}, {Rodriguez Garcia}, {R{\"u}gamer}, {Saito}, {Saito}, {Satalecka}, {Scalzotto}, {Scapin}, {Schultz}, {Schlammer}, {Schmidl}, {Schweizer}, {Shore}, {Sillanp{\"a}{\"a}}, {Sitarek}, {Snidaric}, {Sobczynska}, {Spanier}, {Stamerra}, {Steinbring}, {Storz}, {Strzys}, {Takalo}, {Takami}, {Tavecchio}, {Tejedor}, {Temnikov}, {Terzi{\'c}}, {Tescaro}, {Teshima}, {Thaele}, {Tibolla}, {Torres}, {Toyama}, {Treves}, {Vogler}, {Wetteskind}, {Will}, \& {Zanin}}]{aleksic2016}
{Aleksi{\'c}}, J., {Ansoldi}, S., {Antonelli}, L.~A., {et~al.} 2016, Astroparticle Physics, 72, 76

\bibitem[{{Arlen} {et~al.}(2013){Arlen}, {Aune}, {Beilicke}, {Benbow}, {Bouvier}, {Buckley}, {Bugaev}, {Cesarini}, {Ciupik}, {Connolly}, {Cui}, {Dickherber}, {Dumm}, {Errando}, {Falcone}, {Federici}, {Feng}, {Finley}, {Finnegan}, {Fortson}, {Furniss}, {Galante}, {Gall}, {Griffin}, {Grube}, {Gyuk}, {Hanna}, {Holder}, {Humensky}, {Kaaret}, {Karlsson}, {Kertzman}, {Khassen}, {Kieda}, {Krawczynski}, {Krennrich}, {Maier}, {Moriarty}, {Mukherjee}, {Nelson}, {O'Faol{\'a}in de Bhr{\'o}ithe}, {Ong}, {Orr}, {Park}, {Perkins}, {Pichel}, {Pohl}, {Prokoph}, {Quinn}, {Ragan}, {Reyes}, {Reynolds}, {Roache}, {Saxon}, {Schroedter}, {Sembroski}, {Staszak}, {Telezhinsky}, {Te{\v{s}}i{\'c}}, {Theiling}, {Tsurusaki}, {Varlotta}, {Vincent}, {Wakely}, {Weekes}, {Weinstein}, {Welsing}, {Williams}, {Zitzer}, {VERITAS Collaboration}, {Jorstad}, {MacDonald}, {Marscher}, {Smith}, {Walker}, {Hovatta}, {Richards}, {Max-Moerbeck}, {Readhead}, {Lister}, {Kovalev}, {Pushkarev}, {Gurwell}, {L{\"a}hteenm{\"a}ki}, {Nieppola}, {Tornikoski}, \&
  {J{\"a}rvel{\"a}}}]{arlen2013}
{Arlen}, T., {Aune}, T., {Beilicke}, M., {et~al.} 2013, \apj, 762, 92

\bibitem[{{Atwood} {et~al.}(2009){Atwood}, {Abdo}, {Ackermann}, {Althouse}, {Anderson}, {Axelsson}, {Baldini}, {Ballet}, {Band}, {Barbiellini}, {Bartelt}, {Bastieri}, {Baughman}, {Bechtol}, {B{\'e}d{\'e}r{\`e}de}, {Bellardi}, {Bellazzini}, {Berenji}, {Bignami}, {Bisello}, {Bissaldi}, {Blandford}, {Bloom}, {Bogart}, {Bonamente}, {Bonnell}, {Borgland}, {Bouvier}, {Bregeon}, {Brez}, {Brigida}, {Bruel}, {Burnett}, {Busetto}, {Caliandro}, {Cameron}, {Caraveo}, {Carius}, {Carlson}, {Casandjian}, {Cavazzuti}, {Ceccanti}, {Cecchi}, {Charles}, {Chekhtman}, {Cheung}, {Chiang}, {Chipaux}, {Cillis}, {Ciprini}, {Claus}, {Cohen-Tanugi}, {Condamoor}, {Conrad}, {Corbet}, {Corucci}, {Costamante}, {Cutini}, {Davis}, {Decotigny}, {DeKlotz}, {Dermer}, {de Angelis}, {Digel}, {do Couto e Silva}, {Drell}, {Dubois}, {Dumora}, {Edmonds}, {Fabiani}, {Farnier}, {Favuzzi}, {Flath}, {Fleury}, {Focke}, {Funk}, {Fusco}, {Gargano}, {Gasparrini}, {Gehrels}, {Gentit}, {Germani}, {Giebels}, {Giglietto}, {Giommi}, {Giordano}, {Glanzman},
  {Godfrey}, {Grenier}, {Grondin}, {Grove}, {Guillemot}, {Guiriec}, {Haller}, {Harding}, {Hart}, {Hays}, {Healey}, {Hirayama}, {Hjalmarsdotter}, {Horn}, {Hughes}, {J{\'o}hannesson}, {Johansson}, {Johnson}, {Johnson}, {Johnson}, {Johnson}, {Kamae}, {Katagiri}, {Kataoka}, {Kavelaars}, {Kawai}, {Kelly}, {Kerr}, {Klamra}, {Kn{\"o}dlseder}, {Kocian}, {Komin}, {Kuehn}, {Kuss}, {Landriu}, {Latronico}, {Lee}, {Lee}, {Lemoine-Goumard}, {Lionetto}, {Longo}, {Loparco}, {Lott}, {Lovellette}, {Lubrano}, {Madejski}, {Makeev}, {Marangelli}, {Massai}, {Mazziotta}, {McEnery}, {Menon}, {Meurer}, {Michelson}, {Minuti}, {Mirizzi}, {Mitthumsiri}, {Mizuno}, {Moiseev}, {Monte}, {Monzani}, {Moretti}, {Morselli}, {Moskalenko}, {Murgia}, {Nakamori}, {Nishino}, {Nolan}, {Norris}, {Nuss}, {Ohno}, {Ohsugi}, {Omodei}, {Orlando}, {Ormes}, {Paccagnella}, {Paneque}, {Panetta}, {Parent}, {Pearce}, {Pepe}, {Perazzo}, {Pesce-Rollins}, {Picozza}, {Pieri}, {Pinchera}, {Piron}, {Porter}, {Poupard}, {Rain{\`o}}, {Rando}, {Rapposelli}, {Razzano},
  {Reimer}, {Reimer}, {Reposeur}, {Reyes}, {Ritz}, {Rochester}, {Rodriguez}, {Romani}, {Roth}, {Russell}, {Ryde}, {Sabatini}, {Sadrozinski}, {Sanchez}, {Sander}, {Sapozhnikov}, {Parkinson}, {Scargle}, {Schalk}, {Scolieri}, {Sgr{\`o}}, {Share}, {Shaw}, {Shimokawabe}, {Shrader}, {Sierpowska-Bartosik}, {Siskind}, {Smith}, {Smith}, {Spandre}, {Spinelli}, {Starck}, {Stephens}, {Strickman}, {Strong}, {Suson}, {Tajima}, {Takahashi}, {Takahashi}, {Tanaka}, {Tenze}, {Tether}, {Thayer}, {Thayer}, {Thompson}, {Tibaldo}, {Tibolla}, {Torres}, {Tosti}, {Tramacere}, {Turri}, {Usher}, {Vilchez}, {Vitale}, {Wang}, {Watters}, {Winer}, {Wood}, {Ylinen}, \& {Ziegler}}]{atwood2009}
{Atwood}, W.~B., {Abdo}, A.~A., {Ackermann}, M., {et~al.} 2009, \apj, 697, 1071

\bibitem[{{Ballet} {et~al.}(2020){Ballet}, {Burnett}, {Digel}, \& {Lott}}]{ballet2020}
{Ballet}, J., {Burnett}, T.~H., {Digel}, S.~W., \& {Lott}, B. 2020, arXiv e-prints, arXiv:2005.11208

\bibitem[{{Bellm} {et~al.}(2019){Bellm}, {Kulkarni}, {Graham}, {Dekany}, {Smith}, {Riddle}, {Masci}, {Helou}, {Prince}, {Adams}, {Barbarino}, {Barlow}, {Bauer}, {Beck}, {Belicki}, {Biswas}, {Blagorodnova}, {Bodewits}, {Bolin}, {Brinnel}, {Brooke}, {Bue}, {Bulla}, {Burruss}, {Cenko}, {Chang}, {Connolly}, {Coughlin}, {Cromer}, {Cunningham}, {De}, {Delacroix}, {Desai}, {Duev}, {Eadie}, {Farnham}, {Feeney}, {Feindt}, {Flynn}, {Franckowiak}, {Frederick}, {Fremling}, {Gal-Yam}, {Gezari}, {Giomi}, {Goldstein}, {Golkhou}, {Goobar}, {Groom}, {Hacopians}, {Hale}, {Henning}, {Ho}, {Hover}, {Howell}, {Hung}, {Huppenkothen}, {Imel}, {Ip}, {Ivezi{\'c}}, {Jackson}, {Jones}, {Juric}, {Kasliwal}, {Kaspi}, {Kaye}, {Kelley}, {Kowalski}, {Kramer}, {Kupfer}, {Landry}, {Laher}, {Lee}, {Lin}, {Lin}, {Lunnan}, {Giomi}, {Mahabal}, {Mao}, {Miller}, {Monkewitz}, {Murphy}, {Ngeow}, {Nordin}, {Nugent}, {Ofek}, {Patterson}, {Penprase}, {Porter}, {Rauch}, {Rebbapragada}, {Reiley}, {Rigault}, {Rodriguez}, {van Roestel}, {Rusholme}, {van
  Santen}, {Schulze}, {Shupe}, {Singer}, {Soumagnac}, {Stein}, {Surace}, {Sollerman}, {Szkody}, {Taddia}, {Terek}, {Van Sistine}, {van Velzen}, {Vestrand}, {Walters}, {Ward}, {Ye}, {Yu}, {Yan}, \& {Zolkower}}]{bellm2019}
{Bellm}, E.~C., {Kulkarni}, S.~R., {Graham}, M.~J., {et~al.} 2019, \pasp, 131, 018002

\bibitem[{{Brown} {et~al.}(2013){Brown}, {Baliber}, {Bianco}, {Bowman}, {Burleson}, {Conway}, {Crellin}, {Depagne}, {De Vera}, {Dilday}, {Dragomir}, {Dubberley}, {Eastman}, {Elphick}, {Falarski}, {Foale}, {Ford}, {Fulton}, {Garza}, {Gomez}, {Graham}, {Greene}, {Haldeman}, {Hawkins}, {Haworth}, {Haynes}, {Hidas}, {Hjelstrom}, {Howell}, {Hygelund}, {Lister}, {Lobdill}, {Martinez}, {Mullins}, {Norbury}, {Parrent}, {Paulson}, {Petry}, {Pickles}, {Posner}, {Rosing}, {Ross}, {Sand}, {Saunders}, {Shobbrook}, {Shporer}, {Street}, {Thomas}, {Tsapras}, {Tufts}, {Valenti}, {Vander Horst}, {Walker}, {White}, \& {Willis}}]{brown2013}
{Brown}, T.~M., {Baliber}, N., {Bianco}, F.~B., {et~al.} 2013, \pasp, 125, 1031

\bibitem[{{Burrows} {et~al.}(2005)}]{Burrows2005}
{Burrows}, D.~N. {et~al.} 2005, \ssr, 120, 165

\bibitem[{{Christie} {et~al.}(2019){Christie}, {Petropoulou}, {Sironi}, \& {Giannios}}]{christie2019}
{Christie}, I.~M., {Petropoulou}, M., {Sironi}, L., \& {Giannios}, D. 2019, \mnras, 482, 65

\bibitem[{{Cleary} {et~al.}(2007){Cleary}, {Lawrence}, {Marshall}, {Hao}, \& {Meier}}]{cleary2007}
{Cleary}, K., {Lawrence}, C.~R., {Marshall}, J.~A., {Hao}, L., \& {Meier}, D. 2007, \apj, 660, 117

\bibitem[{{Dermer} \& {Schlickeiser}(1994)}]{1994ApJS...90..945D}
{Dermer}, C.~D. \& {Schlickeiser}, R. 1994, \apjs, 90, 945

\bibitem[{{Donath} {et~al.}(2023){Donath}, {Terrier}, {Remy}, {Sinha}, {Nigro}, {Pintore}, {Kh{\'e}lifi}, {Olivera-Nieto}, {Ruiz}, {Br{\"u}gge}, {Linhoff}, {Contreras}, {Acero}, {Aguasca-Cabot}, {Berge}, {Bhattacharjee}, {Buchner}, {Boisson}, {Carreto Fidalgo}, {Chen}, {de Bony de Lavergne}, {de Miranda Cardoso}, {Deil}, {F{\"u}{\ss}ling}, {Funk}, {Giunti}, {Hinton}, {Jouvin}, {King}, {Lefaucheur}, {Lemoine-Goumard}, {Lenain}, {L{\'o}pez-Coto}, {Mohrmann}, {Morcuende}, {Panny}, {Regeard}, {Saha}, {Siejkowski}, {Siemiginowska}, {Sip{\H{o}}cz}, {Unbehaun}, {van Eldik}, {Vuillaume}, \& {Zanin}}]{donath2023}
{Donath}, A., {Terrier}, R., {Remy}, Q., {et~al.} 2023, \aap, 678, A157

\bibitem[{{Dondi} \& {Ghisellini}(1995)}]{dondi1995}
{Dondi}, L. \& {Ghisellini}, G. 1995, \mnras, 273, 583

\bibitem[{{Escudero Pedrosa} {et~al.}(2024{\natexlab{a}}){Escudero Pedrosa}, {Agudo}, {Morcuende}, {Otero-Santos}, {Bonnoli}, {Piirola}, {Husillos}, {Bernardos}, {L{\'o}pez-Coto}, {Sota}, {Casanova}, {Aceituno}, \& {Santos-Sanz}}]{escudero2024}
{Escudero Pedrosa}, J., {Agudo}, I., {Morcuende}, D., {et~al.} 2024{\natexlab{a}}, \aj, 168, 84

\bibitem[{{Escudero Pedrosa} {et~al.}(2024{\natexlab{b}}){Escudero Pedrosa}, {Morcuende Parrilla}, \& {Otero-Santos}}]{escudero2024_software}
{Escudero Pedrosa}, J., {Morcuende Parrilla}, D., \& {Otero-Santos}, J. 2024{\natexlab{b}}, {IOP4}

\bibitem[{{Evans} {et~al.}(2009){Evans}, {Beardmore}, {Page}, {Osborne}, {O'Brien}, {Willingale}, {Starling}, {Burrows}, {Godet}, {Vetere}, {Racusin}, {Goad}, {Wiersema}, {Angelini}, {Capalbi}, {Chincarini}, {Gehrels}, {Kennea}, {Margutti}, {Morris}, {Mountford}, {Pagani}, {Perri}, {Romano}, \& {Tanvir}}]{Evans2009}
{Evans}, P.~A., {Beardmore}, A.~P., {Page}, K.~L., {et~al.} 2009, \mnras, 397, 1177

\bibitem[{{Fitzpatrick}(1999)}]{Fitzpatrick1999}
{Fitzpatrick}, E.~L. 1999, \pasp, 111, 63

\bibitem[{{Fomin} {et~al.}(1994){Fomin}, {Stepanian}, {Lamb}, {Lewis}, {Punch}, \& {Weekes}}]{fomin1994}
{Fomin}, V.~P., {Stepanian}, A.~A., {Lamb}, R.~C., {et~al.} 1994, Astroparticle Physics, 2, 137

\bibitem[{{Ghisellini} \& {Tavecchio}(2009)}]{ghisellini2009}
{Ghisellini}, G. \& {Tavecchio}, F. 2009, \mnras, 397, 985

\bibitem[{{Ghisellini} {et~al.}(2010){Ghisellini}, {Tavecchio}, {Foschini}, {Ghirlanda}, {Maraschi}, \& {Celotti}}]{ghisellini2010}
{Ghisellini}, G., {Tavecchio}, F., {Foschini}, L., {et~al.} 2010, \mnras, 402, 497

\bibitem[{{Giannios} {et~al.}(2009){Giannios}, {Uzdensky}, \& {Begelman}}]{giannios2009}
{Giannios}, D., {Uzdensky}, D.~A., \& {Begelman}, M.~C. 2009, \mnras, 395, L29

\bibitem[{{Guo} {et~al.}(2021){Guo}, {Li}, {Daughton}, {Li}, {Kilian}, {Liu}, {Zhang}, \& {Zhang}}]{guo2021}
{Guo}, F., {Li}, X., {Daughton}, W., {et~al.} 2021, \apj, 919, 111

\bibitem[{{Guo} {et~al.}(2016){Guo}, {Li}, {Li}, {Daughton}, {Zhang}, {Lloyd-Ronning}, {Liu}, {Zhang}, \& {Deng}}]{guo2016}
{Guo}, F., {Li}, X., {Li}, H., {et~al.} 2016, \apjl, 818, L9

\bibitem[{{Henri} \& {Saug{\'e}}(2006)}]{henri2006}
{Henri}, G. \& {Saug{\'e}}, L. 2006, \apj, 640, 185

\bibitem[{{Ho} {et~al.}(2004){Ho}, {Moran}, \& {Lo}}]{ho2004}
{Ho}, P. T.~P., {Moran}, J.~M., \& {Lo}, K.~Y. 2004, \apjl, 616, L1

\bibitem[{{Jormanainen} {et~al.}(2023){Jormanainen}, {Hovatta}, {Christie}, {Lindfors}, {Petropoulou}, \& {Liodakis}}]{jormanainen2023}
{Jormanainen}, J., {Hovatta}, T., {Christie}, I.~M., {et~al.} 2023, \aap, 678, A140

\bibitem[{{Jorstad} {et~al.}(2022){Jorstad}, {Marscher}, {Raiteri}, {Villata}, {Weaver}, {Zhang}, {Dong}, {G{\'o}mez}, {Perel}, {Savchenko}, {Larionov}, {Carosati}, {Chen}, {Kurtanidze}, {Marchini}, {Matsumoto}, {Mortari}, {Aceti}, {Acosta-Pulido}, {Andreeva}, {Apolonio}, {Arena}, {Arkharov}, {Bachev}, {Banfi}, {Bonnoli}, {Borman}, {Bozhilov}, {Carnerero}, {Damljanovic}, {Ehgamberdiev}, {Els{\"a}sser}, {Frasca}, {Gabellini}, {Grishina}, {Gupta}, {Hagen-Thorn}, {Hallum}, {Hart}, {Hasuda}, {Hemrich}, {Hsiao}, {Ibryamov}, {Irsmambetova}, {Ivanov}, {Joner}, {Kimeridze}, {Klimanov}, {Kn{\"o}tt}, {Kopatskaya}, {Kurtanidze}, {Kurtenkov}, {Kuutma}, {Larionova}, {Leonini}, {Lin}, {Lorey}, {Mannheim}, {Marino}, {Minev}, {Mirzaqulov}, {Morozova}, {Nikiforova}, {Nikolashvili}, {Ovcharov}, {Papini}, {Pursimo}, {Rahimov}, {Reinhart}, {Sakamoto}, {Salvaggio}, {Semkov}, {Shakhovskoy}, {Sigua}, {Steineke}, {Stojanovic}, {Strigachev}, {Troitskaya}, {Troitskiy}, {Tsai}, {Valcheva}, {Vasilyev}, {Vince}, {Waller}, {Zaharieva}, \&
  {Chatterjee}}]{jorstad2022}
{Jorstad}, S.~G., {Marscher}, A.~P., {Raiteri}, C.~M., {et~al.} 2022, \nat, 609, 265

\bibitem[{{Kaspi} {et~al.}(2007){Kaspi}, {Brandt}, {Maoz}, {Netzer}, {Schneider}, \& {Shemmer}}]{kaspi2007}
{Kaspi}, S., {Brandt}, W.~N., {Maoz}, D., {et~al.} 2007, \apj, 659, 997

\bibitem[{{Khatoon} {et~al.}(2024){Khatoon}, {B{\"o}ttcher}, \& {Prince}}]{khatoon2024}
{Khatoon}, R., {B{\"o}ttcher}, M., \& {Prince}, R. 2024, \apj, 974, 233

\bibitem[{{Kochanek} {et~al.}(2017){Kochanek}, {Shappee}, {Stanek}, {Holoien}, {Thompson}, {Prieto}, {Dong}, {Shields}, {Will}, {Britt}, {Perzanowski}, \& {Pojma{\'n}ski}}]{kochanek2017}
{Kochanek}, C.~S., {Shappee}, B.~J., {Stanek}, K.~Z., {et~al.} 2017, \pasp, 129, 104502

\bibitem[{{Konigl}(1981)}]{1981ApJ...243..700K}
{Konigl}, A. 1981, \apj, 243, 700

\bibitem[{{La Mura}(2022)}]{lamura2022}
{La Mura}, G. 2022, The Astronomer's Telegram, 15688, 1

\bibitem[{{Li} \& {Ma}(1983)}]{li1983}
{Li}, T.~P. \& {Ma}, Y.~Q. 1983, \apj, 272, 317

\bibitem[{{Liodakis} {et~al.}(2019){Liodakis}, {Peirson}, \& {Romani}}]{liodakis2019}
{Liodakis}, I., {Peirson}, A.~L., \& {Romani}, R.~W. 2019, \apj, 880, 29

\bibitem[{{Liodakis} {et~al.}(2018){Liodakis}, {Romani}, {Filippenko}, {Kiehlmann}, {Max-Moerbeck}, {Readhead}, \& {Zheng}}]{liodakis2018}
{Liodakis}, I., {Romani}, R.~W., {Filippenko}, A.~V., {et~al.} 2018, \mnras, 480, 5517

\bibitem[{{Liodakis} {et~al.}(2025){Liodakis}, {Zhang}, {Boula}, {Middei}, {Otero-Santos}, {Blinov}, {Agudo}, {B{\"o}ttcher}, {Chen}, {Ehlert}, {Jorstad}, {Kaaret}, {Krawczynski}, {Peirson}, {Romani}, {Tavecchio}, {Weisskopf}, {Kouch}, {Lindfors}, {Nilsson}, {McCall}, {Jermak}, {Steele}, {Myserlis}, {Gurwell}, {Keating}, {Rao}, {Kang}, {Lee}, {Kim}, {Yeon Cheong}, {Jeong}, {Angelakis}, {Kraus}, {Jos{\'e} Aceituno}, {Bonnoli}, {Casanova}, {Escudero}, {Ag{\'\i}s-Gonz{\'a}lez}, {Morcuende}, {Sota}, {Bachev}, {Grishina}, {Kopatskaya}, {Larionova}, {Morozova}, {Savchenko}, {Shishkina}, {Troitskiy}, {Troitskaya}, \& {Vasilyev}}]{liodakis2025}
{Liodakis}, I., {Zhang}, H., {Boula}, S., {et~al.} 2025, \aap, 698, L19

\bibitem[{L\'opez-Coto {et~al.}(2022)}]{lopez-coto2022}
L\'opez-Coto, R. {et~al.} 2022, ASP Conf. Ser., 532, 357

\bibitem[{{MAGIC Collaboration} {et~al.}(2019){MAGIC Collaboration}, {Acciari}, {Ansoldi}, {Antonelli}, {Arbet Engels}, {Baack}, {Babi{\'c}}, {Banerjee}, {Bangale}, {Barres de Almeida}, {Barrio}, {Becerra Gonz{\'a}lez}, {Bednarek}, {Bernardini}, {Berti}, {Besenrieder}, {Bhattacharyya}, {Bigongiari}, {Biland}, {Blanch}, {Bonnoli}, {Carosi}, {Ceribella}, {Cikota}, {Colak}, {Colin}, {Colombo}, {Contreras}, {Cortina}, {Covino}, {D'Elia}, {da Vela}, {Dazzi}, {de Angelis}, {de Lotto}, {Delfino}, {Delgado}, {di Pierro}, {Do Souto Espi{\~n}era}, {Dom{\'\i}nguez}, {Dominis Prester}, {Dorner}, {Doro}, {Einecke}, {Elsaesser}, {Fallah Ramazani}, {Fattorini}, {Fern{\'a}ndez-Barral}, {Ferrara}, {Fidalgo}, {Foffano}, {Fonseca}, {Font}, {Fruck}, {Galindo}, {Gallozzi}, {Garc{\'\i}a L{\'o}pez}, {Garczarczyk}, {Gaug}, {Giammaria}, {Godinovi{\'c}}, {Guberman}, {Hadasch}, {Hahn}, {Hassan}, {Herrera}, {Hoang}, {Hrupec}, {Inoue}, {Ishio}, {Iwamura}, {Kubo}, {Kushida}, {Kuve{\v{z}}di{\'c}}, {Lamastra}, {Lelas}, {Leone}, {Lindfors},
  {Lombardi}, {Longo}, {L{\'o}pez}, {L{\'o}pez-Oramas}, {Maggio}, {Majumdar}, {Makariev}, {Maneva}, {Manganaro}, {Mannheim}, {Maraschi}, {Mariotti}, {Mart{\'\i}nez}, {Masuda}, {Mazin}, {Minev}, {Miranda}, {Mirzoyan}, {Molina}, {Moralejo}, {Moreno}, {Moretti}, {Munar-Adrover}, {Neustroev}, {Niedzwiecki}, {Nievas Rosillo}, {Nigro}, {Nilsson}, {Ninci}, {Nishijima}, {Noda}, {Nogu{\'e}s}, {N{\"o}the}, {Paiano}, {Palacio}, {Paneque}, {Paoletti}, {Paredes}, {Pedaletti}, {Pe{\~n}il}, {Peresano}, {Persic}, {Prada Moroni}, {Prandini}, {Puljak}, {Garcia}, {Rhode}, {Rib{\'o}}, {Rico}, {Righi}, {Rugliancich}, {Saha}, {Saito}, {Satalecka}, {Schweizer}, {Sitarek}, {{\v{S}}nidari{\'c}}, {Sobczynska}, {Somero}, {Stamerra}, {Strzys}, {Suri{\'c}}, {Tavecchio}, {Temnikov}, {Terzi{\'c}}, {Teshima}, {Torres-Alb{\`a}}, {Tsujimoto}, {van Scherpenberg}, {Vanzo}, {Vazquez Acosta}, {Vovk}, {Will}, {Zari{\'c}}, {D'Ammando}, {Hada}, {Jorstad}, {Marscher}, {Mobeen}, {Hovatta}, {Larionov}, {Borman}, {Grishina}, {Kopatskaya}, {Morozova},
  {Nikiforova}, {L{\"a}hteenm{\"a}ki}, {Tornikoski}, \& {Agudo}}]{magic2019}
{MAGIC Collaboration}, {Acciari}, V.~A., {Ansoldi}, S., {et~al.} 2019, \aap, 623, A175

\bibitem[{{MAGIC Collaboration} {et~al.}(2020){MAGIC Collaboration}, {Acciari}, {Ansoldi}, {Antonelli}, {Arbet Engels}, {Baack}, {Babi{\'c}}, {Banerjee}, {Barres de Almeida}, {Barrio}, {Becerra Gonz{\'a}lez}, {Bednarek}, {Bellizzi}, {Bernardini}, {Berti}, {Besenrieder}, {Bhattacharyya}, {Bigongiari}, {Biland}, {Blanch}, {Bonnoli}, {Bosnjak}, {Busetto}, {Carosi}, {Ceribella}, {Chai}, {Cikota}, {Colak}, {Colin}, {Colombo}, {Contreras}, {Cortina}, {Covino}, {D'Elia}, {da Vela}, {Dazzi}, {de Angelis}, {de Lotto}, {Delfino}, {Delgado}, {di Pierro}, {Do Souto Espi{\~n}eira}, {Dominis Prester}, {Donini}, {Dorner}, {Doro}, {Elsaesser}, {Fallah Ramazani}, {Fattorini}, {Fern{\'a}ndez-Barral}, {Ferrara}, {Fidalgo}, {Foffano}, {Fonseca}, {Font}, {Fruck}, {Fukami}, {Gallozzi}, {Garc{\'\i}a L{\'o}pez}, {Garczarczyk}, {Gasparyan}, {Gaug}, {Godinovi{\'c}}, {Green}, {Guberman}, {Hadasch}, {Hahn}, {Herrera}, {Hoang}, {Hrupec}, {Inada}, {Inoue}, {Ishio}, {Iwamura}, {Jouvin}, {Kubo}, {Kushida}, {Lamastra}, {Lelas}, {Leone},
  {Lindfors}, {Lombardi}, {Longo}, {L{\'o}pez}, {L{\'o}pez-Coto}, {L{\'o}pez-Oramas}, {Machado de Oliveira Fraga}, {Maggio}, {Majumdar}, {Makariev}, {Mallamaci}, {Maneva}, {Manganaro}, {Mannheim}, {Maraschi}, {Mariotti}, {Mart{\'\i}nez}, {Masuda}, {Mazin}, {Mi{\'c}anovi{\'c}}, {Miceli}, {Minev}, {Miranda}, {Mirzoyan}, {Molina}, {Moralejo}, {Morcuende}, {Moreno}, {Moretti}, {Munar-Adrover}, {Neustroev}, {Niedzwiecki}, {Nigro}, {Nilsson}, {Ninci}, {Nishijima}, {Noda}, {Nogu{\'e}s}, {N{\"o}the}, {Nozaki}, {Paiano}, {Palacio}, {Palatiello}, {Paneque}, {Paoletti}, {Paredes}, {Pe{\~n}il}, {Peresano}, {Persic}, {Prada Moroni}, {Prandini}, {Puljak}, {Rhode}, {Rib{\'o}}, {Rico}, {Righi}, {Rugliancich}, {Saha}, {Sahakyan}, {Saito}, {Sakurai}, {Satalecka}, {Schweizer}, {Sitarek}, {{\v{S}}nidari{\'c}}, {Sobczynska}, {Somero}, {Stamerra}, {Strom}, {Strzys}, {Suri{\'c}}, {Takahashi}, {Tavecchio}, {Temnikov}, {Terzi{\'c}}, {Teshima}, {Torres-Alb{\`a}}, {Tsujimoto}, {van Scherpenberg}, {Vanzo}, {Vazquez Acosta}, {Vovk},
  {Will}, {Zari{\'c}}, {Fermi-Lat Collaboration}, \& {Hayashida}}]{magic2020}
{MAGIC Collaboration}, {Acciari}, V.~A., {Ansoldi}, S., {et~al.} 2020, \aap, 638, A14

\bibitem[{{Mannheim} \& {Biermann}(1992)}]{mannheim1992}
{Mannheim}, K. \& {Biermann}, P.~L. 1992, \aap, 253, L21

\bibitem[{{Maraschi} {et~al.}(1992){Maraschi}, {Ghisellini}, \& {Celotti}}]{1992ApJ...397L...5M}
{Maraschi}, L., {Ghisellini}, G., \& {Celotti}, A. 1992, \apjl, 397, L5

\bibitem[{{Marrone} \& {Rao}(2008)}]{marrone2008}
{Marrone}, D.~P. \& {Rao}, R. 2008, in Society of Photo-Optical Instrumentation Engineers (SPIE) Conference Series, Vol. 7020, Millimeter and Submillimeter Detectors and Instrumentation for Astronomy IV, ed. W.~D. {Duncan}, W.~S. {Holland}, S.~{Withington}, \& J.~{Zmuidzinas}, 70202B

\bibitem[{{Mattox} {et~al.}(1996){Mattox}, {Bertsch}, {Chiang}, {Dingus}, {Digel}, {Esposito}, {Fierro}, {Hartman}, {Hunter}, {Kanbach}, {Kniffen}, {Lin}, {Macomb}, {Mayer-Hasselwander}, {Michelson}, {von Montigny}, {Mukherjee}, {Nolan}, {Ramanamurthy}, {Schneid}, {Sreekumar}, {Thompson}, \& {Willis}}]{mattox1996}
{Mattox}, J.~R., {Bertsch}, D.~L., {Chiang}, J., {et~al.} 1996, \apj, 461, 396

\bibitem[{{Max-Moerbeck} {et~al.}(2014){Max-Moerbeck}, {Hovatta}, {Richards}, {King}, {Pearson}, {Readhead}, {Reeves}, {Shepherd}, {Stevenson}, {Angelakis}, {Fuhrmann}, {Grainge}, {Pavlidou}, {Romani}, \& {Zensus}}]{max-moerbeck2014}
{Max-Moerbeck}, W., {Hovatta}, T., {Richards}, J.~L., {et~al.} 2014, \mnras, 445, 428

\bibitem[{{Middei} {et~al.}(2023){Middei}, {Liodakis}, {Perri}, {Puccetti}, {Cavazzuti}, {Di Gesu}, {Ehlert}, {Madejski}, {Marscher}, {Marshall}, {Muleri}, {Negro}, {Jorstad}, {Ag{\'\i}s-Gonz{\'a}lez}, {Agudo}, {Bonnoli}, {Bernardos}, {Casanova}, {Garc{\'\i}a-Comas}, {Husillos}, {Marchini}, {Sota}, {Kouch}, {Lindfors}, {Borman}, {Kopatskaya}, {Larionova}, {Morozova}, {Savchenko}, {Vasilyev}, {Zhovtan}, {Casadio}, {Escudero}, {Myserlis}, {Hales}, {Kameno}, {Kneissl}, {Messias}, {Nagai}, {Blinov}, {Bourbah}, {Kiehlmann}, {Kontopodis}, {Mandarakas}, {Romanopoulos}, {Skalidis}, {Vervelaki}, {Masiero}, {Mawet}, {Millar-Blanchaer}, {Panopoulou}, {Tinyanont}, {Berdyugin}, {Kagitani}, {Kravtsov}, {Sakanoi}, {Imazawa}, {Sasada}, {Fukazawa}, {Kawabata}, {Uemura}, {Mizuno}, {Nakaoka}, {Akitaya}, {Gurwell}, {Rao}, {Di Lalla}, {Cibrario}, {Donnarumma}, {Kim}, {Omodei}, {Pacciani}, {Poutanen}, {Tavecchio}, {Antonelli}, {Bachetti}, {Baldini}, {Baumgartner}, {Bellazzini}, {Bianchi}, {Bongiorno}, {Bonino}, {Brez},
  {Bucciantini}, {Capitanio}, {Castellano}, {Ciprini}, {Costa}, {De Rosa}, {Del Monte}, {Di Marco}, {Doroshenko}, {Dov{\v{c}}iak}, {Enoto}, {Evangelista}, {Fabiani}, {Ferrazzoli}, {Garcia}, {Gunji}, {Hayashida}, {Heyl}, {Iwakiri}, {Karas}, {Kitaguchi}, {Kolodziejczak}, {Krawczynski}, {La Monaca}, {Latronico}, {Maldera}, {Manfreda}, {Marin}, {Marinucci}, {Massaro}, {Matt}, {Mitsuishi}, {Ng}, {O'Dell}, {Oppedisano}, {Papitto}, {Pavlov}, {Peirson}, {Pesce-Rollins}, {Petrucci}, {Pilia}, {Possenti}, {Ramsey}, {Rankin}, {Ratheesh}, {Romani}, {Sgr{\'o}}, {Slane}, {Soffitta}, {Spandre}, {Tamagawa}, {Taverna}, {Tawara}, {Tennant}, {Thomas}, {Tombesi}, {Trois}, {Tsygankov}, {Turolla}, {Vink}, {Weisskopf}, {Wu}, {Xie}, \& {Zane}}]{middei2023}
{Middei}, R., {Liodakis}, I., {Perri}, M., {et~al.} 2023, \apjl, 942, L10

\bibitem[{{Miller} {et~al.}(1978){Miller}, {French}, \& {Hawley}}]{miller1978}
{Miller}, J.~S., {French}, H.~B., \& {Hawley}, S.~A. 1978, \apjl, 219, L85

\bibitem[{{Moderski} {et~al.}(2005){Moderski}, {Sikora}, {Coppi}, \& {Aharonian}}]{2005MNRAS.363..954M}
{Moderski}, R., {Sikora}, M., {Coppi}, P.~S., \& {Aharonian}, F. 2005, \mnras, 363, 954

\bibitem[{Moralejo {et~al.}(2025)Moralejo, Lopez-Coto, Vuillaume, Cassol, Linhoff, Priyadarshi, Morcuende, Nozaki, Bernardos, Gliwny, Ruiz, deborahDOR, Dalchenko, yrenier, Saha, Nickel, Aguasca-Cabot, Sitarek, Alispach, de~Bony, Pillera, Láinez, Andres-Baquero, Balbo, Muñoz, Takahashi, sn621, \& yiwamura}]{moralejo2025}
Moralejo, A., Lopez-Coto, R., Vuillaume, T., {et~al.} 2025, cta-observatory/cta-lstchain: v0.10.18 - 2025-02-17

\bibitem[{{Myserlis} {et~al.}(2025){Myserlis}, {Agudo}, {Thum}, {Rao}, {Homan}, {Jorstad}, {Marscher}, {Kraus}, \& {Angelakis}}]{myserlis2025}
{Myserlis}, I., {Agudo}, I., {Thum}, C., {et~al.} 2025, in Highlights of Spanish Astrophysics XII, ed. M.~{Manteiga}, F.~{Gonz{\'a}lez-Galindo}, A.~{Labiano-Ortega}, M.~J. {Mart{\'\i}nez-Gonz{\'a}lez}, N.~{Rea}, M.~{Romero-G{\'o}mez}, A.~{Ulla-Miguel}, G.~{Yepes}, C.~{Rodr{\'\i}guez-L{\'o}pez}, A.~{G{\'o}mez-Garc{\'\i}a}, \& C.~{Dafonte}, 121

\bibitem[{{Nalewajko} {et~al.}(2011){Nalewajko}, {Giannios}, {Begelman}, {Uzdensky}, \& {Sikora}}]{nalewajko2011}
{Nalewajko}, K., {Giannios}, D., {Begelman}, M.~C., {Uzdensky}, D.~A., \& {Sikora}, M. 2011, \mnras, 413, 333

\bibitem[{{Nievas Rosillo} {et~al.}(2025){Nievas Rosillo}, {Acero}, {Otero-Santos}, {Vazquez Acosta}, {Terrier}, {Morcuende}, \& {Arbet-Engels}}]{nievas2025}
{Nievas Rosillo}, M., {Acero}, F., {Otero-Santos}, J., {et~al.} 2025, \aap, 693, A287

\bibitem[{Nigro {et~al.}(2023)Nigro, Sitarek, Gliwny, Sanchez, Craig, Vuillaume, Viale, \& Maniadakis}]{nigro2023}
Nigro, C., Sitarek, J., Gliwny, P., {et~al.} 2023, agnpy

\bibitem[{{Nigro} {et~al.}(2022){Nigro}, {Sitarek}, {Gliwny}, {Sanchez}, {Tramacere}, \& {Craig}}]{nigro2022}
{Nigro}, C., {Sitarek}, J., {Gliwny}, P., {et~al.} 2022, \aap, 660, A18

\bibitem[{{Nilsson} {et~al.}(2018){Nilsson}, {Lindfors}, {Takalo}, {Reinthal}, {Berdyugin}, {Sillanp{\"a}{\"a}}, {Ciprini}, {Halkola}, {Hein{\"a}m{\"a}ki}, {Hovatta}, {Kadenius}, {Nurmi}, {Ostorero}, {Pasanen}, {Rekola}, {Saarinen}, {Sainio}, {Tuominen}, {Villforth}, {Vornanen}, \& {Zaprudin}}]{nilsson2018}
{Nilsson}, K., {Lindfors}, E., {Takalo}, L.~O., {et~al.} 2018, \aap, 620, A185

\bibitem[{{Norris} {et~al.}(1996){Norris}, {Nemiroff}, {Bonnell}, {Scargle}, {Kouveliotou}, {Paciesas}, {Meegan}, \& {Fishman}}]{norris1996}
{Norris}, J.~P., {Nemiroff}, R.~J., {Bonnell}, J.~T., {et~al.} 1996, \apj, 459, 393

\bibitem[{{Peirson} {et~al.}(2023){Peirson}, {Negro}, {Liodakis}, {Middei}, {Kim}, {Marscher}, {Marshall}, {Pacciani}, {Romani}, {Wu}, {Di Marco}, {Di Lalla}, {Omodei}, {Jorstad}, {Agudo}, {Kouch}, {Lindfors}, {Aceituno}, {Bernardos}, {Bonnoli}, {Casanova}, {Garc{\'\i}a-Comas}, {Ag{\'\i}s-Gonz{\'a}lez}, {Husillos}, {Marchini}, {Sota}, {Casadio}, {Escudero}, {Myserlis}, {Sievers}, {Gurwell}, {Rao}, {Imazawa}, {Sasada}, {Fukazawa}, {Kawabata}, {Uemura}, {Mizuno}, {Nakaoka}, {Akitaya}, {Cheong}, {Jeong}, {Kang}, {Kim}, {Lee}, {Angelakis}, {Kraus}, {Cibrario}, {Donnarumma}, {Poutanen}, {Tavecchio}, {Antonelli}, {Bachetti}, {Baldini}, {Baumgartner}, {Bellazzini}, {Bianchi}, {Bongiorno}, {Bonino}, {Brez}, {Bucciantini}, {Capitanio}, {Castellano}, {Cavazzuti}, {Chen}, {Ciprini}, {Costa}, {De Rosa}, {Del Monte}, {Di Gesu}, {Doroshenko}, {Dov{\v{c}}iak}, {Ehlert}, {Enoto}, {Evangelista}, {Fabiani}, {Ferrazzoli}, {Garcia}, {Gunji}, {Hayashida}, {Heyl}, {Iwakiri}, {Kaaret}, {Karas}, {Kitaguchi}, {Kolodziejczak},
  {Krawczynski}, {La Monaca}, {Latronico}, {Madejski}, {Maldera}, {Manfreda}, {Marin}, {Marinucci}, {Massaro}, {Matt}, {Mitsuishi}, {Muleri}, {Ng}, {O'Dell}, {Oppedisano}, {Papitto}, {Pavlov}, {Perri}, {Pesce-Rollins}, {Petrucci}, {Pilia}, {Possenti}, {Puccetti}, {Ramsey}, {Rankin}, {Ratheesh}, {Roberts}, {Sgr{\'o}}, {Slane}, {Soffitta}, {Spandre}, {Swartz}, {Tamagawa}, {Taverna}, {Tawara}, {Tennant}, {Thomas}, {Tombesi}, {Trois}, {Tsygankov}, {Turolla}, {Vink}, {Weisskopf}, {Xie}, \& {Zane}}]{peirson2023}
{Peirson}, A.~L., {Negro}, M., {Liodakis}, I., {et~al.} 2023, \apjl, 948, L25

\bibitem[{{Peirson} \& {Romani}(2019)}]{peirson2019}
{Peirson}, A.~L. \& {Romani}, R.~W. 2019, \apj, 885, 76

\bibitem[{{Petropoulou} {et~al.}(2016){Petropoulou}, {Giannios}, \& {Sironi}}]{petropoulou2016}
{Petropoulou}, M., {Giannios}, D., \& {Sironi}, L. 2016, \mnras, 462, 3325

\bibitem[{{Poole} {et~al.}(2008){Poole}, {Breeveld}, {Page}, {Landsman}, {Holland}, {Roming}, {Kuin}, {Brown}, {Gronwall}, {Hunsberger}, {Koch}, {Mason}, {Schady}, {vanden Berk}, {Blustin}, {Boyd}, {Broos}, {Carter}, {Chester}, {Cucchiara}, {Hancock}, {Huckle}, {Immler}, {Ivanushkina}, {Kennedy}, {Marshall}, {Morgan}, {Pandey}, {de Pasquale}, {Smith}, \& {Still}}]{Poole2008}
{Poole}, T.~S., {Breeveld}, A.~A., {Page}, M.~J., {et~al.} 2008, \mnras, 383, 627

\bibitem[{{Raiteri} {et~al.}(2009){Raiteri}, {Villata}, {Capetti}, {Aller}, {Bach}, {Calcidese}, {Gurwell}, {Larionov}, {Ohlert}, {Nilsson}, {Strigachev}, {Agudo}, {Aller}, {Bachev}, {Ben{\'\i}tez}, {Berdyugin}, {B{\"o}ttcher}, {Buemi}, {Buttiglione}, {Carosati}, {Charlot}, {Chen}, {Dultzin}, {Forn{\'e}}, {Fuhrmann}, {G{\'o}mez}, {Gupta}, {Heidt}, {Hiriart}, {Hsiao}, {Jel{\'\i}nek}, {Jorstad}, {Kimeridze}, {Konstantinova}, {Kopatskaya}, {Kostov}, {Kurtanidze}, {L{\"a}hteenm{\"a}ki}, {Lanteri}, {Larionova}, {Leto}, {Latev}, {Le Campion}, {Lee}, {Ligustri}, {Lindfors}, {Marscher}, {Mihov}, {Nikolashvili}, {Nikolov}, {Ovcharov}, {Principe}, {Pursimo}, {Ragozzine}, {Robb}, {Ros}, {Sadun}, {Sagar}, {Semkov}, {Sigua}, {Smart}, {Sorcia}, {Takalo}, {Tornikoski}, {Trigilio}, {Uckert}, {Umana}, {Valcheva}, \& {Volvach}}]{Raiteri2009}
{Raiteri}, C.~M., {Villata}, M., {Capetti}, A., {et~al.} 2009, \aap, 507, 769

\bibitem[{{Raiteri} {et~al.}(2013){Raiteri}, {Villata}, {D'Ammando}, {Larionov}, {Gurwell}, {Mirzaqulov}, {Smith}, {Acosta-Pulido}, {Agudo}, {Ar{\'e}valo}, {Bachev}, {Ben{\'\i}tez}, {Berdyugin}, {Blinov}, {Borman}, {B{\"o}ttcher}, {Bozhilov}, {Carnerero}, {Carosati}, {Casadio}, {Chen}, {Doroshenko}, {Efimov}, {Efimova}, {Ehgamberdiev}, {G{\'o}mez}, {Gonz{\'a}lez-Morales}, {Hiriart}, {Ibryamov}, {Jadhav}, {Jorstad}, {Joshi}, {Kadenius}, {Klimanov}, {Kohli}, {Konstantinova}, {Kopatskaya}, {Koptelova}, {Kimeridze}, {Kurtanidze}, {Larionova}, {Larionova}, {Ligustri}, {Lindfors}, {Marscher}, {McBreen}, {McHardy}, {Metodieva}, {Molina}, {Morozova}, {Nazarov}, {Nikolashvili}, {Nilsson}, {Okhmat}, {Ovcharov}, {Panwar}, {Pasanen}, {Peneva}, {Phipps}, {Pulatova}, {Reinthal}, {Ros}, {Sadun}, {Schwartz}, {Semkov}, {Sergeev}, {Sigua}, {Sillanp{\"a}{\"a}}, {Smith}, {Stoyanov}, {Strigachev}, {Takalo}, {Taylor}, {Thum}, {Troitsky}, {Valcheva}, {Wehrle}, \& {Wiesemeyer}}]{Raiteri2013}
{Raiteri}, C.~M., {Villata}, M., {D'Ammando}, F., {et~al.} 2013, \mnras, 436, 1530

\bibitem[{{Roming} {et~al.}(2005)}]{Roming2005}
{Roming}, P. W.~A. {et~al.} 2005, \ssr, 120, 95

\bibitem[{{Sahakyan} \& {Giommi}(2022)}]{sahakyan2022}
{Sahakyan}, N. \& {Giommi}, P. 2022, \mnras, 513, 4645

\bibitem[{{Saldana-Lopez} {et~al.}(2021){Saldana-Lopez}, {Dom{\'\i}nguez}, {P{\'e}rez-Gonz{\'a}lez}, {Finke}, {Ajello}, {Primack}, {Paliya}, \& {Desai}}]{saldana-lopez2021}
{Saldana-Lopez}, A., {Dom{\'\i}nguez}, A., {P{\'e}rez-Gonz{\'a}lez}, P.~G., {et~al.} 2021, \mnras, 507, 5144

\bibitem[{{Schlafly} \& {Finkbeiner}(2011)}]{S&F2011}
{Schlafly}, E.~F. \& {Finkbeiner}, D.~P. 2011, \apj, 737, 103

\bibitem[{{Shappee} {et~al.}(2014){Shappee}, {Prieto}, {Grupe}, {Kochanek}, {Stanek}, {De Rosa}, {Mathur}, {Zu}, {Peterson}, {Pogge}, {Komossa}, {Im}, {Jencson}, {Holoien}, {Basu}, {Beacom}, {Szczygie{\l}}, {Brimacombe}, {Adams}, {Campillay}, {Choi}, {Contreras}, {Dietrich}, {Dubberley}, {Elphick}, {Foale}, {Giustini}, {Gonzalez}, {Hawkins}, {Howell}, {Hsiao}, {Koss}, {Leighly}, {Morrell}, {Mudd}, {Mullins}, {Nugent}, {Parrent}, {Phillips}, {Pojmanski}, {Rosing}, {Ross}, {Sand}, {Terndrup}, {Valenti}, {Walker}, \& {Yoon}}]{shappee2014}
{Shappee}, B.~J., {Prieto}, J.~L., {Grupe}, D., {et~al.} 2014, \apj, 788, 48

\bibitem[{{Sironi} {et~al.}(2016){Sironi}, {Giannios}, \& {Petropoulou}}]{sironi2016}
{Sironi}, L., {Giannios}, D., \& {Petropoulou}, M. 2016, \mnras, 462, 48

\bibitem[{{Stickel} {et~al.}(1991){Stickel}, {Padovani}, {Urry}, {Fried}, \& {Kuehr}}]{stickel1991}
{Stickel}, M., {Padovani}, P., {Urry}, C.~M., {Fried}, J.~W., \& {Kuehr}, H. 1991, \apj, 374, 431

\bibitem[{{Tavecchio} {et~al.}(2011){Tavecchio}, {Becerra-Gonzalez}, {Ghisellini}, {Stamerra}, {Bonnoli}, {Foschini}, \& {Maraschi}}]{tavecchio2011}
{Tavecchio}, F., {Becerra-Gonzalez}, J., {Ghisellini}, G., {et~al.} 2011, \aap, 534, A86

\bibitem[{{Tavecchio} \& {Ghisellini}(2016)}]{tavecchio2016}
{Tavecchio}, F. \& {Ghisellini}, G. 2016, \mnras, 456, 2374

\bibitem[{{Tavecchio} {et~al.}(2010){Tavecchio}, {Ghisellini}, {Bonnoli}, \& {Ghirlanda}}]{tavecchio2010}
{Tavecchio}, F., {Ghisellini}, G., {Bonnoli}, G., \& {Ghirlanda}, G. 2010, \mnras, 405, L94

\bibitem[{{Teraesranta} {et~al.}(1998){Teraesranta}, {Tornikoski}, {Mujunen}, {Karlamaa}, {Valtonen}, {Henelius}, {Urpo}, {Lainela}, {Pursimo}, {Nilsson}, {Wiren}, {Laehteenmaeki}, {Korpi}, {Rekola}, {Heinaemaeki}, {Hanski}, {Nurmi}, {Kokkonen}, {Keinaenen}, {Joutsamo}, {Oksanen}, {Pietilae}, {Valtaoja}, {Valtonen}, \& {Koenoenen}}]{teraesanta1998}
{Teraesranta}, H., {Tornikoski}, M., {Mujunen}, A., {et~al.} 1998, \aaps, 132, 305

\bibitem[{{Urry} \& {Padovani}(1995)}]{urry1995}
{Urry}, C.~M. \& {Padovani}, P. 1995, \pasp, 107, 803

\bibitem[{{Wagner} {et~al.}(2022){Wagner}, {Burd}, {Dorner}, {Mannheim}, {Buson}, {Gokus}, {Madejski}, {Scargle}, {Arbet-Engels}, {Baack}, {Balbo}, {Biland}, {Bretz}, {Buss}, {Elsaesser}, {Eisenberger}, {Hildebrand}, {Iotov}, {Kalenski}, {Neise}, {Noethe}, {Paravac}, {Rhode}, {Schleicher}, {Sliusar}, \& {Walter}}]{wagner2022}
{Wagner}, S.~M., {Burd}, P., {Dorner}, D., {et~al.} 2022, in 37th International Cosmic Ray Conference, 868

\bibitem[{{Weaver} {et~al.}(2020){Weaver}, {Williamson}, {Jorstad}, {Marscher}, {Larionov}, {Raiteri}, {Villata}, {Acosta-Pulido}, {Bachev}, {Baida}, {Balonek}, {Ben{\'\i}tez}, {Borman}, {Bozhilov}, {Carnerero}, {Carosati}, {Chen}, {Damljanovic}, {Dhiman}, {Dougherty}, {Ehgamberdiev}, {Grishina}, {Gupta}, {Hart}, {Hiriart}, {Hsiao}, {Ibryamov}, {Joner}, {Kimeridze}, {Kopatskaya}, {Kurtanidze}, {Kurtanidze}, {Larionova}, {Matsumoto}, {Matsumura}, {Minev}, {Mirzaqulov}, {Morozova}, {Nikiforova}, {Nikolashvili}, {Ovcharov}, {Rizzi}, {Sadun}, {Savchenko}, {Semkov}, {Slater}, {Smith}, {Stojanovic}, {Strigachev}, {Troitskaya}, {Troitsky}, {Tsai}, {Vince}, {Valcheva}, {Vasilyev}, {Zaharieva}, \& {Zhovtan}}]{Weaver2020}
{Weaver}, Z.~R., {Williamson}, K.~E., {Jorstad}, S.~G., {et~al.} 2020, \apj, 900, 137

\bibitem[{Wilks(1938)}]{wilks1938}
Wilks, S.~S. 1938, The Annals of Mathematical Statistics, 9, 60

\bibitem[{{Zech} \& {Lemoine}(2021)}]{zech2021}
{Zech}, A. \& {Lemoine}, M. 2021, \aap, 654, A96

\bibitem[{{Zhang} {et~al.}(2024){Zhang}, {B{\"o}ttcher}, \& {Liodakis}}]{zhang2024}
{Zhang}, H., {B{\"o}ttcher}, M., \& {Liodakis}, I. 2024, \apj, 967, 93

\bibitem[{{Zhang} {et~al.}(2022){Zhang}, {Li}, {Giannios}, {Guo}, {Thiersen}, {B{\"o}ttcher}, {Lewis}, \& {Venters}}]{zhang2022}
{Zhang}, H., {Li}, X., {Giannios}, D., {et~al.} 2022, \apj, 924, 90

\bibitem[{{Zhang} {et~al.}(2023){Zhang}, {Marscher}, {Guo}, {Giannios}, {Li}, \& {Negro}}]{zhang2023}
{Zhang}, H., {Marscher}, A.~P., {Guo}, F., {et~al.} 2023, \apj, 949, 71

\bibitem[{{Zhang} {et~al.}(1999){Zhang}, {Celotti}, {Treves}, {Chiappetti}, {Ghisellini}, {Maraschi}, {Pian}, {Tagliaferri}, {Tavecchio}, \& {Urry}}]{zhang1999}
{Zhang}, Y.~H., {Celotti}, A., {Treves}, A., {et~al.} 1999, \apj, 527, 719

\end{thebibliography}

\begin{appendix} 

\onecolumn

\section{Night-wise LST-1 data analysis results}\label{appendix_a}
In Table \ref{tab:observation_details_lst} we compile and summarise the night-wise results of the LST-1 data analysis of BL Lac, that is, the effective observing time after data quality selection, the significance, and the flux of the source above 200 GeV.

\begin{table*}[h]
\begin{center}
\caption{Effective observation time, significance and flux level above 200 GeV of BL Lac's emission for each night of observation.}
\begin{tabular}{@{}ccccccccc@{}}
\cline{1-4} \cline{6-9} 
\multirow{2}{*}{Day} & Observation & Significance & Flux$^{a}$ & &\multirow{2}{*}{Day} & Observation & Significance & Flux$^{a}$ \\ 
 & Time [h] & [$\sigma$] & [$\times 10^{-11}$ cm$^{-2}$ s$^{-1}$] & & & Time [h] & [$\sigma$] & [$\times 10^{-11}$ cm$^{-2}$ s$^{-1}$] \\   \cline{1-4} \cline{6-9} 
21/09 & 1.71 & 0.32 & $ <1.2 $ & & 27/10 & 1.36 & 6.48 & $ 10.7 \pm 1.4 $ \\ \cline{1-4} \cline{6-9} 
01/10 & 1.65 & 1.22 & $ <2.1 $ & & 29/10 & 0.72 & 6.19 & $ 2.9 \pm 1.2 $ \\ \cline{1-4} \cline{6-9} 
03/10 & 2.86 & 2.28 & $ 1.1 \pm 0.5 $ & & 31/10 & 1.73 & 5.08 & $ 4.7 \pm 0.8 $ \\ \cline{1-4} \cline{6-9} 
15/10 & 1.31 & 7.55 & $ 6.5 \pm 1.0 $ & & 01/11 & 1.35 & 5.22 & $ 2.3 \pm 0.7 $ \\ \cline{1-4} \cline{6-9} 
17/10 & 2.14 & 10.77 & $ 6.3 \pm 0.7 $ & & 13/11 & 1.51 & 30.47 & $ 38.7 \pm 1.3 $ \\ \cline{1-4} \cline{6-9} 
18/10 & 2.48 & 1.14 & $ <2.7 $ & & 15/11 & 0.76 & 2.20 & $ 8.1 \pm 2.7 $ \\ \cline{1-4} \cline{6-9} 
19/10 & 2.67 & 3.26 & $ 3.1 \pm 1.0 $ & & 16/11 & 0.88 & 0.54 & $ <6.7 $ \\ \cline{1-4} \cline{6-9} 
20/10 & 2.50 & 48.73 & $ 23.8 \pm 1.0 $ & & 17/11 & 0.87 & 3.01 & $ 5.0 \pm 2.4$ \\ \cline{1-4} \cline{6-9} 
21/10 & 1.87 & 0.70 & $ <1.6 $ & & 18/11 & 1.01 & 0.68 & $ <8.3 $ \\ \cline{1-4} \cline{6-9} 
23/10 & 1.41 & 1.70 & $ <5.1 $ & & 19/11 & 0.57 & 0.66 & $ <6.6 $ \\ \cline{1-4} \cline{6-9} 
25/10 & 2.25 & 0.13 & $ <1.9 $ & & 25/11 & 0.76 & 10.53 & $ 10.8 \pm 1.7 $ \\ \cline{1-4} \cline{6-9} 
26/10 & 1.23 & 4.67 & $ 3.3 \pm 1.1 $  & &  &  & \\ \cline{1-4} 
\end{tabular}
\label{tab:observation_details_lst}
\end{center}
\vspace{-0.4cm}
\tablefoot{
$^{a}$For the observations in which the signal has a significance <$2\sigma$ (TS$<$4), the 95\% confidence level flux upper limits are reported.
}
\end{table*}

\vspace{-0.3cm}

\section{X-ray data results}\label{appendix_c}

\begin{figure}
\centering
\includegraphics[width=0.52\textwidth]{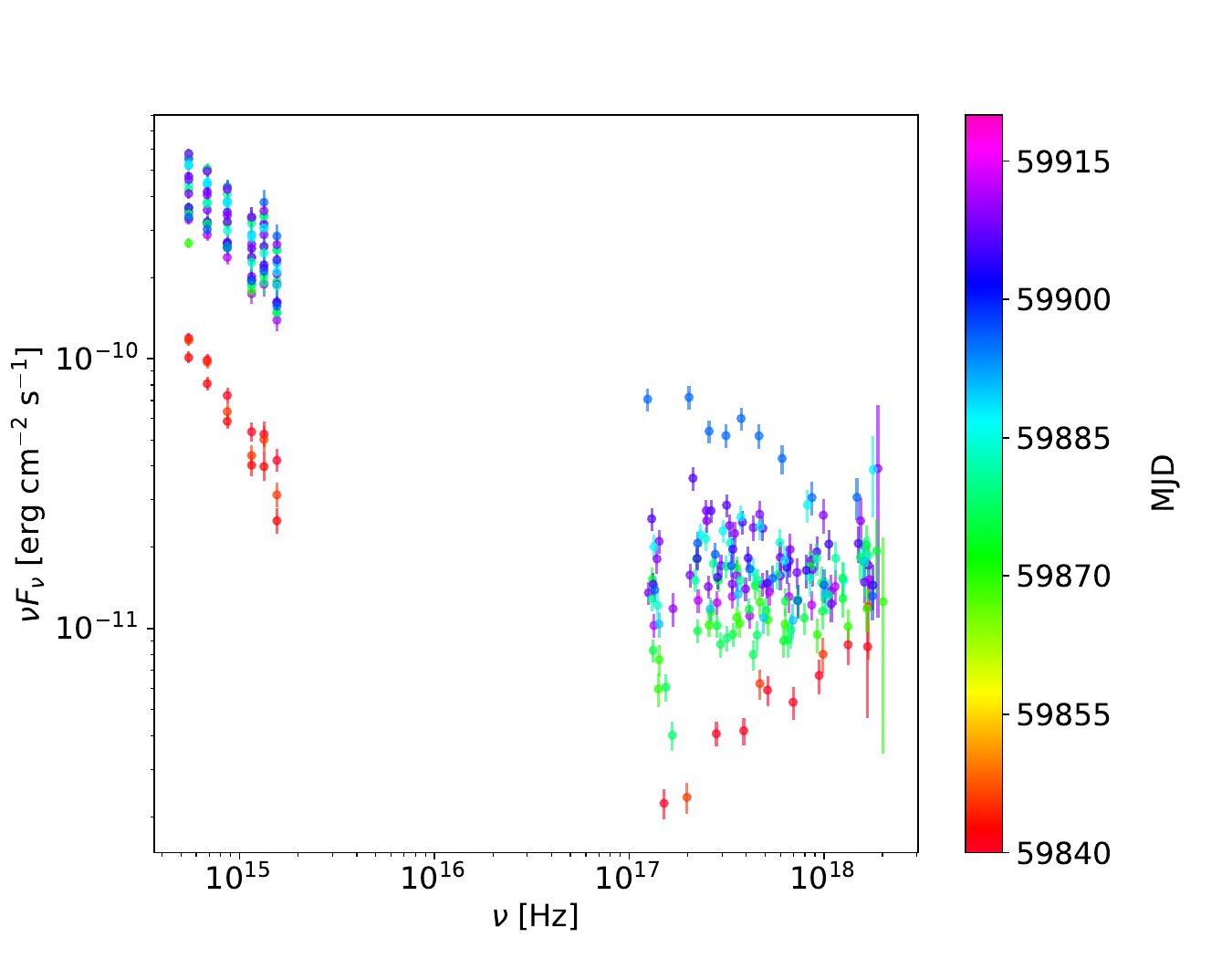}
\includegraphics[width=0.475\textwidth]{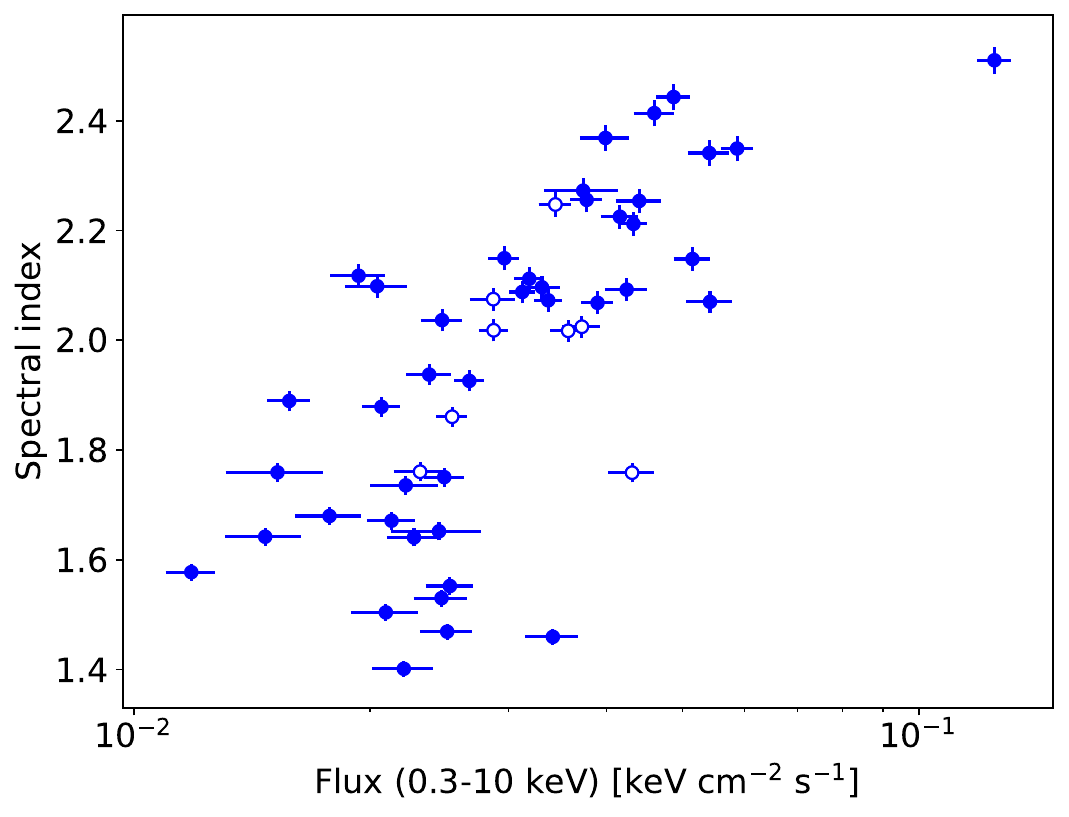}
\caption{Results of the X-ray and UV \textit{Swift} data analysis. \textit{Left:} X-ray and optical-UV SEDs obtained from the \textit{Swift}-XRT/UVOT analyses between September and November 2022. The colour scale corresponds to the MJD of the observation. \textit{Right:} X-ray power-law spectral index as function of the 0.3-10 keV flux. Observations for which a log-parabola shape was preferred  over a power law with a significance $\geq$3$\sigma$ are shown with open markers.}
\label{fig:xrt_uv_spectra}
\end{figure}

In Table \ref{tab:appendix_table_xrt_results} we report the summary of the results of the \textit{Swift}-XRT analysis, presented in Sect. \ref{subsec:XRT_analysis}.
For the observations in which the log-parabola model was preferred over the power-law one with a $\geq$3$\sigma$ significance, Table \ref{tab:appendix_table_xrt_results_pt2} reports the best-fit parameters of the log-parabola model. The {\fontfamily{cmtt}\selectfont logpar} implementation in XSPEC was employed in the fit procedures. The derived SEDs are represented in Fig.~\ref{fig:xrt_uv_spectra}, along with those derived from the UVOT analysis in the optical-UV range.

\begin{table}
\centering
\tiny
\caption{Results of the spectral analysis of the \textit{Swift}-XRT observations of BL Lac during September to November 2022.}
\label{tab:appendix_table_xrt_results}
\begin{tabular}{ccccccccccc}
\hline
\multirow{2}{*}{Day} & 
\multirow{2}{*}{MJD} &
\multirow{2}{*}{ObsID} &
\multicolumn{1}{c}{Exposure} &
\multicolumn{1}{c}{Significance [$\sigma$]} &
\multicolumn{1}{c}{Flux $(0.3-10\,\text{keV})$} &
\multicolumn{1}{c}{Flux $(2.0-10\,\text{keV})$} &
\multirow{2}{*}{Spectral index} \\
\multicolumn{1}{c}{} & 
\multicolumn{1}{c}{} & 
\multicolumn{1}{c}{} & 
\multicolumn{1}{c}{[s]} & 
\multicolumn{1}{c}{Log-parabola model} & 
\multicolumn{1}{c}{[$\times$10$^{-12}\,\text{erg}\,\text{cm}^{-2}\,\text{s}^{-1}$]} &
\multicolumn{1}{c}{[$\times$10$^{-12}\,\text{erg}\,\text{cm}^{-2}\,\text{s}^{-1}$]} &
\multicolumn{1}{c}{} \\
\hline
09/16 & 59838.6 & 00014925014 & 1981 & 1.7 & $53 \pm 5$ & $24 \pm 2$ & $1.81 \pm 0.02$ \\ \hline
09/18 & 59840.5 & 00014925015 & 1994 & 1.9 & $19 \pm 2$ & $11 \pm 1$ & $1.63 \pm 0.02$ \\ \hline
09/20 & 59842.3 & 00014925016 & 572 & 0.9 & $26 \pm 3$ & $13 \pm 3$ & $1.82 \pm 0.02$ \\ \hline
09/23 & 59845.5 & 00014925017 & 705 & 0.3 & $24 \pm 3$ & $14 \pm 2$ & $1.70 \pm 0.02$ \\ \hline
10/17 & 59869.8 & 00014925018 & 1102 & 1.5 & $42 \pm 3$ & $18 \pm 2$ & $2.10 \pm 0.02$ \\ \hline
10/17 & 59869.9 & 00014925019 & 1412 & 0.4 & $40 \pm 3$ & $18 \pm 2$ & $2.00 \pm 0.02$ \\ \hline
10/22 & 59874.4 & 00014925021 & 1978 & 3.7 & $77 \pm 5$ & $24 \pm 3$ & $2.33 \pm 0.02$ \\ \hline
10/23 & 59875.8 & 00014925020 & 1832 & 4.3 & $56 \pm 5$ & $26 \pm 2$ & $1.92 \pm 0.02$ \\ \hline
10/25 & 59877.0 & 00014925023 & 1008 & 3.8 & $55 \pm 6$ & $27 \pm 5$ & $1.82 \pm 0.02$ \\ \hline
10/26 & 59878.3 & 00096990012 & 1034 & 1.9 & $37 \pm 3$ & $23 \pm 2$ & $1.70 \pm 0.02$ \\ \hline
10/29 & 59881.1 & 00014925024 & 1661 & 3.4 & $78 \pm 5$ & $27 \pm 3$ & $2.09 \pm 0.02$ \\ \hline
11/01 & 59884.1 & 00014925026 & 951 & 1.1 & $72 \pm 3$ & $27 \pm 2$ & $2.16 \pm 0.02$ \\ \hline
11/04 & 59887.3 & 00014925027 & 1240 & 2.3 & $87 \pm 5$ & $32 \pm 3$ & $2.22 \pm 0.02$ \\ \hline
11/07 & 59890.1 & 00014925028 & 1035 & 4.2 & $60 \pm 5$ & $24 \pm 3$ & $2.15 \pm 0.02$ \\ \hline
11/12 & 59895.4 & 00096990013 & 849 & 1.9 & $219 \pm 11$ & $50 \pm 3$ & $2.60 \pm 0.03$ \\ \hline
11/13 & 59896.3 & 00014925030 & 1278 & 2.2 & $64 \pm 3$ & $21 \pm 2$ & $2.33 \pm 0.02$ \\ \hline
11/21 & 59904.8 & 00014925032 & 1681 & 2.4 & $66 \pm 3$ & $26 \pm 2$ & $2.14 \pm 0.02$ \\ \hline
11/23 & 59906.0 & 00014925033 & 1447 & 2.7 & $124 \pm 8$ & $34 \pm 3$ & $2.43 \pm 0.02$ \\ \hline
11/25 & 59908.4 & 00096990014 & 865 & 2.0 & $93 \pm 5$ & $27 \pm 2$ & $2.43 \pm 0.02$ \\ \hline
11/25 & 59908.7 & 00014925034 & 2021 & 2.5 & $69 \pm 6$ & $24 \pm 3$ & $2.14 \pm 0.02$ \\ \hline
11/26 & 59909.5 & 00096990015 & 905 & 2.6 & $71 \pm 6$ & $27 \pm 3$ & $2.14 \pm 0.02$ \\ \hline
11/27 & 59910.2 & 00096990016 & 834 & 2.8 & $97 \pm 15$ & $29 \pm 6$ & $2.35 \pm 0.02$ \\ \hline
11/28 & 59911.9 & 00089562001 & 1473 & 3.7 & $61 \pm 5$ & $24 \pm 2$ & $2.09 \pm 0.02$ \\ \hline
11/29 & 59912.2 & 00096990017 & 826 & 2.7 & $98 \pm 8$ & $31 \pm 5$ & $2.33 \pm 0.02$ \\ \hline
11/30 & 59913.4 & 00096990018 & 895 & 3.7 & $84 \pm 13$ & $47 \pm 10$ & $1.82 \pm 0.02$ \\ \hline
12/01 & 59914.1 & 00096990019 & 960 & 3.2 & $80 \pm 8$ & $32 \pm 5$ & $2.09 \pm 0.02$ \\
\hline
\end{tabular}
\tablefoot{For each observation, the following quantities are reported: day, MJD, \textit{Swift} ObsID, {exposure,} intrinsic flux reconstructed in the $0.3-10\,\text{keV}$ and $2.0-10\,\text{keV}$ bands and photon index from the fit assuming a power-law model for the intrinsic source spectrum.
For those observations in which the log-parabola model was preferred over the power-law {at a significance greater than} 3$\sigma$, the reported fluxes were reconstructed from the best-fit log-parabola model.}
\end{table}

\newpage

\begin{table}
\tiny
\centering
\caption{
Results of the spectral analyses of \textit{Swift}-XRT data assuming a log-parabola model for the intrinsic spectrum.}
\label{tab:appendix_table_xrt_results_pt2}
\begin{tabular}{ccccccccccc}
\hline
\multirow{2}{*}{Day} & 
\multirow{2}{*}{MJD} &
\multirow{2}{*}{ObsID} &
\multicolumn{1}{c}{Significance [$\sigma$]} &
\multicolumn{1}{c}{$E_0$} & 
\multirow{2}{*}{$\alpha$} & 
\multirow{2}{*}{$\beta$} &
\multicolumn{1}{c}{$N_0$}
\\
\multicolumn{1}{c}{} & 
\multicolumn{1}{c}{} & 
\multicolumn{1}{c}{} & 
\multicolumn{1}{c}{Log-parabola model} & 
\multicolumn{1}{c}{[keV]} &
\multicolumn{1}{c}{} &
\multicolumn{1}{c}{} &
\multicolumn{1}{c}{[$\times$10$^{-3}\,\text{keV}^{-1}\,\text{cm}^{-2}\,\text{s}^{-1}$]} \\
\hline
10/22 & 59874.4 & 00014925021 & 3.7 & 1 & $2.68 \pm 0.03$ & $-0.720 \pm 0.007$ & $11.7 \pm 0.1$ \\ \hline
10/23 & 59875.8 & 00014925020 & 4.3 & 1 & $2.39 \pm 0.02$ & $-0.808 \pm 0.008$ & $7.3 \pm 0.1$ \\ \hline
10/25 & 59877.0 & 00014925023 & 3.8 & 1 & $2.41 \pm 0.02$ & $-0.993 \pm 0.010$ & $6.2 \pm 0.1$ \\ \hline
10/29 & 59881.1 & 00014925024 & 3.4 & 1 & $2.40 \pm 0.02$ & $-0.587 \pm 0.006$ & $11.2 \pm 0.1$ \\ \hline
11/07 & 59890.1 & 00014925028 & 4.2 & 1 & $2.63 \pm 0.03$ & $-0.918 \pm 0.009$ & $9.1 \pm 0.1$ \\ \hline
11/28 & 59911.9 & 00089562001 & 3.7 & 1 & $2.45 \pm 0.02$ & $-0.658 \pm 0.007$ & $9.0 \pm 0.1$ \\ \hline
11/30 & 59913.4 & 00096990018 & 3.7 & 1 & $2.31 \pm 0.02$ & $-0.812 \pm 0.008$ & $11.6 \pm 0.1$ \\ \hline
12/01 & 59914.1 & 00096990019 & 3.2 & 1 & $2.46 \pm 0.02$ & $-0.686 \pm 0.007$ & $11.6 \pm 0.1$ \\ \hline
\end{tabular}
\tablefoot{The pivot energy was fixed to $E_0=1\,\text{keV}$. Only the observations in which the log-parabola model was preferred over the power-law {at a significance greater than 3$\sigma$} are included in this table. For these observations, the best-fit values of the spectral index $\alpha$, curvature parameter $\beta$ and normalisation $N_0$ are reported.}
\end{table}

The right panel of Fig. \ref{fig:xrt_uv_spectra} shows the scatter plot of the X-ray intrinsic power-law index $\Gamma_{\text{X}}$ and flux $F_{\text{X}}$ reconstructed from the 50 \textit{Swift}-XRT observations of BL Lac from March 2022 to January 2023 (Sect.~\ref{subsec:XRT_analysis}). A softer-when-brighter trend can be highlighted, that is, higher fluxes correspond to softer spectra. The correlation between $\Gamma_{\text{X}}$ and $\log_{10} F_{\text{X}}$ yields a Pearson coefficient of $\rho = 0.70$, with a $p$-value under the null hypothesis of non-correlation of $p = 3 \times 10^{-8}$. Hence, a positive correlation is statistically significant. In addition, we observe that the X-ray photon index transitions from values $\sim$2.5, up to $\sim$1.4, indicating a spectrum dominated by the high-energy emission when $\Gamma_{\text{X}} \lesssim 2.0$, and indicating the emergence of an X-ray synchrotron component for $\Gamma_{\text{X}} > 2.0$. The transition is indeed visible in the SED models presented in Sect.~\ref{sec6} and Appendix~\ref{appendix_e}, as well as in the left panel of Fig.~\ref{fig:xrt_uv_spectra}, where the X-ray spectrum is found in the increasing part of the high-energy SED peak during low emission states, whereas it corresponds to the decreasing part of the synchrotron peak for high-flux periods. For some of the observations, concave log-parabola models were preferred over power-law ones with a significance $\gtrsim$3$\sigma$ to describe the intrinsic X-ray spectrum of BL Lac (Table~\ref{tab:appendix_table_xrt_results_pt2}). This indicates that the $0.3-10$\,keV band hosts the valley resulting from the transition between the low-energy and high-energy radiative components (see left panel of Fig. \ref{fig:xrt_uv_spectra}).

\section{\textit{Swift}-UVOT data results}\label{appendix_swift_uvot}

In Table \ref{tab:appendix_table_uvot_results}, we report the results of the analyses of \textit{Swift}-UVOT data from the observations carried out from September to November 2022.
For each observation, the flux densities in the \textit{Swift}-UVOT photometric bands are reported, corrected for both Galactic extinction and host galaxy contribution (see Sect. \ref{sec2.4} for more details). The optical-UV SEDs over this period are shown in the left panel of Fig.~\ref{fig:xrt_uv_spectra}.

\newpage

\begin{table}[h]
\tiny
\caption{Results of the \textit{Swift}-UVOT photometry of BL Lac in the observations carried out from September to November 2022.}
\label{tab:appendix_table_uvot_results}
\begin{center}
\begin{tabular}{cccccccccc}
\hline
\multirow{2}{*}{Day} & 
\multirow{2}{*}{MJD} &
\multirow{2}{*}{ObsID} &
\multicolumn{1}{c}{Exposure} &
\multicolumn{1}{c}{V} &
\multicolumn{1}{c}{B} &
\multicolumn{1}{c}{U} &
\multicolumn{1}{c}{W1} &
\multicolumn{1}{c}{M2} &
\multicolumn{1}{c}{W2} \\
\multicolumn{1}{c}{} &
\multicolumn{1}{c}{} &
\multicolumn{1}{c}{} &
\multicolumn{1}{c}{[s]} &
\multicolumn{1}{c}{[mJy]} &
\multicolumn{1}{c}{[mJy]} &
\multicolumn{1}{c}{[mJy]} &
\multicolumn{1}{c}{[mJy]} &
\multicolumn{1}{c}{[mJy]} &
\multicolumn{1}{c}{[mJy]} \\
\hline
09/16 & 59838.6 & 00014925014 & 1948 & $26.8\pm1.4$ & $19.5\pm1.1$ & $12.3\pm0.8$ & $6.7\pm0.6$ & $6.2\pm0.7$ & $4.0\pm0.4$ \\ \hline
09/18 & 59840.5 & 00014925015 & 1898 & $18.8\pm1.0$ & $13.2\pm0.8$ & $8.1\pm0.6$ & $4.6\pm0.4$ & $3.9\pm0.4$ & $2.7\pm0.3$ \\ \hline
09/20 & 59842.3 & 00014925016 & 547 & $15.5\pm1.0$ & $10.5\pm0.7$ & $6.4\pm0.5$ & $3.5\pm0.3$ & $3.0\pm0.4$ & $1.6\pm0.2$ \\ \hline
09/23 & 59845.5 & 00014925017 & 680 & $18.5\pm1.1$ & $12.9\pm0.8$ & $7.0\pm0.5$ & $3.8\pm0.4$ & $3.7\pm0.5$ & $2.0\pm0.2$ \\ \hline
10/17 & 59869.8 & 00014925018 & 1450 & $46.0\pm2.0$ & - & - & - & - & - \\ \hline
10/17 & 59869.9 & 00014925019 & 1453 & - & - & - & $15.5\pm1.3$ & - & - \\ \hline
10/22 & 59874.4 & 00014925021 & 1925 & $99.3\pm4.7$ & $72.7\pm3.8$ & $49.6\pm3.3$ & $28.5\pm2.4$ & $25.5\pm2.8$ & $16.2\pm1.6$ \\ \hline
10/23 & 59875.8 & 00014925020 & 1758 & $74.6\pm3.5$ & $54.3\pm2.8$ & $36.5\pm2.4$ & $20.4\pm1.7$ & $19.5\pm2.2$ & $12.4\pm1.2$ \\ \hline
10/25 & 59877.0 & 00014925023 & 961 & $63.4\pm3.0$ & $45.4\pm2.6$ & $30.0\pm2.0$ & $16.4\pm1.4$ & $15.2\pm1.7$ & $9.5\pm1.0$ \\ \hline
10/26 & 59878.3 & 00096990012 & 1009 & $59.8\pm2.9$ & $45.0\pm2.5$ & $29.5\pm2.0$ & $16.2\pm1.4$ & $14.4\pm1.6$ & $9.5\pm1.0$ \\ \hline
10/29 & 59881.1 & 00014925024 & 1610 & $102.1\pm4.8$ & $72.0\pm3.7$ & $46.9\pm3.1$ & $27.5\pm2.3$ & $25.3\pm2.8$ & $16.2\pm1.6$ \\ \hline
11/01 & 59884.1 & 00014925026 & 928 & $78.2\pm3.7$ & $53.8\pm3.0$ & $34.2\pm2.3$ & $19.7\pm1.7$ & $18.5\pm2.0$ & $12.0\pm1.2$ \\ \hline
11/04 & 59887.2 & 00014925027 & 1188 & $92.9\pm4.4$ & $65.0\pm3.6$ & $43.2\pm2.9$ & $24.4\pm2.1$ & $22.8\pm2.5$ & $13.5\pm1.4$ \\ \hline
11/07 & 59890.1 & 00014925028 & 990 & $92.0\pm4.3$ & $63.8\pm3.6$ & $44.0\pm2.9$ & $25.0\pm2.2$ & $22.8\pm2.5$ & $14.5\pm1.5$ \\ \hline
11/12 & 59895.4 & 00096990013 & 826 & $97.4\pm4.6$ & $72.0\pm4.0$ & $49.6\pm3.3$ & $29.0\pm2.5$ & $28.5\pm3.2$ & $18.3\pm1.9$ \\ \hline
11/13 & 59896.3 & 00014925030 & 1254 & $58.1\pm2.8$ & $42.9\pm2.3$ & $29.5\pm2.0$ & $16.8\pm1.5$ & $15.8\pm1.7$ & $10.1\pm1.0$ \\ \hline
11/21 & 59904.8 & 00014925032 & 1655 & $63.4\pm3.0$ & $45.8\pm2.4$ & $30.6\pm2.0$ & $17.0\pm1.5$ & $16.7\pm1.8$ & $10.3\pm1.0$ \\ \hline
11/23 & 59906.0 & 00014925033 & 1397 & $83.7\pm4.0$ & $59.7\pm3.1$ & $40.1\pm2.6$ & $22.2\pm1.9$ & $23.7\pm2.6$ & $15.0\pm1.5$ \\ \hline
11/25 & 59908.4 & 00096990014 & 839 & $71.8\pm3.4$ & $50.9\pm2.9$ & $36.8\pm2.4$ & $20.6\pm1.8$ & $19.5\pm2.2$ & $12.1\pm1.2$ \\ \hline
11/25 & 59908.6 & 00014925034 & 1985 & $61.0\pm2.9$ & $45.0\pm2.4$ & $30.9\pm2.0$ & $17.5\pm1.5$ & $16.2\pm1.8$ & $10.4\pm1.1$ \\ \hline
11/26 & 59909.5 & 00096990015 & 879 & $81.3\pm3.8$ & $58.0\pm3.3$ & $39.0\pm2.6$ & $23.0\pm2.0$ & $21.6\pm2.4$ & $13.3\pm1.3$ \\ \hline
11/27 & 59910.2 & 00096990016 & 809 & $102.1\pm4.8$ & $71.4\pm4.0$ & $48.7\pm3.2$ & $29.0\pm2.5$ & $26.4\pm2.9$ & $17.0\pm1.7$ \\ \hline
11/28 & 59911.9 & 00089562001 & 1446 & $57.0\pm2.7$ & $40.9\pm2.2$ & $27.1\pm1.8$ & $15.1\pm1.3$ & $14.1\pm1.6$ & $8.9\pm0.9$ \\ \hline
11/29 & 59912.2 & 00096990017 & 801 & $62.8\pm3.0$ & $43.7\pm2.5$ & $29.5\pm2.0$ & $17.5\pm1.5$ & $15.6\pm1.8$ & $9.5\pm1.0$ \\ \hline
11/30 & 59913.4 & 00096990018 & 870 & $69.8\pm3.3$ & $48.1\pm2.7$ & $32.3\pm2.1$ & $18.6\pm1.6$ & $17.3\pm1.9$ & $10.6\pm1.1$ \\ \hline
12/01 & 59914.1 & 00096990019 & 914 & $98.3\pm4.6$ & $68.7\pm3.9$ & $47.8\pm3.2$ & $27.5\pm2.4$ & $25.7\pm2.8$ & $15.8\pm1.6$ \\ \hline
\end{tabular}
\end{center}
\end{table}

\section{Multi-wavelength light curve analysis}\label{sec3}

\subsection{Variability}\label{sec3.1}
The HE and VHE $\gamma$-ray flare studied here and shown in Fig.~\ref{fig:mwl_lcs} corresponds to one of the three highest $\gamma$-ray activities from this source, together with the roughly equally bright outbursts {that occurred} on August 2021 and October 2024. The night-wise VHE variability ranges from non-detections during which the emission of the source was low, up to a flux of $\sim$1.9 Crab Units (C.U.), a factor $\sim$10 with respect to the average emission of 0.23~C.U. over the whole period (see Table~\ref{tab:observation_details_lst} for the night-wise results). This variability becomes even more evident when evaluating the flux on a shorter timescale (see Sect.~\ref{sec4} for more details), reaching a maximum emission of 2.8~C.U. above 200~GeV for the night of November 13, and 4.4~C.U. above 100~GeV for October 20.

Extreme variability is also present in the HE emission of BL Lac, with several rises and decays of the flux over time. The LAT data show an extraordinarily bright period, with the flux changing almost a factor 70 from the minimum to the maximum emission detected, and reaching a flux of $\sim$$1.2 \times 10^{-5}$~cm$^{-2}$~s$^{-1}$ on October 16, being amongst the brightest blazar flares ever detected by this instrument. {The LST-1 light curve does not reveal a similar VHE flux increase around October 16,} however we note that the LAT light curve is based on a 12-hour binning and the highest state occurs during day time for LST-1, therefore after the VHE $\gamma$-ray observations. {Hence, we can not exclude the presence of a fast VHE flare such as those observed during October 20 and November 13, with a flux above 200 GeV that significantly increased after the LST-1’s observations.}

As for the X-ray flux between 0.3 and 10 keV, we observe an energy flux variation of a factor $\sim$11, similar to that in the VHE band, ranging from {22 to 2, in units of $10^{-11}$~erg~cm$^{-2}$~s$^{-1}$}. However, we stress that the observing coverage is relatively low especially for the first half of the {campaign}. Nevertheless, during the second half, the X-ray emission seems to follow the same variability pattern as that seen in HE $\gamma$ rays. Remarkably, the highest X-ray emission is detected during a period in which the HE $\gamma$-ray emission is moderate. {Unfortunately, no simultaneous LST-1 observation is available to compare the most energetic $\gamma$-ray emission with the high-energy end of the synchrotron component, located in the X-ray band for this observation, see Sect.~\ref{sec6} and Appendix~\ref{appendix_swift_uvot}.}

{Large} variability is {also} observed in the optical-UV {bands}. Taking the UVOT $w2$ UV band as reference, we measure an average emission of $\sim$8.5 mJy, varying a factor 20 between $\sim$0.9 and 18.8 mJy. The same increase is measured in the optical $R$ band, where the flux density goes from $\sim$8.6 mJy at the lowest state, up to $\sim$192.9 mJy at the peak of the flare. Finally, as commonly observed for $\gamma$-ray blazars, the radio bands show the lowest variability across all bands. At a frequency of 37 GHz, the emission changes from 5.47 and 7.83 Jy, only a factor 1.4, contrasting with the large variability observed at other wavelengths.

\subsection{Correlations}\label{sec3.2}
As shown by Fig.~\ref{fig:mwl_lcs}, the multi-wavelength emission seems to follow similar variability patterns in several bands. In order to {quantify} the possible correlations between the different bands, we have computed the Spearman's linear correlation coefficient $\rho$ for each wavelength with respect to the $R$ band. We use strictly simultaneous data within $\pm$0.1~days to avoid biasing the results as much as possible, with the exception of LAT data, for which we consider the $R$-band observations within each 12-hour bin. 

No significant correlation is observed between the VHE and optical light curves, for which we measure a coefficient $\rho_{R-VHE}=0.31$ {($p$-value $p=0.08$)}. This contrasts with a mild correlation between the HE and VHE $\gamma$-ray bands, that show $\rho_{HE-VHE}=0.53$ {($p$-value $p=0.01$)}. A very high correlation degree is observed between the optical and UV bands, for which we obtain coefficients $\rho_{R-V}=0.99$ and $\rho_{R-w2}=0.98$ ($p$-values $p=4.8\times 10^{-25}$ and $p=6.3\times 10^{-22}$), respectively (using the $V$ and $w2$ bands in each case). A high correlation is also observed between the optical and HE $\gamma$-ray emission, with $\rho_{R-HE}=0.79$ ($p$-value $p=2.5\times 10^{-7}$). The X-ray emission on the other hand shows a slightly smaller value, indicating a hint of correlation with the optical emission. The estimated coefficient for this pair of light curves is $\rho_{R-X-ray}=0.64$ ($p=4.0 \times 10^{-4}$). We have evaluated the cause of this somewhat lower correlation with respect to the optical-UV bands by calculating the correlation coefficient for the soft (0.3-2 keV) and hard (2-10~keV) X-ray bands. The former shows again a very high degree of correlation with the $R$ band ($\rho_{R-soft}=0.81$, $p=6.1 \times 10^{-6}$), whereas the latter shows a lower correlation degree ($\rho_{R-hard}=0.52$, $p=0.002$). This can be attributed to the transition of the X-ray emission from the high-energy to the synchrotron SED bump. The hard X-ray band hosts more frequently the valley between both peaks compared to the soft X-ray band (see Appendix~\ref{appendix_swift_uvot}), 
hence lowering the level of correlation with the optical emission. Finally, the radio emission shows the lowest correlation degree, with $\rho_{R-radio}=0.41$ ($p=0.038$). This is not surprising since, as reported in Appendix~\ref{sec3.1}, the radio emission is the least variable, therefore not following exactly the same variability features as the optical band.

\section{Joint HE-VHE SED fits}\label{appendix_d}
\begin{figure*}[h]
\centering
\includegraphics[width=0.49\columnwidth]{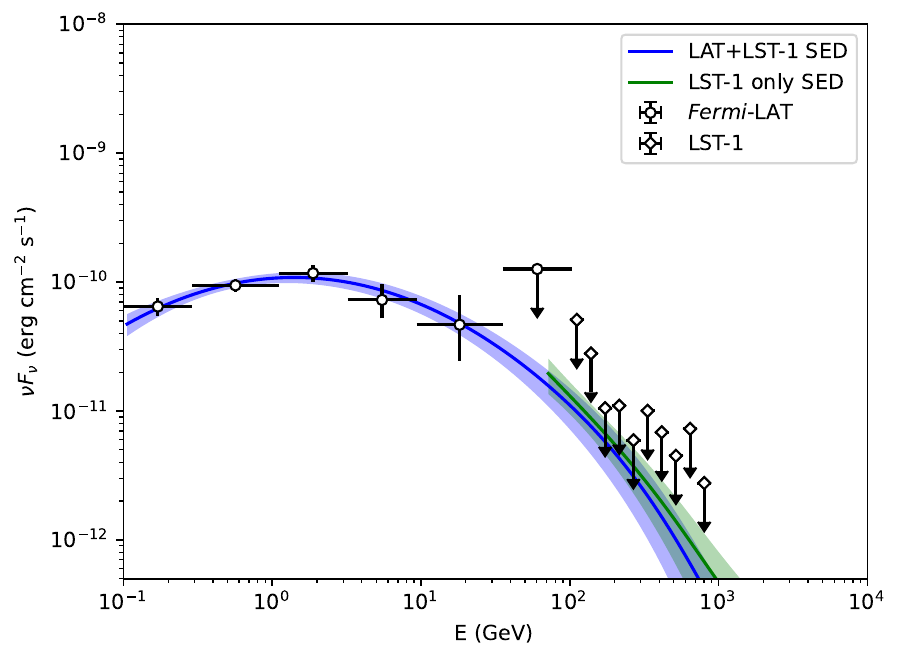}
\includegraphics[width=0.49\columnwidth]{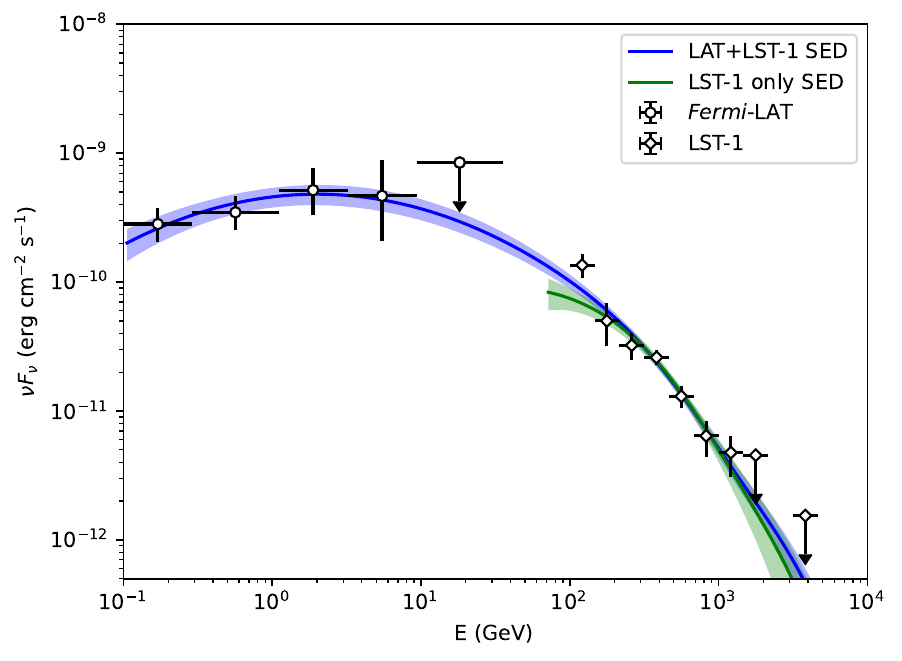}
\includegraphics[width=0.49\columnwidth]{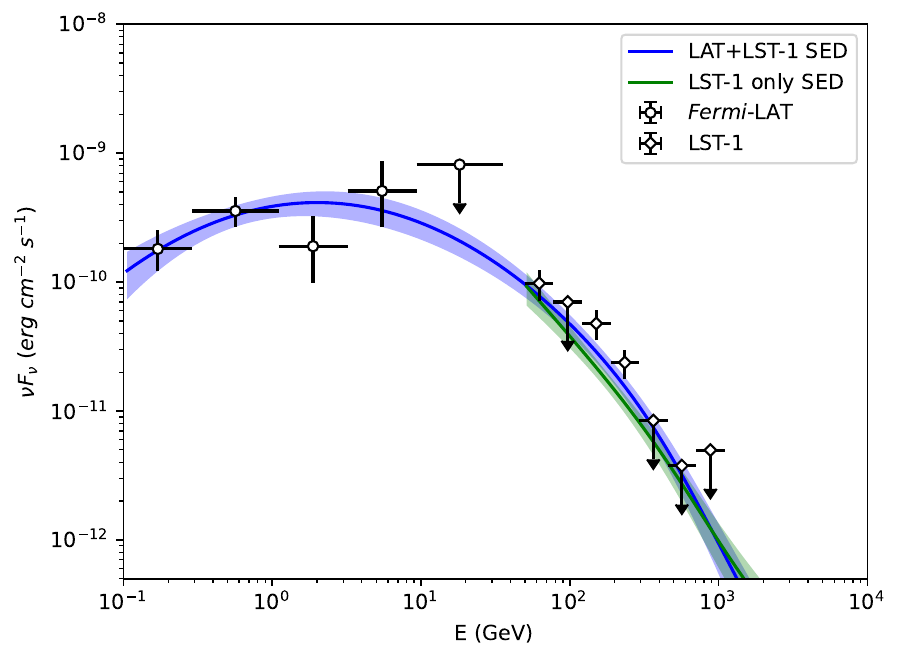}
\includegraphics[width=0.49\columnwidth]{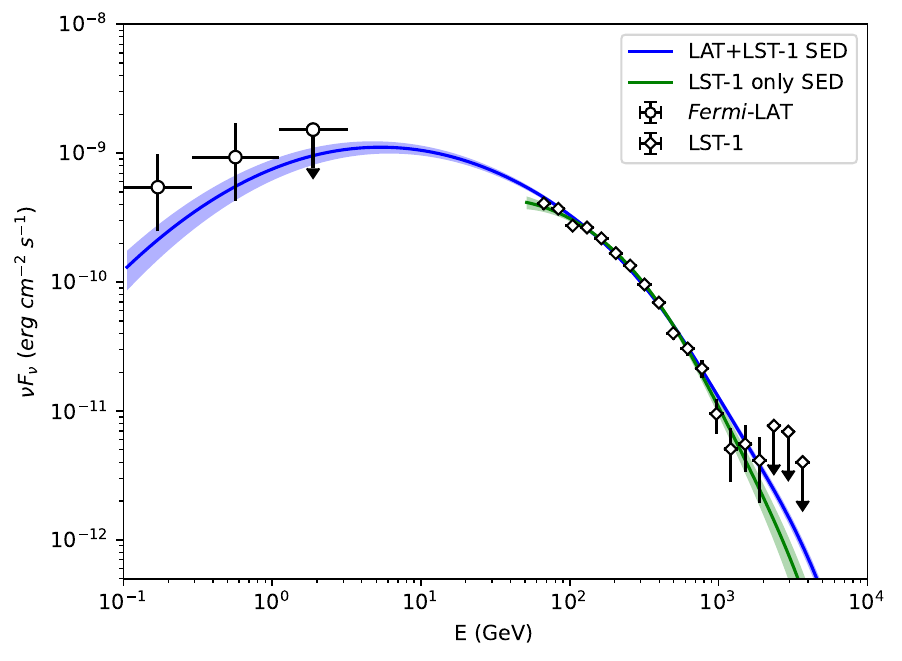}
\caption{Joint LAT-LST-1 SED using a compound log-parabola plus the EBL model from \cite{saldana-lopez2021} for BB1 (top left), BB2 (top right), BB3 (bottom left) and BB4 (bottom right). Same description as Fig.~\ref{fig:joint_fit_bb4}.}
\label{fig:joint_sed_all1}
\end{figure*}

\begin{figure*}[h]
\centering
\includegraphics[width=0.49\columnwidth]{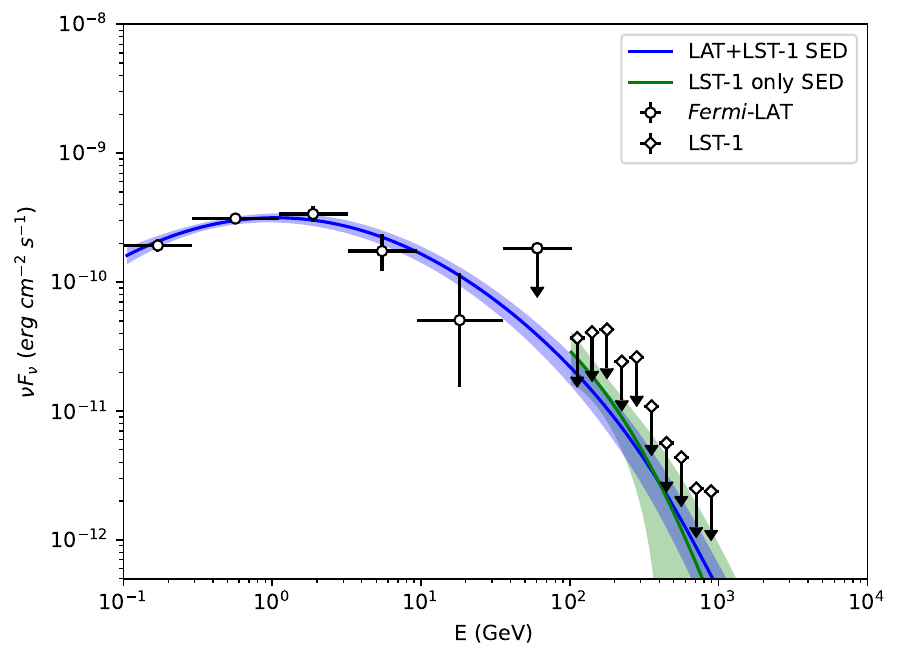}
\includegraphics[width=0.49\columnwidth]{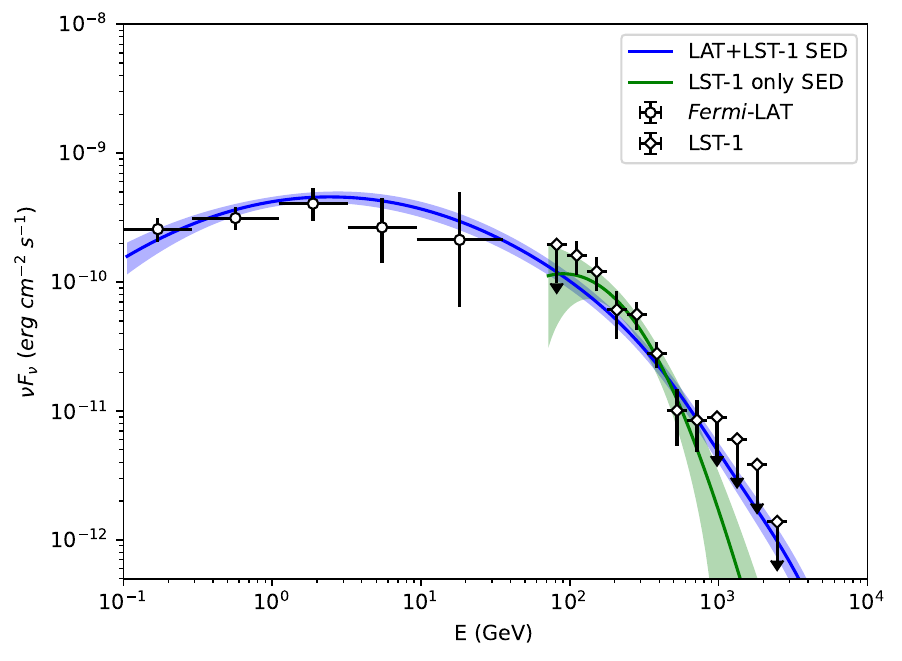}
\includegraphics[width=0.49\columnwidth]{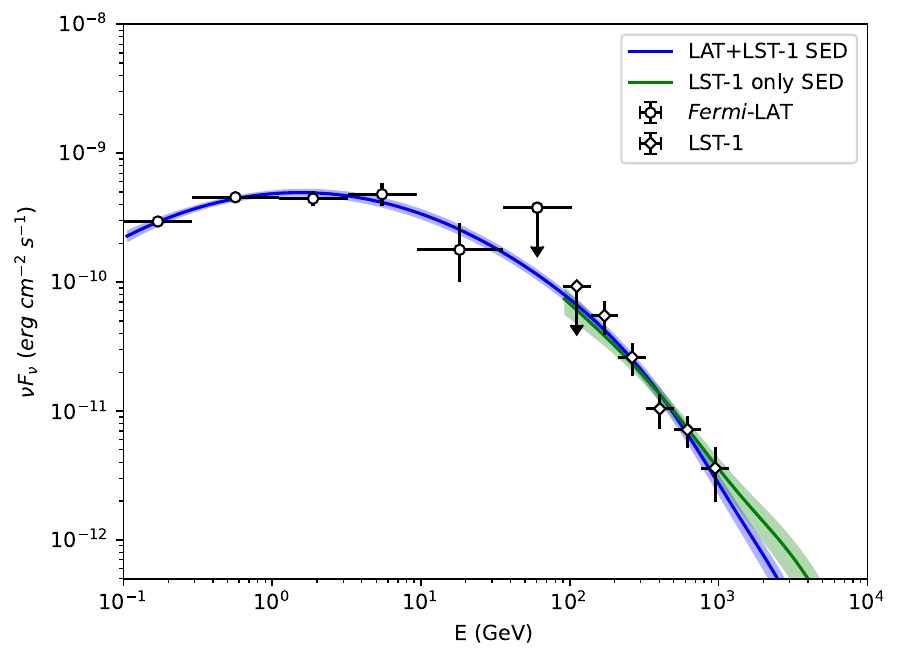}
\includegraphics[width=0.49\columnwidth]{figures/joint_fits/BLLac_BB8_joint_LAT_LST_new_analysis_moon_ok2.pdf}
\includegraphics[width=0.49\columnwidth]{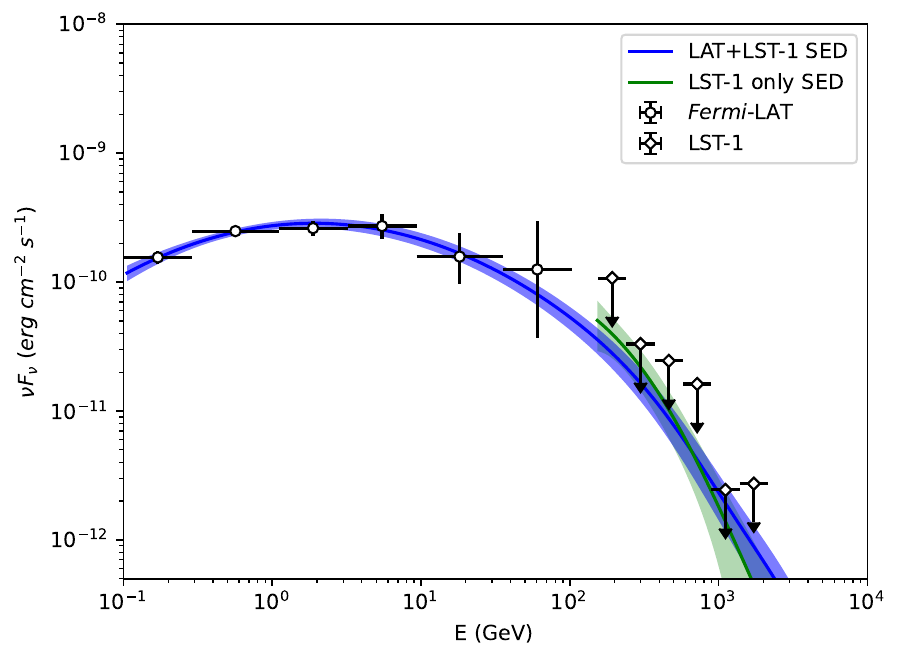}
\includegraphics[width=0.49\columnwidth]{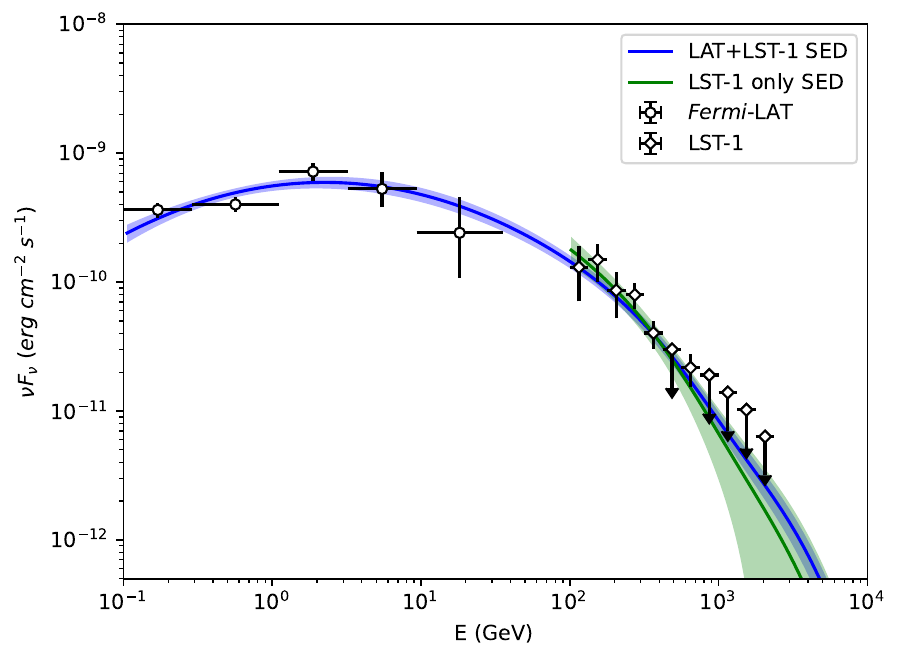}
\caption{Joint LAT-LST-1 SED using a compound log-parabola plus the EBL model from \cite{saldana-lopez2021} for BB5 (top left), BB6 (top right), BB7 (middle left), BB8 (middle right), BB9 (bottom left) and BB10 (bottom right). Same description as Fig.~\ref{fig:joint_fit_bb4}.}
\label{fig:joint_sed_all2}
\end{figure*}

In Figs.~\ref{fig:joint_sed_all1} and \ref{fig:joint_sed_all2} we show the results of the joint LAT-LST-1 spectral characterisation described in Sect.~\ref{sec5}
for all the BBs defined in our analysis. These SEDs correspond to the spectra whose parameters are summarised in Table~\ref{tab:joint_spectra}. The spectral shape of the joint analysis at VHE $\gamma$-ray energies is also compared with that obtained from an LST-only analysis.

\newpage

\newpage

\section{SED models}\label{appendix_e}
\begin{figure*}[h]
\centering
\includegraphics[width=0.47\columnwidth]{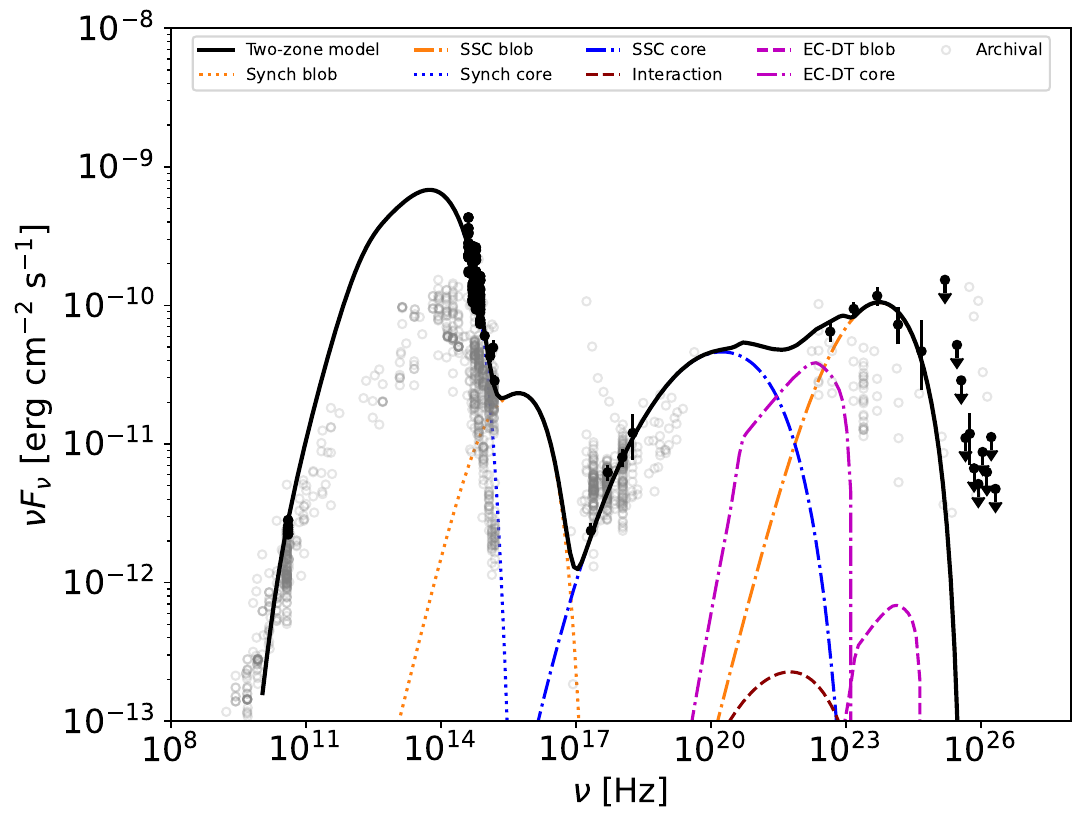}
\includegraphics[width=0.47\columnwidth]{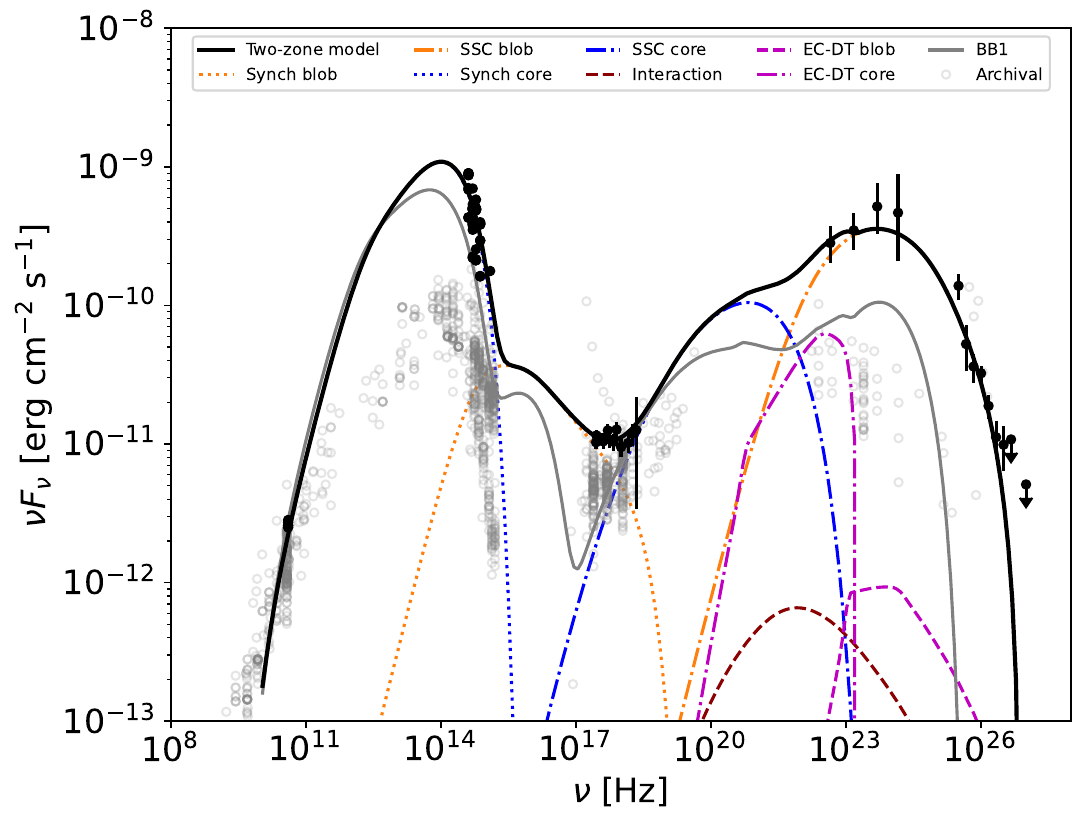}
\includegraphics[width=0.47\columnwidth]{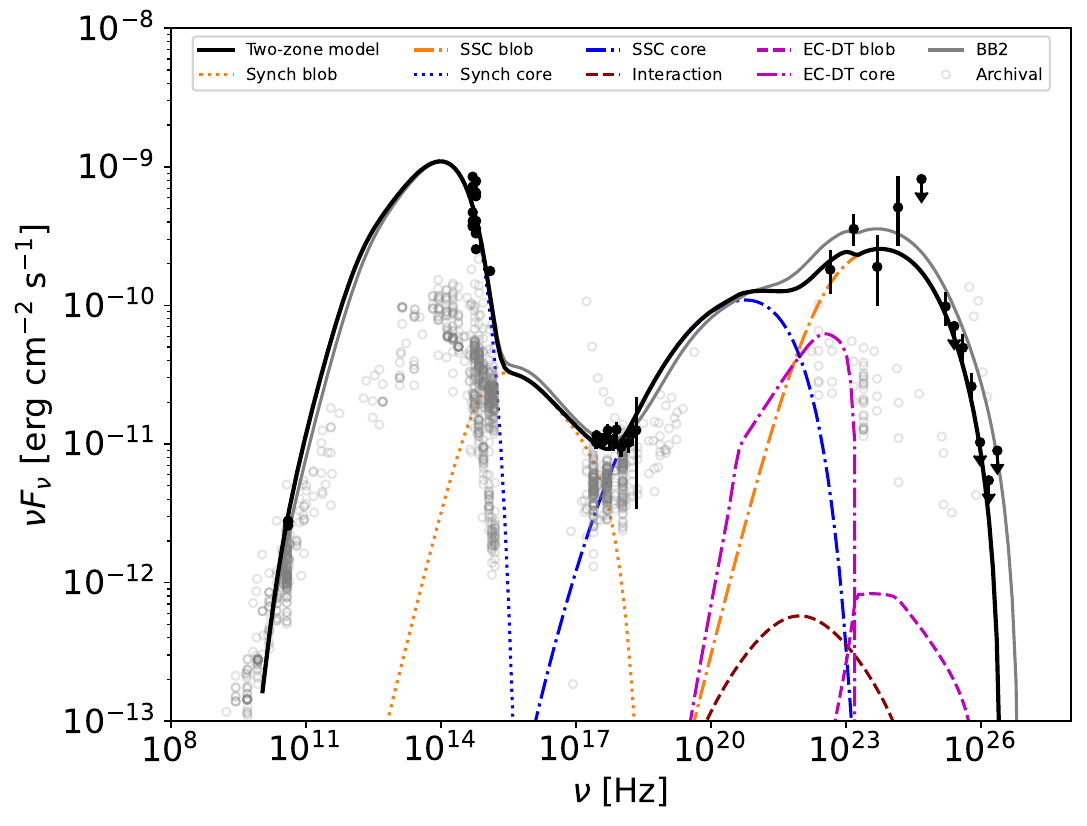}
\includegraphics[width=0.47\columnwidth]{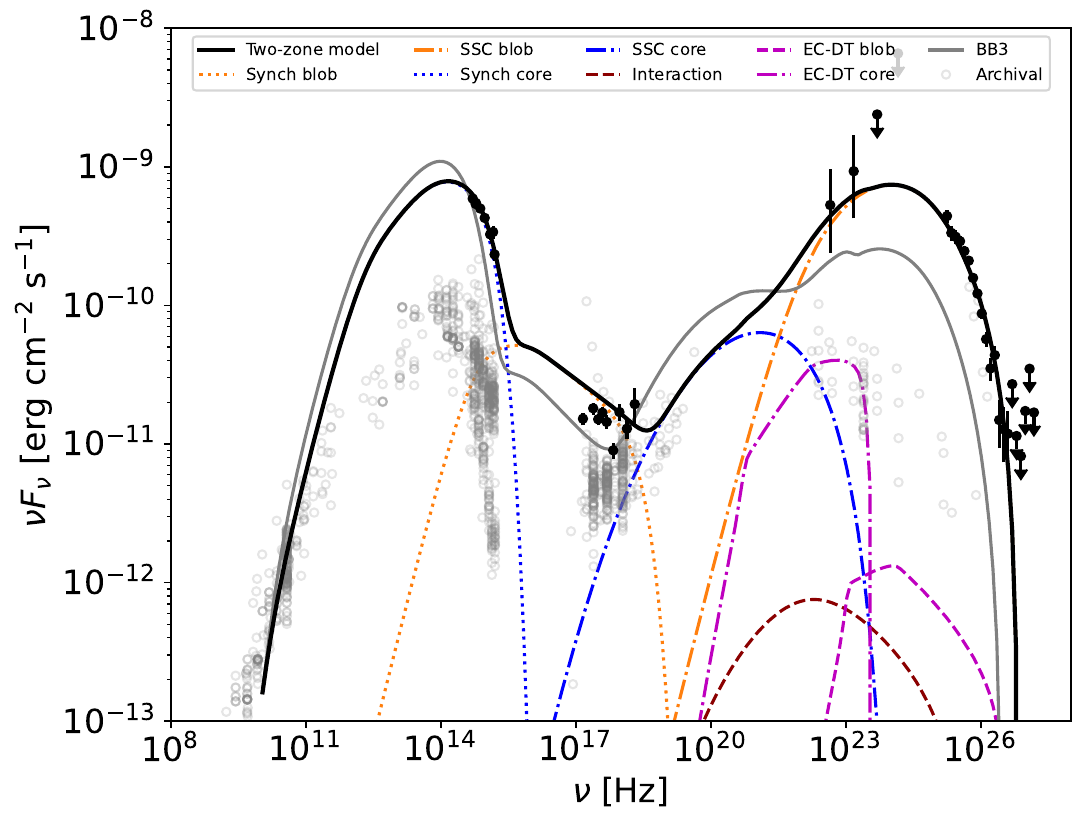}
\includegraphics[width=0.47\columnwidth]{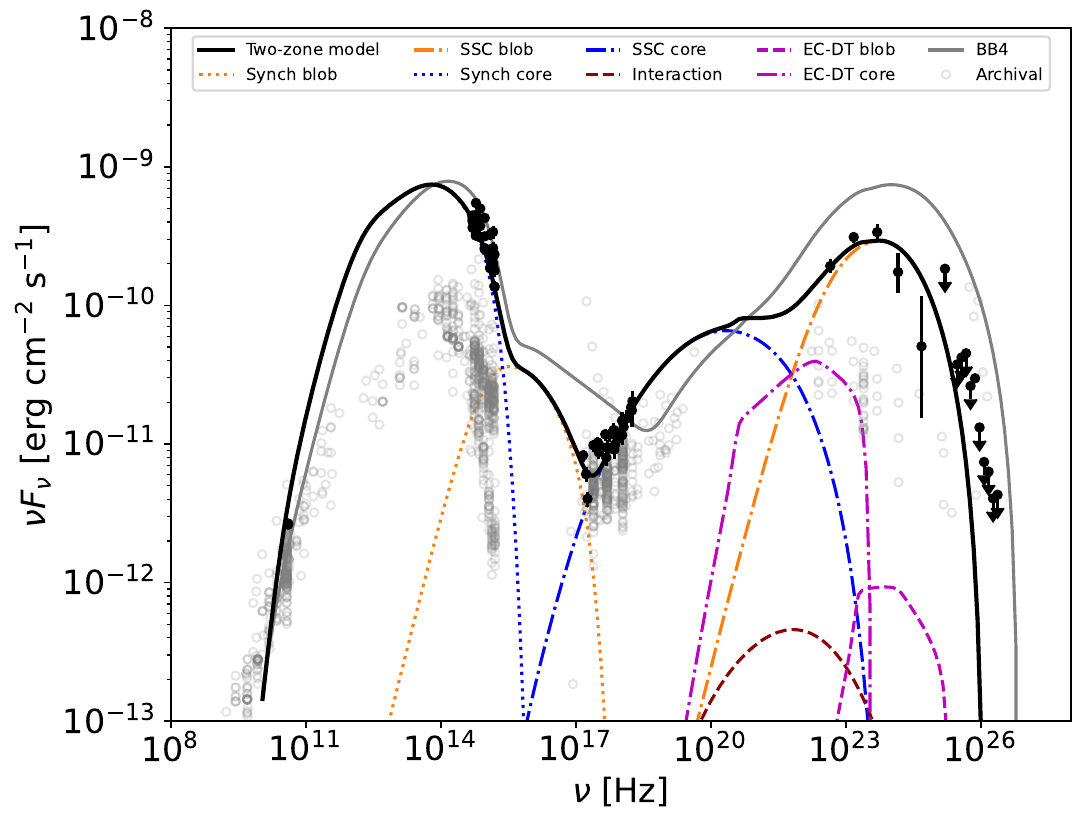}
\includegraphics[width=0.47\columnwidth]{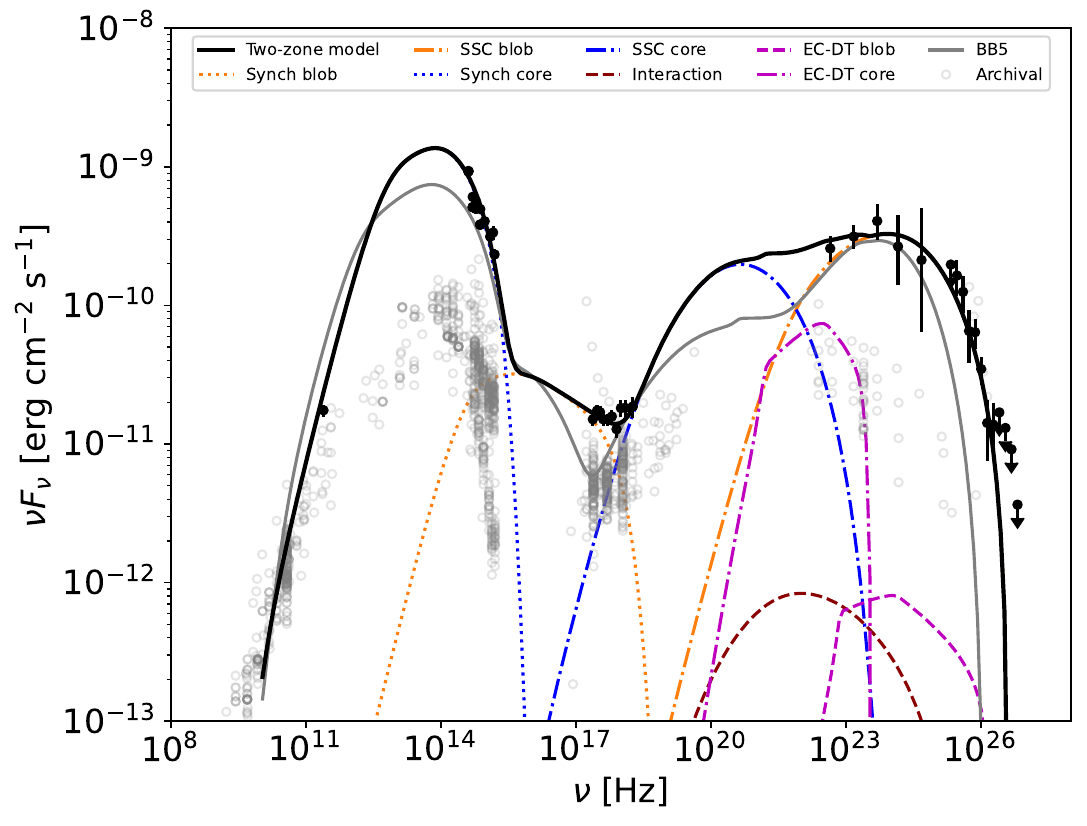}
\caption{Broadband SEDs of BL Lac for BB1 (top left), BB2 (top right), BB3 (middle left), BB4 (middle right), BB5 (bottom left) and BB6 (bottom right). Same description as Fig.~\ref{fig:sed_model_bb8}.}
\label{fig:sed_models}
\end{figure*}

\begin{figure*}
\centering
\includegraphics[width=0.47\columnwidth]{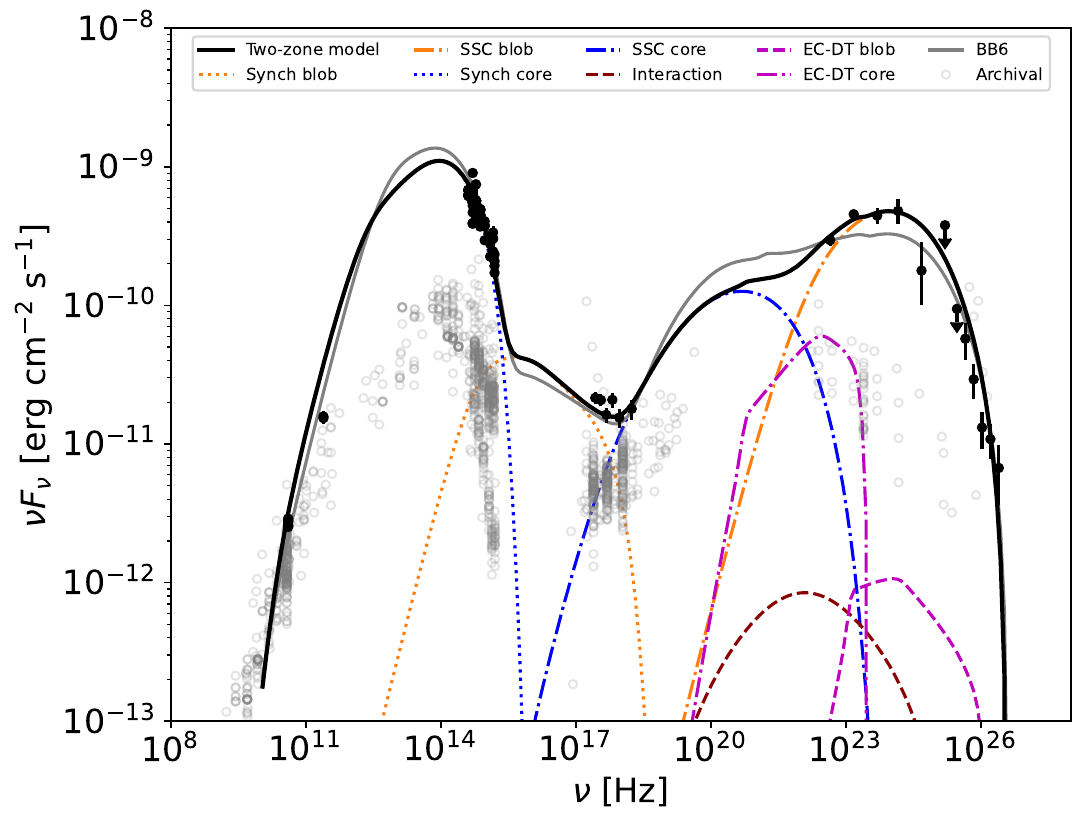}
\includegraphics[width=0.47\columnwidth]{figures/SED_models/final_two_zone_modeling_EC_BLLac_BB8_B0.5.pdf}
\includegraphics[width=0.47\columnwidth]{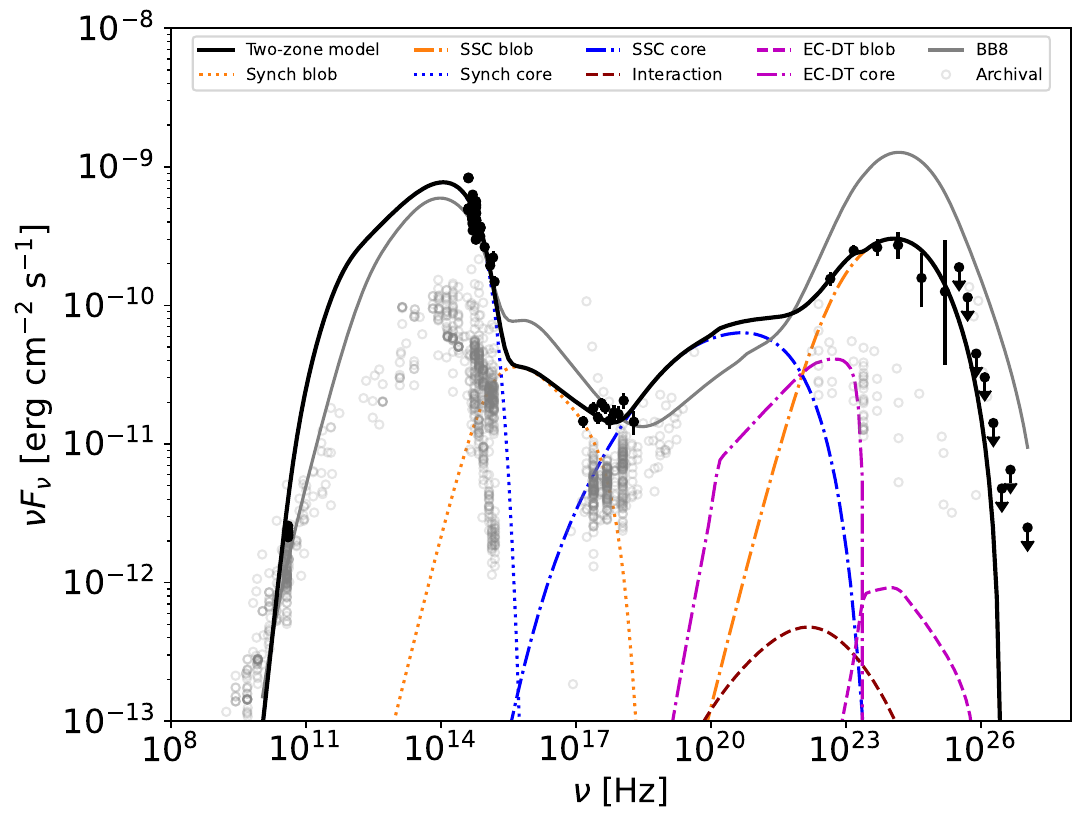}
\includegraphics[width=0.47\columnwidth]{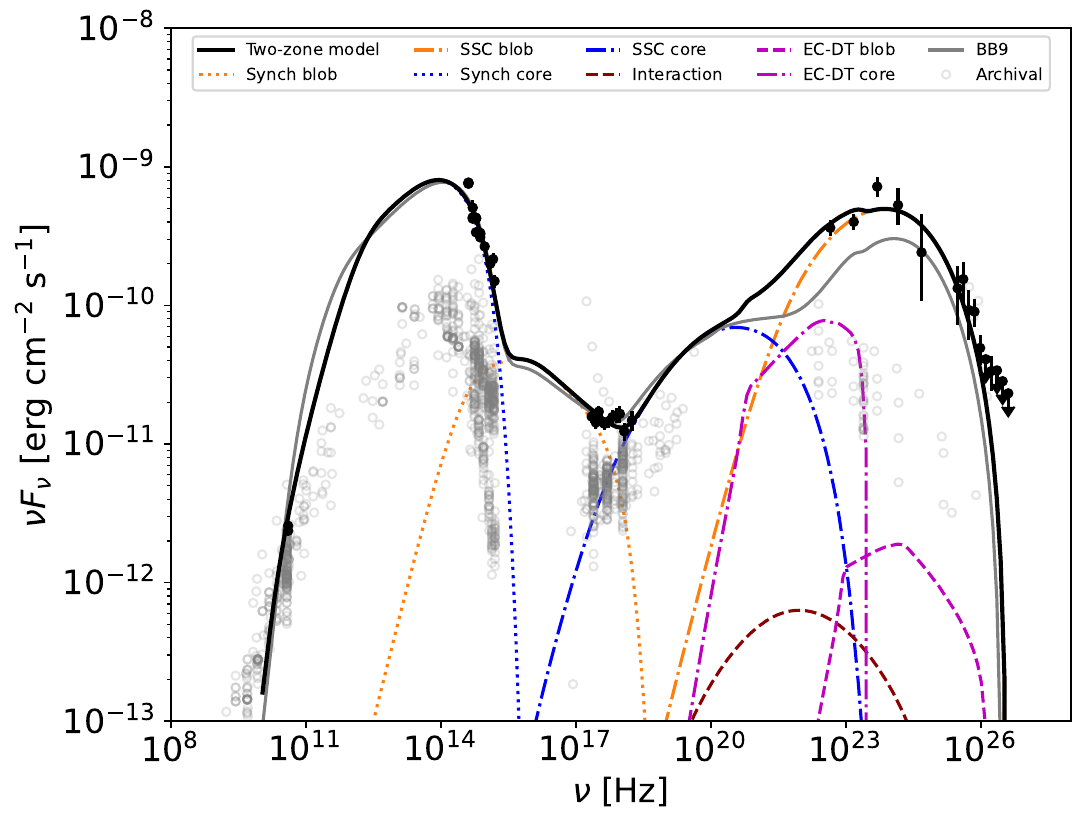}
\caption{Broadband SEDs of BL Lac for BB7 (top left), BB8 (top right), BB9 (bottom left) and BB10 (bottom right). Same description as Fig.~\ref{fig:sed_model_bb8}.}
\label{fig:sed_models2}
\end{figure*}

This appendix gathers the SED models for all 10 BBs defined in Table \ref{tab:joint_spectra}. The fits are represented in Figs.~\ref{fig:sed_models} and \ref{fig:sed_models2}, and correspond to the models defined by the parameters reported in Table~\ref{tab:sedparams}.

\newpage

\section*{Acknowledgements}
\textit{Acknowledgements:} We sincerely thank the anonymous referee for their thorough review of the manuscript and helpful comments.
We gratefully acknowledge financial support from the following agencies and organisations:
Conselho Nacional de Desenvolvimento Cient\'{\i}fico e Tecnol\'{o}gico (CNPq) Grant 309053/2022-6 and Funda\c{c}\~{a}o de Amparo \`{a} Pesquisa do Estado do Rio de Janeiro (FAPERJ) Grants E-26/200.532/2023 and E-26/211.342/2021, Funda\c{c}\~{a}o de Amparo \`{a} Pesquisa do Estado de S\~{a}o Paulo (FAPESP), Funda\c{c}\~{a}o de Apoio \`{a} Ci\^encia, Tecnologia e Inova\c{c}\~{a}o do Paran\'a - Funda\c{c}\~{a}o Arauc\'aria, Ministry of Science, Technology, Innovations and Communications (MCTIC), Brasil;
Ministry of Education and Science, National RI Roadmap Project DO1-153/28.08.2018, Bulgaria;
Croatian Science Foundation (HrZZ) Project IP-2022-10-4595, Rudjer Boskovic Institute, University of Osijek, University of Rijeka, University of Split, Faculty of Electrical Engineering, Mechanical Engineering and Naval Architecture, University of Zagreb, Faculty of Electrical Engineering and Computing, Croatia;
Ministry of Education, Youth and Sports, MEYS  LM2023047, EU/MEYS CZ.02.1.01/0.0/0.0/16\_013/0001403, CZ.02.1.01/0.0/0.0/18\_046/0016007, CZ.02.1.01/0.0/0.0/16\_019/0000754, CZ.02.01.01/00/22\_008/0004632 and CZ.02.01.01/00/23\_015/0008197 Czech Republic;
CNRS-IN2P3, the French Programme d’investissements d’avenir and the Enigmass Labex, 
This work has been done thanks to the facilities offered by the Univ. Savoie Mont Blanc - CNRS/IN2P3 MUST computing center, France;
Max Planck Society, German Bundesministerium f{\"u}r Forschung, Technologie und Raumfahrt (Verbundforschung / ErUM), the Deutsche Forschungsgemeinschaft (SFB 1491) and the Lamarr-Institute for Machine Learning and Artificial Intelligence, Germany;
Istituto Nazionale di Astrofisica (INAF), Istituto Nazionale di Fisica Nucleare (INFN), Italian Ministry for University and Research (MUR), and the financial support from the European Union -- Next Generation EU under the project IR0000012 - CTA+ (CUP C53C22000430006), announcement N.3264 on 28/12/2021: ``Rafforzamento e creazione di IR nell’ambito del Piano Nazionale di Ripresa e Resilienza (PNRR)'';
ICRR, University of Tokyo, JSPS, MEXT, Japan;
JST SPRING - JPMJSP2108;
Narodowe Centrum Nauki, grant number 2023/50/A/ST9/00254, Poland;
The Spanish groups acknowledge the Spanish Ministry of Science and Innovation and the Spanish Research State Agency (AEI) through the government budget lines
PGE2022/28.06.000X.711.04,
28.06.000X.411.01 and 28.06.000X.711.04 of PGE 2023, 2024 and 2025,
and grants PID2019-104114RB-C31,  PID2019-107847RB-C44, PID2019-105510GB-C31, PID2019-104114RB-C33, PID2019-107847RB-C43, PID2019-107847RB-C42, PID2019-107988GB-C22, PID2021-124581OB-I00, PID2021-125331NB-I00, PID2022-136828NB-C41, PID2022-137810NB-C22, PID2022-138172NB-C41, PID2022-138172NB-C42, PID2022-138172NB-C43, PID2022-139117NB-C41, PID2022-139117NB-C42, PID2022-139117NB-C43, PID2022-139117NB-C44, PID2022-136828NB-C42, PID2024-155316NB-I00, PDC2023-145839-I00 funded by the Spanish MCIN/AEI/10.13039/501100011033 and by ERDF/EU and NextGenerationEU PRTR; CSIC PIE 202350E189; the "Centro de Excelencia Severo Ochoa" program through grants no. CEX2020-001007-S, CEX2021-001131-S, CEX2024-001442-S; the "Unidad de Excelencia Mar\'ia de Maeztu" program through grants no. CEX2019-000918-M, CEX2020-001058-M; the "Ram\'on y Cajal" program through grants RYC2021-032991-I  funded by MICIN/AEI/10.13039/501100011033 and the European Union “NextGenerationEU”/PRTR and RYC2020-028639-I; the "Juan de la Cierva-Incorporaci\'on" program through grant no. IJC2019-040315-I and "Juan de la Cierva-formaci\'on"' through grant JDC2022-049705-I; the “Viera y Clavijo” postdoctoral program of Universidad de La Laguna, funded by the Agencia Canaria de Investigaci\'on, Innovaci\'on y Sociedad de la Informaci\'on. They also acknowledge the "Atracci\'on de Talento" program of Comunidad de Madrid through grant no. 2019-T2/TIC-12900; “MAD4SPACE: Desarrollo de tecnolog\'ias habilitadoras para estudios del espacio en la Comunidad de Madrid" (TEC-2024/TEC-182) project, Doctorado Industrial (IND2024/TIC34250) and Ayudas para la contrataci\'on de personal investigador predoctoral en formación (PIPF-2023/TEC-29694) funded by Comunidad de Madrid; the La Caixa Banking Foundation, grant no. LCF/BQ/PI21/11830030; Junta de Andaluc\'ia under Plan Complementario de I+D+I (Ref. AST22\_0001) and Plan Andaluz de Investigaci\'on, Desarrollo e Innovaci\'on as research group FQM-322; Project ref. AST22\_00001\_9 with funding from NextGenerationEU funds; the “Ministerio de Ciencia, Innovaci\'on y Universidades”  and its “Plan de Recuperaci\'on, Transformaci\'on y Resiliencia”; “Consejer\'ia de Universidad, Investigaci\'on e Innovaci\'on” of the regional government of Andaluc\'ia and “Consejo Superior de Investigaciones Cient\'ificas”, Grant CNS2023-144504 funded by MICIU/AEI/10.13039/501100011033 and by the European Union NextGenerationEU/PRTR,  the European Union's Recovery and Resilience Facility-Next Generation, in the framework of the General Invitation of the Spanish Government's public business entity Red.es to participate in talent attraction and retention programmes within Investment 4 of Component 19 of the Recovery, Transformation and Resilience Plan; Junta de Andaluc\'{\i}a under Plan Complementario de I+D+I (Ref. AST22\_00001), Plan Andaluz de Investigaci\'on, Desarrollo e Innovación (Ref. FQM-322). ``Programa Operativo de Crecimiento Inteligente" FEDER 2014-2020 (Ref.~ESFRI-2017-IAC-12), Ministerio de Ciencia e Innovaci\'on, 15\% co-financed by Consejer\'ia de Econom\'ia, Industria, Comercio y Conocimiento del Gobierno de Canarias; the "CERCA" program and the grants 2021SGR00426 and 2021SGR00679, all funded by the Generalitat de Catalunya; and the European Union's NextGenerationEU (PRTR-C17.I1). This work is funded/Co-funded by the European Union (ERC, MicroStars, 101076533). This research used the computing and storage resources provided by the Port d'Informaci\'o Cient\'ifica (PIC) data center.
State Secretariat for Education, Research and Innovation (SERI) and Swiss National Science Foundation (SNSF), Switzerland;
The research leading to these results has received funding from the European Union's Seventh Framework Programme (FP7/2007-2013) under grant agreements No~262053 and No~317446;
This project is receiving funding from the European Union's Horizon 2020 research and innovation programs under agreement No~676134;
ESCAPE - The European Science Cluster of Astronomy \& Particle Physics ESFRI Research Infrastructures has received funding from the European Union’s Horizon 2020 research and innovation programme under Grant Agreement no. 824064.
H.Z. is supported by NASA under award number 80GSFC24M0006.
Some of the data are based on observations collected at the Observatorio de Sierra Nevada; which is owned and operated by the Instituto de Astrof\'isica de Andaluc\'ia (IAA-CSIC).
This article is based on observations made with the IAC80 operated on the island of Tenerife by the Instituto de Astrofísica de Canarias in the Spanish Observatorio del Teide.
This work makes use of observations from the Las Cumbres Observatory global telescope network.
This work makes use of data from the All-Sky Automated Survey for Supernovae (ASAS-SN).
This work has made use of data from the Joan Oró Telescope (TJO) of the Montsec Observatory (OdM), which is owned by the Catalan Government and operated by the Institute for Space Studies of Catalonia (IEEC).
Based on observations obtained with the Samuel Oschin Telescope 48-inch and the 60-inch Telescope at the Palomar Observatory as part of the Zwicky Transient Facility project. ZTF is supported by the National Science Foundation under Grants No. AST-1440341 and AST-2034437 and a collaboration including current partners Caltech, IPAC, the Oskar Klein Center at Stockholm University, the University of Maryland, University of California, Berkeley , the University of Wisconsin at Milwaukee, University of Warwick, Ruhr University, Cornell University, Northwestern University and Drexel University. Operations are conducted by COO, IPAC, and UW.
This publication makes use of data obtained at the Mets\"ahovi Radio Observatory, operated by the Aalto University.
The Submillimeter Array is a joint project between the Smithsonian Astrophysical Observatory and the Academia Sinica Institute of Astronomy and Astrophysics and is funded by the Smithsonian Institution and the Academia Sinica. Maunakea, the location of the SMA, is a culturally important site for the indigenous Hawaiian people; we are privileged to study the cosmos from its summit. \\

\noindent \textit{Author contributions:} D. Cerasole: \textit{Swift}-XRT/UVOT data analysis, SED modelling and discussion, paper edition. 
G. Emery: LST-1 data analysis, systematic evaluation, paper edition.
D. Morcuende: LST-1 data analysis, paper edition. 
J.~Otero-Santos: project coordination, LST-1, \textit{Fermi}-LAT and multi-wavelength data analysis, SED modelling and interpretation, paper edition.
H.~Zhang: theoretical interpretation, paper edition.
The rest of the authors have contributed in one or several of the following ways: design, construction, maintenance and operation of the instrument(s) used to acquire the data; preparation and/or evaluation of the observation proposals; data acquisition, processing, calibration and/or reduction; production of analysis tools and/or related Monte Carlo simulations; discussion and approval of the contents of the draft.

\end{appendix}

\end{document}